\pdfoutput=1
\documentclass[11pt]{article}
\usepackage{jheppub}
\usepackage{amsthm}
\usepackage{amsmath}
\usepackage{amssymb}
\usepackage{epsfig,amssymb,amsmath,psfrag,hyperref}
\usepackage[dvipsnames]{xcolor}
\usepackage{float}
\usepackage{mathtools}
\usepackage[enableskew]{youngtab} 
\usepackage{ytableau} 
\usepackage{slashed} 
\numberwithin{equation}{section}
\usepackage{caption}
\usepackage{booktabs}
\usepackage{mwe}
\usepackage{subfig}
\usepackage{gensymb}
\usepackage{ulem}
\usepackage{cleveref}
\usepackage{xcolor}

\def\gsim{\raise0.3ex\hbox{$\;>$\kern-0.75em\raise-1.1ex\hbox{$\sim\;$}}}

\graphicspath{{./Figs/}}
\begin{document}

$\left. \right.$   \vspace{-18cm}

\title{$\left. \right.$ \vspace{17cm} \\ Gauge/gravity dual dynamics for the strongly coupled sector of composite Higgs models}
\author[a]{Johanna Erdmenger,}
\author[b]{Nick Evans,}
\author[a]{Werner Porod,}
\author[b]{and Konstantinos S.~Rigatos}
\affiliation[a]{Institute for Theoretical Physics and Astrophysics,
Julius-Maximilians-Universit{\"a}t W{\"u}rzburg,\\
97074 W{\"u}rzburg, Germany.}
\affiliation[b]{School of Physics \& Astronomy and STAG Research Centre, University of Southampton, Highfield,\\ Southampton  SO$17$ $1$BJ, UK.}
\emailAdd{erdmenger@physik.uni-wuerzburg.de}
\emailAdd{n.j.evans@soton.ac.uk}
\emailAdd{porod@physik.uni-wuerzburg.de}
\emailAdd{k.c.rigatos@soton.ac.uk}
\abstract{A holographic model of chiral symmetry breaking is used to study the dynamics plus the meson and baryon spectrum of the underlying strong dynamics in composite Higgs models. The model is inspired by top-down D-brane constructions. We introduce this model  by applying it to $N_f=2$ QCD. We compute meson masses, decay constants and the nucleon mass.  The spectrum  is improved  by including higher dimensional operators to reflect the UV physics of QCD. Moving to composite Higgs models,  we impose  perturbative running  for the anomalous dimension of the quark condensate in a variety of theories with varying number of colors and flavours. We compare our results in detail to lattice simulations for the following theories: $SU(2)$ gauge theory with two Dirac fundamentals; $Sp(4)$ gauge theory with fundamental and sextet matter; and $SU(4)$ gauge theory with fundamental and sextet  quarks. In each case, the holographic results  are encouraging since they are close to lattice results for masses and decay constants. Moreover, our models allow us to compute additional observables not yet computed on the lattice, to relax the quenched approximation and move to the precise fermion content of more realistic composite Higgs models not possible on the lattice. We also provide a new holographic description of the top partners including their masses and structure functions. With the addition of higher dimension operators, we show the top Yukawa coupling can be made of order one, to generate the observed top mass.  Finally, we predict the spectrum for the full set of models with top partners proposed by Ferretti and Karateev.}
\maketitle
\setcounter{page}{1}\setcounter{footnote}{0}
\section{Introduction}

The AdS/CFT correspondence was first proposed for conformal field theories such as the ${\cal N}=4$ Super Yang Mills (SYM) theory \cite{Maldacena:1997re, Witten:1998qj}. However, it has created a new paradigm for an effective description of gauge theory, through a five-dimensional gravitational dual, even beyond the conformal case. Non-conformal gauge/gravity dual models have been used extensively to describe theories similar to QCD. For example, chiral symmetry breaking \cite{Babington:2003vm,Kruczenski:2003uq,Sakai:2004cn}, meson masses \cite{Erlich:2005qh,DaRold:2005mxj,Erdmenger:2007cm,Arean:2012mq} and baryon masses\cite{deTeramond:2005su} have all been addressed. This modelling has been more successful than one would expect with sensible predictions of the spectrum and couplings possible at least at the 15\% level, or even better. 
Moreover, the comparison to lattice studies turns out to be convincing. An example of this is the quark mass dependence of QCD, as realized for instance in the dependence of the $\rho$ meson mass on the $\pi$ meson mass \cite{Erdmenger:2007cm,Bali:2007kt,Erdmenger:2014fxa}.
The holographic techniques for QCD described above  can  be extended to other non-abelian gauge theories \cite{Jarvinen:2011qe, Evans:2013vca, Alho:2013dka}. It is natural to apply them to strongly coupled models of physics Beyond the Standard Model (BSM) that have been proposed. For example, holographic work on technicolour includes \cite{Hong:2006si,Hirn:2006nt,Carone:2006wj,Agashe:2007mc,Haba:2008nz,Alho:2013dka,Belyaev:2018jse,Belyaev:2019ybr}.

Another class of BSM models that have generated considerable study are composite Higgs models \cite{Kaplan:1983fs,ArkaniHamed:2002qy}  (the idea that the Standard Model (SM) fields might be composite has a long history - see for example \cite{Terazawa:1976xx}) -  in these models we study the Higgs emerges as a bound 
state of a strongly coupled gauge theory at the 1-5 TeV scale. The composite nature of the Higgs removes the huge levels of fine tuning  in the SM hierarchy problem. In this paper, we will apply holographic methods to survey the full set of gauge theories that may underpin composite Higgs models including \cite{Barnard:2013zea,Ferretti:2014qta} and the exhaustive listing of \cite{Ferretti:2013kya}. We  predict the models' meson spectrum and investigate the properties of top partner baryons. We build on the work in our earlier, short paper \cite{Erdmenger:2020lvq}, expanding the analysis to a much wider set of quantities in the gauge theories previously studied and hugely enlarging the set of gauge theories considered.  The holographic model we use moves beyond simple holographic models such as the Randall-Sundrum \cite{Randall:1999ee}
approach of \cite{Contino:2003ve,Agashe:2004rs} by directly including the running dynamics of a particular UV completion of the model.
 
Holographic models of QCD-like theories split into two types: so-called top down models use the precise tools of the AdS/CFT dictionnary to study deformed versions of ${\cal N}=4$ SYM that display confinement and chiral symmetry breaking. Quark fields have been rigorously included in ${\cal N}=4$ SYM by adding probe D7-branes \cite{Karch:2002sh, Kruczenski:2003be, Erdmenger:2007cm}. These models are usually highly predictive, yet an actual rigorous string dual of QCD does not exist, in particular due to the large $N$ limit involved in holography. Thus the gravity dual theories only exemplify aspects of the dynamics. There are also bottom up models that have been constructed (often called AdS/QCD \cite{Erlich:2005qh,DaRold:2005mxj}) which apply the basic tools of holography but are less rigorous. These models typically contain more free parameters - for example the early models imposed chiral symmetry breaking by hand and the quark condensate was a fitted parameter. More elaborate constructions such as \cite{Gursoy:2008bu} address many of these issues and fit QCD well. 

The model we will use here, Dynamic AdS/YangMills \cite{Alho:2013dka}, lies between the two extremes of top down and bottom up. The action is based on the Dirac Born Infeld (DBI) action of a D7 brane in AdS$_5$ which describes a quenched quark in the top down models. In examples with chiral symmetry breaking based on this action, deformations of the supersymmetric set up induce a running anomalous dimension for the quark condensate, which shows as a radially dependent mass for the scalar that describes the embedding \cite{Alvares:2012kr,Erdmenger:2014fxa}. In the IR the Breitenlohner-Freedman bound \cite{Breitenlohner:1982jf}   is violated and this scalar develops a vaccum expectation value dual to the quark condensate, which is therefore dynamically determined. The DBI action then naturally predicts the spectrum and couplings of a variety of  bosonic and fermionic exitations/states. It is very natural to use this DBI action to describe the quark/meson physics for more complex models by simply feeding it the running anomalous dimension appropriate for those models - although one loses the prediction of the form of this running the spectrum remains a prediction. We will use the two loop running of the couplings in theses theories extended (beyond their formal regime of validity) into the non-perturbative regime to provide sensible ansatz for the runnings in all possible gauge groups and with quarks in all representations. The Dynamic AdS/YM theory can therefore make predictions for the spectrum of the full set of asymptotically free gauge theories proposed as composite Higgs models. A small number of previous holographic analyses of composite Higgs models exist \cite{Contino:2003ve,Agashe:2004rs,Croon:2015wba,Dillon:2018wye} but they do not attempt to include the particular $N_c$ and $N_f$ dependent runnings of the theories in the dynamics. 

Recent work has also shown that it is straightforward to include higher dimension operators (HDOs), such as Nambu-Jona-Lasinio operators \cite{Nambu:1961tp}, into the Dynamic AdS/YM model \cite{Evans:2016yas, Jarvinen:2015ofa}. This is achieved by using Witten's double trace prescription \cite{Witten:2001ua}. We will review this mechanism and explore the role of higher dimension operators in our theories.  In particular we will present a section where we study N$_f=2$ QCD to allow the reader to understand the ball-park success of the holographic model in a familiar setting. Here, to introduce the HDO work, we introduce, in the spirit of \cite{Evans:2005ip,Evans:2006ea},  many HDOs to ``perfect'' the predictions. This should be compared to perfecting a lattice action as introduced by L\"uscher and Hasenfratz long ago \cite{Luscher:1984zf,Hasenfratz:1993sp}. We introduce a UV cut-off corresponding to the scale where QCD transitions to the strong-coupling regime from the perturbative UV. Note that  the gravity dual should be strongly coupled in the region where QCD becomes perturbative above this cut off. We show that HDOs, reflecting the matching at that scale, can improve the spectrum predictions, although with a growing loss of predictivity. 

We will then turn to using our holographic model for composite Higgs models. The key component of composite Higgs models is that a strongly coupled gauge theory causes chiral symmetry breaking in the quark sector, generating four or more Nambu-Goldstone bosons
 \cite{ArkaniHamed:2002qy}. 
By weakly gauging the global chiral symmetries the four then pseudo-Nambu Goldstone bosons (pNGBs) can be placed in the two-dimensional representation of $SU(2)_L$ to become the complex Higgs field. This strong dynamics is expected to happen at a scale of roughly 1-5 TeV. The Higgs Yukawa couplings must be formed by higher dimension operators from a flavour scale above the strong dynamics scale. It has been argued, for example in \cite{Golterman:2015zwa}, that the electroweak gauge fields and top Yukawa interactions in the low energy effective theory of the pNGBs generates the standard model (SM) Higgs potential. We will not address the generation of the Higgs potential here, concentrating instead on the dynamics and spectrum of the strong coupled theory one level higher. The need for higher dimensional operators to give mass to the SM fermions motivates
the study of related operators with a particular focus of their impact on the spectrum of the composite states.

Three theories have had particular focus in the literature. Note we will generically refer to fermions transforming under the strongly coupled gauge theory, in any representation, as quarks, in analogy to QCD (elsewhere they are referred to as hyper-quarks etc).  Firstly, an $SU(2)$ model with two fundamental Dirac fermions breaks an $SU(4)/SO(6)$ global symmetry to $Sp(4)/SO(5)$ generating five Goldstones \cite{Appelquist:1999dq,Lewis:2011zb}. Secondly, an $Sp(4)$ gauge theory with fundamental quarks has the same symmetry breaking pattern \cite{Barnard:2013zea}. Thirdly, an $SU(4)$ theory with five quarks in the sextet representation breaks $SU(5)$ to $SO(5)$ generating fourteen Goldstone modes \cite{Ferretti:2014qta}. We will study these cases in detail and compare to lattice simulations of these theories,  quenched versions or versions with slightly different fermionic content. The comparison is very favourable and leads us to place some trust in our model's predictions as flavours are unquenched or flavours added to make the precise content needed by composite Higgs models. Here we see the huge benefit of holographic models where the field content can be changed rapidly, albeit without the rigour of the lattice. 

The generation of the top quark Yukawa coupling in composite Higgs models is difficult since it is so large. A possible mechanism to enhance it is for the strong dynamics to have baryons with the same symmetries as the chiral top quarks which they mix with via flavour higher dimension operators \cite{Kaplan:1991dc}.  In the $Sp(4)$ model this can be achieved by adding quarks in the sextet representation \cite{Barnard:2013zea}; and in the $SU(4)$ model by adding quarks in the fundamental representation \cite{Ferretti:2014qta} as we will explore in detail. These baryons naturally have order one couplings to the Higgs (pNGBs) generated by the strong dynamics. Even here a Yukawa coupling of order one is hard to achieve requiring anomalously light baryons (phenomenologically they must lie above 800-900 GeV or so \cite{Cacciapaglia:2019zmj,Brooijmans:2020yij},
 dependent on the precise decay channels) and or large structure functions. Here we will investigate this dynamics using holography. The D7-brane action, extended to its fermionic sector, naturally describes baryons (super-partners of the mesons) consisting of three fermions (a quark an anti-quark and an adjoint fermion in the root ${\cal N}=2$ theory) as fermionic fields in the DBI action \cite{Kruczenski:2003be,Kirsch:2006he,Abt:2019tas,Nakas:2020hyo}.  We phenomenologically extend this description to describe the top partners which are also usually constructed from three constituents.  We do indeed find it hard to generate a large top Yukawa coupling in the base theories. Here, as in \cite{Erdmenger:2020lvq}, we propose a novel mechanism of adding an additional new  higher dimension operator that can reduce the top partner masses.  We explore the impact of this operators showing that a physical Yukawa coupling can be achieved by reducing the top partner mass relative to the vector meson mass along with simultaneous enlargements of the relevant structure functions. If the strong coupling scale is $>1$ TeV then the top partner masses are still likely compatible with LHC constraints yet with a top Yukawa coupling of order one.

In particular our new results in these full theories include: meson decay constants beyond lattice analysis to date for the $SU(2)$ model; the meson spectrum and decay constants for the $Sp(4)$ model in the unquenched theory which has not been studied on the lattice; first computations of the axial meson and scalar ($\sigma$ or $S$) meson sectors in an $SU(4)$ theory where other observables have been studied on the lattice (the theory has four Weyl sextet quarks and two  flavours of Dirac fundamental quarks); and the full unquenched spectrum of the true proposed $SU(4)$ composite Higgs model (with five Weyl sextet quarks and three flavours of Dirac fundamental quarks), a theory that is beyond lattice study currently.

We will further exploit the power of holography by computing the predicted spectrum for the full class of twenty six models in the classification of \cite{Ferretti:2013kya}.
Note, that we find that some of these models lie, at least based on the ansatz of the two loop running of the coupling, in the conformal window \cite{Appelquist:1996dq,Dietrich:2006cm} with an infra-red (IR) fixed point that is too small to break chiral symmetries. The scalar meson  mass is particularly sensitive to the rate of running of the coupling in any given theory and some of these proposed models are walking theories with very low scalar masses (as expected from \cite{Evans:2013vca}; but also see \cite{Elander:2020csd,Elander:2020ial,Pomarol:2019aae} for an important discussion of the possible role of mixing with the glueball sector). The ability to see these effects is straightforward holographically but on the lattice needs both unquenched simulations and a wide separation of scales. 

This is a long paper, however it is designed so that the reader can drop in to the self-contained sections of interest to them:

Those interested in the specfics of the holographic model should read \cref{sec: QCD} - we review the model, including determining the vacuum, the meson spectrum computation and the baryon spectrum computation (supplemented by \cref{app: A} where we present the first analysis of the Dirac operator in these models) and describe the addition of higher dimensional operators (HDOs). The reader should note that the base model, without HDOs, only has the parameters $N_c, N_f$ and the quark mass so parameter spaces are directly analogous to the true theories.

In \cref{sec: two_flav_qcd} we apply the model to two flavour QCD. We compute the $\rho$, $a$, $\sigma$, $\pi$ and nucleon masses and decay constants including the quark mass dependence and the dependence on radial excitation number $n$. We apply the idea of perfecting the model through a UV cut off and HDOs. We compare these results to observed values and lattice simulations - the model can match the data at the 10$\%$ level in most observables.

\Cref{sec: composite_higgs_models} is where we turn to composite Higgs models but as before the reader may only be interested in the sub-sections for particular models which are again self-contained. We generically introduce these models and our notation for the classification of  their field content. 

In particular, \cref{sec: cfSU(2)} provides our results for the $SU(2)$ gauge theory, including the  $V$($\rho$), $A$($a$), $S$ ($\sigma$), $\pi$ masses and decay constants including the quark mass dependence. The results correspond well to the  lattice data.

\Cref{sec: sp4 model} provides results for the $Sp(4)$ model including the  $V$, $A$, $S$, $\pi$ and top partner baryon masses and decay constants including the quark mass dependence in each of the fundamental and sextet quark sectors. The holographic results in the quenched theory sit well next to quenched lattice simulations for both representations of quarks, and including the quark mass dependence. We unquench and observe the scalar mass decrease by up to 60\% as the weaker running of the coupling is included. Here we study the role of a HDO in bringing the top partners mass down and observe that the top Yukawa coupling can reasonably be raised towards one by this method.

\Cref{sec: su4_model} repeats this analysis for the $SU(4)$ model. Lattice data only exists for a theory with two flavours of both the fundamental and sextet quarks, but it is unquenched. We study that theory and again find sensible agreement with the lattice in both types of quark sectors. Returning to the correct flavour content for the composite Higgs scenario we find the running is a little slower and the gap between the two quark representations sectors is enhanced a little, with lighter scalar fields. We again see how a HDO can help generate a large top quark Yukawa coupling. 

In \cref{app: catalogue_results} we work through all the remaining models proposed in \cite{Ferretti:2013kya}. We work out whether, at the level of the two loop running, they are expected to break chiral symmetries or not and for those that do present the holographic predictions for the spectrum.

\Cref{sec: pheno_implications} briefly discusses some of the immediate phenomenological implications of our studies and results. 

Finally, we draw our work together and discuss future projects in \cref{sec: discussion}.
\Cref{app: A} explores the Dirac operator in Dynamic AdS/YM and \cref{app: B} lists group theoretic factors for the models.

\newpage

\section{Dynamic AdS/YM } \label{sec: QCD}
In this section we introduce the holographic model that we will use. The model was first suggested in \cite{Alho:2013dka}. Here, we refer to this model as {\it Dynamic AdS/YM}  (Anti-de Sitter/Yang-Mills)   to emphasise that it can be used to holographically describe the chiral symmetry breaking dynamics of any gauge theory (not just QCD), including with quarks in several, potentially inequivalent, representations. 

The action for the model is inspired by  the DBI (Dirac-Born-Infeld) action of a holographic top-down model involving a D7-brane embedded in AdS$_5$ or in a perturbed AdS$_5$ geometry. The DBI action is expanded to quadratic order in the embedding function $X$. A detailed description of this expansion in particular cases  is described in \cite{Alvares:2012kr,Erdmenger:2014fxa}. We also add an axial gauge field in the natural fashion familiar from AdS/QCD models \cite{Erlich:2005qh}. We may  think of this model as describing a single quark in the background of the gauge fields, which may include the contribution to the dynamics from any other quarks even in the probe limit. Note here that the origin of the model at large $N_c$ means the $U(1)_A$ flavour symmetry is not anomalous so the pNGB and so forth form part of the same $U(N_f)$ multiplet along with other flavours. In any case  by placing  fields in the adjoint representation of a flavour symmetry, and by  tracing over the action,  multiple mass degenerate quarks can be included directly. 

In particular, the model has a  field of dimension one, in terms of the gauge theory conformal scalings, for each of the relevant gauge invariant operators. For instance, $X$ is dual to the complex quark bilinear - in QCD this is the operator $\bar{q} q$ but it can be any such dimension three, gauge invariant operator of the theory, as we will shortly expand on. The fluctuations of this field are dual to the $\sigma$ (or $S$ for scalar) and $\pi$ mesons of the theory. We will write it as $X= L e^{i\pi}$.  $A^\mu_L$ and $A^\mu_R$  are dual in QCD to the operators $\bar{q} \gamma^\mu q$ which generates the vector (the $V$ or $\rho$) mesons and $\bar{q} \gamma^\mu \gamma_5 q$ which generates the axial vector ($A$) mesons, respectively. 

Note in theories with quarks in real representations one forms a Majorana spinor from each flavour $\Psi_M=(\psi, -i \sigma^2 \psi^*)$. The gauge invariant and Lorentz invariant condensates are then  as in QCD written $X= \bar{\Psi}_M\Psi_M$ and one still inserts the appropriate gamma matrix structure into the operator $X$ to describe vector and axial vector states - the former carry no charge under the broken symmetry, whilst the latter are charged. Apart from this change in meaning for $X$ the spirit of the gravity description is then the same as in QCD.

  The gravity action of Dynamic AdS/YM is
\begin{align} \label{eq: general_action}
\begin{aligned}
S_{boson} = \int d^5 x ~ \rho^3 &\left( \frac{1}{r^2} (D^M X)^{\dagger} (D_M X) + \frac{\Delta m^2}{\rho^2} |X|^2 + \right. \\
    &\left. \frac{1}{2 g^2_5} \left(\vphantom{\frac{1}{2}} F_{L,MN}F_{L}^{MN} + (L \leftrightarrow R) \vphantom{\frac{1}{2}} \right) \right) \, .
\end{aligned}
\end{align}
The five-dimensional coupling may be obtained by matching to the UV vector-vector correlator \cite{Erlich:2005qh}, and  is given by 
\begin{equation}
g_5^2 = \frac{24 \pi^2}{d(R)~N_{f}(R)} \, ,
\end{equation}
where $d(R)$ is the dimension of the quark's representation and $N_f$(R) is the number of flavours in that representation.

The model lives in a five-dimensional asymptotically AdS (AAdS) spacetime, which is given by 
\begin{align}  \label{thing2}
ds^2 = r^2 dx^2_{(1,3)} + \frac{d \rho^2}{r^2} \, ,
\end{align} 
with $r^2 = \rho^2 +L^2$ the holographic radial direction corresponding to the energy scale, and with the AdS radius set to one. Note in D7 brane models \cite{Karch:2002sh, Kruczenski:2003be, Erdmenger:2007cm} $r$ is the RG scale of the gauge fields and $\rho$ that for quark physics. The factors of $\rho$ and $L$ in the action and metric are  implemented directly from the top-down analysis of the D3/probe-D7 brane system - there $L$ corresponds to the direction perpendicular to the D7 in the 10 dimensional space. The factors ensure appropriate UV behaviour, such that the metric returns to pure AdS at the boundary, but also an IR behaviour where the fluctuations know about any chiral symmetry breaking through a non-zero value of $L$.
 From a bottom up perspective it is natural for $L$ to enter with $\rho$ since $\rho$ and $L$ are both dimension one from the field theory perspective - in a sense \cref{thing2} includes the backreaction of the geometry to the formation of the quark condensate.  $dx^2_{(1,3)}$  is a four-dimensional Minkowski spacetime.

\subsection{The running anomalous dimension \& the vacuum}

The dynamics of a particular gauge theory, including quark contributions to any running coupling, are included through the choice of $\Delta m^2$ in the action \cref{eq: general_action}.
In order to find the vacuum of the theory, with a non-zero chiral condensate,  we set all fields to zero except for $|X| = L(\rho)$. For $\Delta m^2$ a constant, the equation of motion  obtained from \cref{eq: general_action}  is
\begin{align}
\partial_{\rho} (\rho^3 \partial_{\rho} L) - \rho ~ \Delta m^2 L = 0 \label{eq: vacuum qcd} \, .
\end{align}
When  $\Delta m^2=0$, near the boundary of the AAdS space which corresponds to the UV,  the solution  is given asymptotically  by  $L(\rho) = m + c/\rho^2$, with $c=\langle \bar{q}q \rangle$ of dimension three and $m$, the mass, of dimension one (note again $L$ and $\rho$ have dimension one).  For non-zero $\Delta m^2$, the solution  takes the form $L(\rho) = m \rho^{-\gamma} + c\rho^{\gamma-2}$, with 
\begin{equation} 
\Delta m^2 = \gamma (\gamma -2) \label{gammarel}. 
\end{equation}
Here $\gamma$ is precisely the anomalous dimension of the quark mass.  
The Breitenlohner-Freedman (BF) bound below which an instability occurs is given by $\Delta m^2=-1$. 

In the gauge theory, we expect $\gamma$ to run. Therefore we impose this running at the level of \cref{eq: vacuum qcd} by allowing $\Delta m^2$ to depend on $\rho$. Our starting point is the perturbative results for the running of $\gamma$. Expanding \cref{gammarel} at small $\gamma$ gives
\begin{align} \label {dm}
\Delta m^2 = - 2 \gamma.
\end{align}
We proceed by determining $\gamma$ from the gauge theory. Note that this relation means that the holographic model determines a theory to break chiral symmetry if the input form of $\gamma$ passes through 1/2, when the BF bound is violated - we will use this criteria below (matching the assumptions in \cite{Appelquist:1996dq}).

Since  the true running of $\gamma$  is not known non-perturbatively,  we allow ourselves to extend the perturbative results as a function of renormalization group (RG) scale $\mu$ to the non-perturbative regime. We will directly set the field theory RG scale $\mu$ equal to the holographic RG scale $r= \sqrt{\rho^2+L^2}$. Note it is important that we let $\Delta m^2$ depend on $L$ for the following reason. Chirally symmetry breaking occurs in the IR because the $L=0$ state has a BF bound violation at small $\rho$. $L$ then becomes non-zero, the condensate switches on, until the BF bound is not violated any more and the state becomes stable. However, if we did not have $L$ in $\Delta m^2$ then the BF bound would remain violated even as $L$ switches on and  $L$ would grow indefinitely. This mechanism happens naturally in the top down probe D7 systems.  We consider the two-loop results for the running because this ansatz includes the possibility of conformal windows \cite{Appelquist:1996dq,Dietrich:2006cm} for ranges of $N_f$. 

The two-loop result for the running coupling in a gauge theory with multi-representational matter is given by
\begin{align}
\mu \frac{d \alpha}{d \mu} = - b_0  \alpha^2 - b_1  \alpha^3 \, ,
\end{align} 
with 
\begin{align}   \label{running}
\begin{aligned}
b_0 &= \frac{1}{6 \pi} \left(11 C_{2}(G) - 2 \sum_{R} T(R)N_f(R) \right) \, ,\\
b_1 &= \frac{1}{24 \pi^2} \left(34 C^2_{2}(G) - \sum_{R} \left(10 C_{2}(G) + 6 C_{2}(R) \right) T(R) N_f(R) \right) \, .
\end{aligned}
\end{align}
Here we have written the results for the number of Weyl fermion flavours in a given representation.
To find the running of $\gamma$ we then use the one-loop anomalous dimension 
\begin{align} \label{grun}
\gamma = \frac{3~C_2(R)}{2 \pi}~\alpha.
\end{align}
Note we do not go beyond one loop here, since the running at large $\alpha$ is already a guess and moving beyond one loop in $\gamma$ does not provide further features (again we are following the conventions of \cite{Appelquist:1996dq} here).

Now for a given theory we numerically solve \cref{eq: vacuum qcd} with our ansatz for $\Delta m^2$ for the function $L(\rho)$ that defines the vacuum. To do so, we need IR boundary conditions that we again import from the D3/probe D7 brane system. The initial conditions that we use are 
\begin{align} \label{vacIR}
L(\rho)|_{\rho=\rho_{IR}} = \rho_{\text{IR}} \, ,&& \partial_{\rho} L(\rho)|_{\rho=\rho_{IR}} = 0 \, .
\end{align}
The first of these corresponds to an on-shell mass condition: once the IR mass, determined by $L(\rho)|_{\rho=\rho_{IR}}   = L_{\mathrm{IR}} $, equals the energy scale $\rho = \rho_{IR}$, we stop the evolution  of $L(\rho)$ to lower scales, since the quarks should now be integrated out.  Geometrically, $\rho_{\rm{IR}}$ corresponds to the value of $\rho$ at which the function $L(\rho)$ crosses a line at
45$\degree$  in the $L$ - $\rho$ plane. The value of $\rho_{IR}$ is fixed  in each particular theory and for each choice of UV quark mass:  we numerically vary $\rho_{IR}$ until the value of $L$ at  the boundary is the desired quark mass.  We refer to the corresponding configuration  that describes the vacuum (for a given $N_c,N_f,$ and quark mass) as  $L_0(\rho)$ with IR value $L_{IR}$ (this is effectively the constituent quark mass) at the IR cut off $\rho_{IR}$. 

Note at this point we observe a crucial difference between our approach and previous papers on holgraphic composite Higgs models \cite{Contino:2003ve,Agashe:2004rs},  which use the boundary conditions to impose chiral symmetry breaking.  Here though it is not the IR boundary conditions that cause the dynamics that we report. In our case the dynamics results from the BF bound violation (or not)  for $L$ in  the bulk and the IR boundary conditions simply provide IR regularity independent of the model's dynamics.

If there are quarks in multiple representations, then we will simply  replicate \cref{eq: general_action} for each representation. This ignores mixing between the mesons made of quarks in different representations, though different representations are still aware of each other through the choices of $\Delta m^2$. We will discuss such cases and their subtleties in more specific models below.

\subsection{The meson sectors  \label{bsection}}

The mesons of the theory can be found by solving the equations of motion for fluctuations in the various fields of the model in \cref{eq: general_action}. In each case a fluctuation is written as $F(\rho) e^{-ik.x}, ~ M^2 = -k^2$ and IR boundary conditions $F(L_{IR})=1, F'(L_{IR})=0$ used. One seeks the values of $M^2$ where the UV solution falls to zero, so there is only a  fluctuation in the vev of operator and not the source in the UV. 

The fluctuations of $L(\rho)$ give rise to scalar mesons. They are obtained by writing  $L= L_0+S$,  and where to linear order $r^2=\rho^2 + L_0^2$.
 The equation of motion for the fluctuation reads
\begin{equation} \label{eq: eqn of motion_scalar}
\partial_{\rho} (\rho^3 \partial_{\rho} S(\rho)) - \rho (\Delta m^2) S(\rho) - \rho L_{0}(\rho) S(\rho) \frac{\partial \Delta m^2}{\partial L} |_{L_{0}} + M^2 \frac{\rho^3}{r^{4}} S(\rho) = 0.
\end{equation}

The vector-mesons are obtained from fluctuations of the gauge fields $V=A_L+A_R$ around the vacuum value of zero and satisfy the equation of motion
\begin{align} \label{eq: eqn of motion_vector}
\partial_{\rho} (\rho^3 \partial_{\rho}V(\rho)) + M^2_{V} \frac{\rho^3}{r^{4}} V(\rho) = 0.
\end{align}
To obtain a canonically normalized kinetic term for the vector meson we must impose
\begin{equation}  
\int d\rho~ {\rho^3 \over g_5^2 r^4} V^2 = 1. \label{Vnorm}
\end{equation}

The dynamics of the axial-mesons ($A=A_L-A_R$) is described by the $\vec{x},t$ components of $A^N$ by the equations
\begin{equation}  \label{eq: eqn of motion_axial}
\partial_{\rho} (\rho^3 \partial_{\rho} A(\rho)) - g_5^2 \frac{\rho^3 L^2_{0}}{r^2} A(\rho) + \frac{\rho^3 M^2_{A} }{r^{4}} A(\rho) = 0\, . 
\end{equation}
The difference between the $V$ and $A$ equations reflect that $L$ carries axial charge so couples to $A$.

To compute decay constants, we must couple the meson to an external source. Those sources are described as fluctuations with a non-normalizable UV asymptotic form. Again we need to fix the coefficient of these solutions
by matching to the gauge theory in the UV. External currents are associated with the non-normalizable modes of the fields in AdS. In the UV we expect 
$L_0(\rho) \sim 0$ and we can solve
the equations of motion for the scalar, $L= K_S(\rho) e^{-i q.x}$, vector $V^\mu= \epsilon^\mu K_V(\rho) e^{-i q.x}$, and  axial $A^\mu= \epsilon^\mu K_A(\rho) e^{-i q.x}$ fields. Each satisfies the same UV asymptotic equation
\begin{equation}  \label{thing}
\partial_\rho [ \rho^3 \partial_\rho K] - {q^2 \over \rho} K= 0\,. 
\end{equation}
The solution  is
\begin{equation} \label{Ks}
K_i = N_i \left( 1 + {q^2 \over 4 \rho^2} \ln (q^2/ \rho^2) \right),\quad (i=S,V,A),
\end{equation}
where $N_i$ are normalization constants that are not fixed by the linearized equation of motion.
Substituting these solutions back into the action gives the scalar correlator $\Pi_{SS}$, the vector correlator $\Pi_{VV}$ and axial vector correlator $\Pi_{AA}$. Performing the usual matching to the UV gauge theory  requires us to set \cite{Erlich:2005qh,Alho:2013dka} 
\begin{equation}  \label{eq: match}
N_S^2 = {d(R) ~ N_f(R) \over 48 \pi^2 }, \hspace{0.5cm} N_V^2 = N_A^2 = {g_5^2 ~ d(R) ~ N_f(R) \over 48 \pi^2 }. 
\end{equation}
where  $d(R)$ is the dimension of the representation (note here again we write for Weyl fermions so for 2 Dirac flavours $N_f=4$) .

The vector meson decay constant is then given by the overlap term between the meson and the external source
\begin{equation} F_V^2 = \int d \rho {1 \over g_5^2} \partial_\rho \left[- \rho^3 \partial_\rho V\right] K_V(q^2=0)\,.
\label{rhodecay}
\end{equation}

Note here that we are using the notation common in the AdS/QCD literature that the dimension two coupling between the meson and its source is called $F_V^2$. It is common in the phenomenology and lattice literature to call this quantity $\tilde{F}_V M_V$ (see for example \cite{Bennett:2017kga}). Below where we compare to lattice results we must fix this choice. We have converted the lattice results  to our definition of $F_V$ in \cref{rhodecay} which seems a purer statement of the strength of that coupling independent of the prediction of the mass. 
The axial meson normalization and decay constant are  given by  \cref{Vnorm} and \cref{rhodecay} with replacement $V\rightarrow A$.

The pion decay constant can be extracted from the expectation that $\Pi_{AA} = f_\pi^2$, with 
\begin{equation} f_\pi^2 = \int d \rho {1 \over g_5^2}  \partial_\rho \left[  \rho^3 \partial_\rho K_A(q^2=0)\right] K_A(q^2=0)\,.
\end{equation}

To compute the pion mass in the presence of a quark mass we should formally work in the $A_\rho =0$ gauge and write $A_\mu = A_{\mu\perp} + \partial_\mu \phi$. The $\phi$ and $\pi$ fields (the phase of $X$)  mix to describe the pion - we have 
\begin{equation} \label{eq: pion_full}
\begin{split}
\partial_{\rho} (\rho^3 \partial_{\rho} \phi (\rho))  - g_5^2 \frac{\rho^3 L_0^2}{r^4} (\pi(\rho) - \phi(\rho)) &=0\, , \\
q^2 \partial_{\rho} \phi(\rho) - g_5^2 L_0^2 \partial_{\rho} \pi(\rho) &= 0\, .
\end{split}
\end{equation}
Here we shoot out from the IR with $\phi(L_{IR})=1$, $\phi'(L_{IR})=0$, and then vary $\pi(L_{IR})$ and $q^2=-M_\pi^2$ to find solutions where both $\phi, \pi$ vanish in the UV. This is numerically very intensive. Below for  the non-zero quark mass cases, we will neglect the axial meson field to simplify the analysis.  When  substituting the lower equation of \cref{eq: pion_full} into the upper one, we find
\begin{equation}
\partial_{\rho} \left( \rho^3 ~ L_0^2 ~ \partial_{\rho} \pi \right) + M^2_{\pi} \frac{\rho^3~L_0^2}{r^4} \left( \pi - \phi \right) = 0  \, .\label{eq: 4}
\end{equation}
We then assume $\phi \ll \pi$ and neglect the mixing, such that there is only the single equation for $\pi$ to solve as for the other fluctuations. This is the natural description of the pion mass in the D3/probe D7 system before we  added the axial field by hand. As we will see, the results below suggest that this is a sensible approximation.

In a particular SU(4) model we will study below, lattice studies have identified an additional spin zero hadron (a tetraquark). Generically spinless states with UV dimension $\Delta$ can be described by adding to the action an additional scalar field $S$,
\begin{equation}
S = S_{boson} + S_{J} \, , \hspace{1cm} {\rm  with} \hspace{1cm}
S_{J} = \frac{1}{2} ~ \int d^5 x ~ \rho^3 ~ \left( \nabla_{M} \nabla^{M} S + m^2 S^2  \right)\, .
\end{equation}
Fluctuations of this scalar $S= f(\rho) ~ e^{ik \cdot x}$ in the background lead to the equation of motion 
\begin{equation}  \label{eq: eqn of motion_spinless}
\partial^{2}_{\rho} f(\rho) + 5~\frac{\rho + L_0~\partial_{\rho}L_0}{r^2} \partial_{\rho} f(\rho) + \frac{M^2}{r^4} f(\rho) - \frac{\Delta (\Delta-4)}{r^2} f(\rho) = 0, 
\end{equation}
where $\Delta$ is the conformal dimension of the operator that we consider.   

\subsection{The fermionic sector}

One of the first new additions of this work is that we wish to allow for the inclusion of baryonic states in the Dynamic AdS/QCD theory. Here we are motivated by the mass of top partners in composite Higgs models which we will explore more below. Of course, in true large $N_c$ holography baryons made of $N_c$ quarks are very heavy stringy modes (for example described by a wrapped D5 with $N_c$ strings attached in the basic AdS/CFT Correspondence \cite{Witten:1998xy}). However, there are fermionic bound states (baryons) described by the supergravity limit of the duality. In the D3/probe D7 system some fermionic superpartners of the mesons are made of a quark, an anti-quark and a gaugino and are described by fermionic excitations of the D7 world volume theory \cite{Kruczenski:2003be,Kirsch:2006he}. Indeed in a preparatory paper we carefully worked through the D3/probe D7 system example \cite{Abt:2019tas} (see also \cite{Nakas:2020hyo}) and that work will lead us here.  That a three fermion bound state can have such a description in a top down model suggests phenomenologically  a proton made of 3 quarks in QCD or the top partners in the models we will discuss below could reasonable be modelled by simply placing a fermion in the bulk. The work of \cite{deTeramond:2005su} has already trialled this in AdS/QCD with some phenomenological success.

In \cref{app: A} we provide a full derivation for placing a fermion in first AdS and then the Dynamic AdS/YM background. Here we simply summarize the results. We add to the action
\begin{equation} 
S = S_{boson} + S_{1/2}\, , \hspace{1cm} {\rm  with} \hspace{1cm}
S_{1/2} = \int d^5 x ~\rho^3~ \bar{\Psi} \left( \slashed{D}_{\text{AAdS}} - m \right) \Psi\, .
\end{equation}

The four component fermion satisfies the second order equation
\begin{align} \label{eq: eqn of motion_fermions}
\begin{aligned}
\left( \partial_{\rho}^2 + \mathcal{P}_{1} \partial_{\rho} + \frac{M^2_{B}}{r^4} + \mathcal{P}_2 \frac{1}{r^4} -  \frac{m^2}{r^2} - \mathcal{P}_3 \frac{m}{r^3} ~ \gamma^{\rho}   \right) \psi = 0 \, ,
\end{aligned}
\end{align}
where $M_B$ is the baryon mass and the pre-factors are given by
\begin{equation}
\begin{aligned}
\mathcal{P}_1 &= \frac{6}{r^2} \left( \rho + L_0~ \partial_{\rho}L_0 \right)\, , \\
\mathcal{P}_2 &= 2 \left( (\rho^2 + L_0^2) L \partial^2_{\rho} L_0 + (\rho^2 + 3L_0^2) (\partial_{\rho}L_0)^2 + 4 \rho L_0 \partial_{\rho}L_0 + 3 \rho^2 + L_0^2  \right)\, ,\\
\mathcal{P}_3 &= \left( \rho + L_0 ~ \partial_{\rho}L_0   \right)\, .
\end{aligned}
\end{equation}
In  five dimensions for the states of UV dimension $9/2$, as appropriate for a three quark state, the bulk fermion mass is $m=5/2$. 

The four component spinor can then be written in terms of eigenstates of $\gamma_\rho$ such that $\psi = \psi_+ \alpha_+ + \psi_- \alpha_-$ where  $\gamma_\rho \alpha_{\pm} = \pm \alpha_{\pm}$. The equation then becomes two equations, one for $\psi_+$ and one for $\psi_-$, obtained by replacing $\gamma_\rho$ in \cref{eq: eqn of motion_fermions} by $\pm 1$ respectively. The two equations are though copies of the same dynamics with explicit relations between the solutions as we describe in Appendix A. Thus one need solve one only and from the UV boundary behaviour extract the source $\mathcal{J}$ and operator $\mathcal{O}$ values. The UV asymptotic form of the solutions are given by 
\begin{align} 
\begin{split}
\psi_{+} &\sim \mathcal{J} \sqrt{\rho} + \mathcal{O} \frac{M_{B}}{6} {\rho}^{-11/2}\, , \\
\psi_{-} &\sim \mathcal{J} \frac{M_{B}}{4} \frac{1}{\sqrt{\rho}} + \mathcal{O} {\rho}^{-9/2} \, .
\end{split}
\end{align} 

The full solution must be found numerically - here we use the D3/probe D7 system as a guide to impose the IR boundary conditions
\begin{align}
\begin{aligned}
\psi_{+}(\rho = L_{IR}) &= 1, &&& \partial_{\rho} \psi_{+}(\rho = L_{IR}) &= 0\, ,\\
\psi_{-}(\rho = L_{IR}) &= 0, &&& \partial_{\rho} \psi_{-}(\rho = L_{IR}) &= \frac{1}{L_{IR}}\, .
\end{aligned}
\end{align}
Note that we impose these boundary conditions at $\rho=L_{IR}$ rather than at $\rho=0$ as in the supersymmetric case in \cite{Abt:2019tas,Nakas:2020hyo}.

\subsection{Higher dimensional operators} \label{sec:higherdim}

Another key ingredient we wish to explore here is the inclusion of higher dimension quark operators using Witten's double trace prescription \cite{Witten:2001ua,Evans:2016yas}. This prescription amounts to introducing a cut-off at some scale $\Lambda_{UV}$ in the gauge theory or an upper boundary in AdS at $\rho = \Lambda_{UV}$. In the field theory for some operator $\mathcal{O}$ we include a ``double trace''  higher dimensional operator (HDO) by
\begin{equation} {\cal L}_{UV} = G \mathcal{O}^\dagger \mathcal{O}, , \label{ff} \end{equation}
where $G$ is a dimensionful coupling. 
Now were $\mathcal{O}$ to acquire a vacuum expectation value then via \cref{ff} there would be an effective source at the boundary
\begin{equation}  \mathcal{J} = G \langle \mathcal{O}^\dagger \rangle\, . \label{source} \end{equation}
Note that the analysis of \cite{Witten:2001ua,Evans:2016yas} shows that adding the HDO as a boundary term in AdS and then minimizing the bulk and
boundary action naturally reproduces \cref{source}.

Until now we have considered a sourceless theory and in any computation of the background ($L_0(\rho)$) or any fluctuation we have only allowed solutions where the appropriate source vanish. For example, it is precisely this prescription that picks out discrete values of the bound state masses. Now though we will allow all of the solutions with non-zero $\mathcal{J}$ and re-interpret them as part of the source free theory but with the HDO present: asymptotically we read off $\mathcal{J}, \mathcal{O}$ and then use \cref{source} to compute $G$. Now we can sort through these solutions and find the masses of bound states which match the boundary condition for a particular $G$.

The operators we will consider in Dynamic AdS/YM, which we will explore below,  are 
\begin{equation}  \label{qcdhdo} 
{g^2_S \over \Lambda_{UV}^2} |\bar{q} q|^2\, , \hspace{1.5cm} 
{g^2_V \over \Lambda_{UV}^2} |\bar{q} \gamma^\mu q|^2\, , \hspace{1.5cm} 
{g^2_A \over \Lambda_{UV}^2} |\bar{q} \gamma^\mu \gamma_5 q|^2\, , \hspace{1.5cm} 
{g^2_{\rm B} \over \Lambda_{UV}^5} |q q q|^2\, , 
\end{equation}
where the $g_i$ are dimensionless couplings.   \newpage

\section{Two-flavour QCD} \label{sec: two_flav_qcd}
To demonstrate the Dynamic AdS/YM model and the role of HDOs, we begin with a study of $N_c=3, N_f=2$ QCD. We first  determine the vacuum of the theory for the massless theory by finding the function $L(\rho)$ using \cref{eq: vacuum qcd}.
Then we  compute the spectrum of the model by looking at fluctuations, study the quark mass dependence and the $n$ dependence of excited states.  Finally we consider introducing a cut off where the theory runs to a perturbative regime and include HDOs at that scale to improve the IR description.

\begin{figure}[H]
\centering
\includegraphics[scale=0.6]{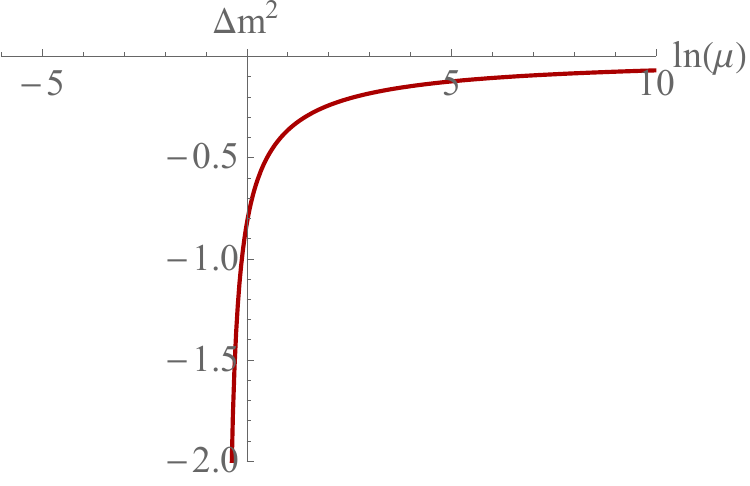} 
\centering
\includegraphics[scale=0.6]{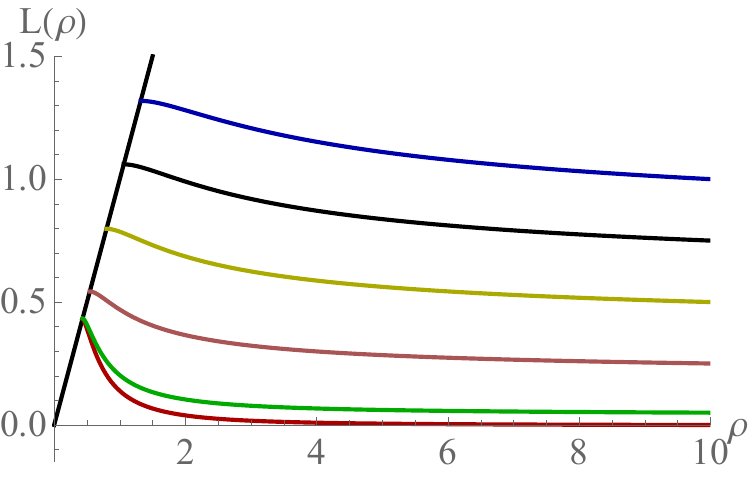} 
\caption{\label{fig: QCD_embedding} The $N_c=3, N_f=2$ QCD model: on the left we display the running of the AdS scalar mass $\Delta m^2$ against log RG scale (we use $\mu=\sqrt{\rho^2+L^2}$ in the holographic model). On the right we show the the vacuum solution for $|X|=L(\rho)$ against  $\rho$.  The 45$\degree$ line is where we apply the on mass shell IR boundary condition in \cref{vacIR}. The $L(\rho)$ with a massless UV quark has $L_{IR}= 0.43$. The quark masses from top to bottom are 1, 0.75, 0.5, 0.25, 0.05, 0. Here units are set by $\alpha(\rho=1)=0.65$.
}
\label{vacuumQCD}
\end{figure}

The key input for any theory we study is the form of $\gamma$ we input in \cref{dm}. The formulae for the one and two-loop coefficients of the $\beta$-function and the one-loop anomalous dimension for QCD are, with $N_f$ the number of Weyl flavours in the fundamental and $\bar{N}_f$ the number in the anti-fundamental representations
\begin{equation} \label{eq: renormalization_qcd}
\begin{aligned}
b_0 &= \frac{1}{6 \pi} \left(11 N_c - (N_f+\bar{N}_f) \right)\, ,  \\
b_1 &= \frac{1}{24 \pi^2} \left(34 {N_c}^2 -5 N_c (N_f+\bar{N}_f) - {3 \over 2} \frac{{N_c}^2-1}{N_c} (N_f+\bar{N}_f)\right)\, , \\
\gamma &= \frac{3({N_c}^2-1)}{4 N_c \pi} \alpha\, .
\end{aligned}
\end{equation}
We choose an initial value for $\alpha(\mu=1)=0.65$ for the numerical analysis but will set the scale with the $\rho$-meson mass below. The resulting running of $\Delta m^2$ in the Dynamic AdS/QCD model is shown in \cref{vacuumQCD} on the left - the BF bound is violated close to the scale $r=\mu=1$.

We can now compute the vacuum for the theory by solving \cref{eq: vacuum qcd} subject to the boundary conditions in \cref{vacIR}. We solve the equation numerically and show the results on the right in  \cref{vacuumQCD} for different asymptotics of $L(\rho)$ corresponding to different UV masses.

\newpage
\subsection{The meson and baryon spectrum of QCD  \label{asection} }

To compute the meson masses, we must set $g_5$ in \cref{eq: general_action} by matching to the UV vector-vector correlator in perturbative QCD
\begin{equation}
g^2_5 = \frac{48 \pi^2}{N_c (N_f+ \bar{N}_f)} \, .
\end{equation}

Having found the massless vacuum, we can now study the spectrum as described in Section 2. We set all sources to zero in the UV. The results for the ground states in each channel are shown at the top of  \Cref{tab: results_qcd_coupling_g5_perturbative} using the $\rho$-meson mass to set the scale. Note we begin to use notation we will use later - labelling the holographic model as AdS/SU(3) to indicate the gauge group and 2F 2 $\bar{F}$ to show there are 2 Weyl fermions in the fundamental and two in the anti-fundamental representation (ie 2 Dirac fermions in the fundamental). Comparing to the physically measured QCD values for the ground states, we see the $\rho$- and $A$-meson sectors are reasonably described but the pion decay constant is low (although we have not yet included a UV quark mass). The $\sigma$ (S) mass is high, but possibly should be compared to the f$_0$(980) if the f$_0$(500) is a pion bound state \cite{Londergan:2013dza} (in which case it fits well). The proton mass is clearly too high though.

\begin{figure}[H]
\includegraphics[scale=0.6]{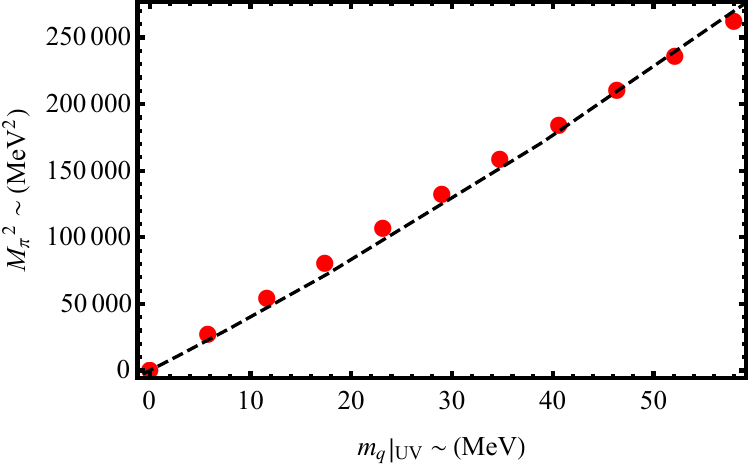} 
\includegraphics[scale=0.6]{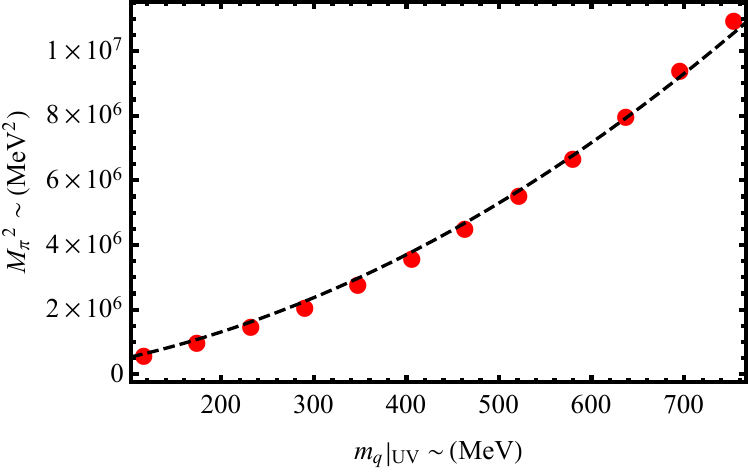}  
\includegraphics[scale=0.65]{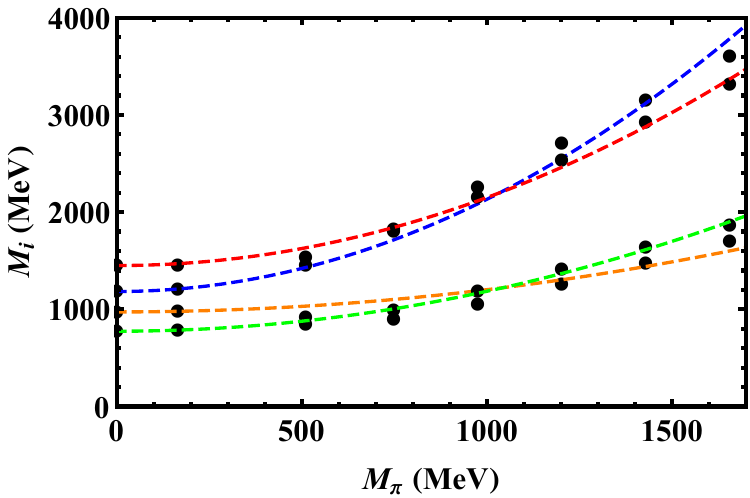} $\left. \right.$ \hspace{8cm}  \vspace{-5.2cm} 

$\left. \right.$ \hspace{8.5cm} 
\begin{tabular}{cl} 
\color{green}$M_{\rho}$ &\color{green}= $\color{green}775 + \color{green}0.00023 ~ \color{green}M^2_{\pi}\, ,$  \\
\color{green}$M_{\rho ~{\rm lat}}$ &\color{green}= $\color{green}770 + \color{green}0.00064 ~ \color{green}M^2_{\pi}\, , $  \\
\color{blue}$M_{A}$ &\color{blue}= $\color{blue}1183.5 + \color{blue}0.00095 ~ \color{blue}M^2_{\pi}\, ,$ \\
\color{blue}$M_{A ~{\rm lat}}$ &\color{blue}= $\color{blue}1230 + \color{blue}0.0015 ~ \color{blue}M^2_{\pi}\, ,$ \\
\color{red}$M_{B}$ &\color{red}= $\color{red}1450.6 + \color{red}0.0007 ~ \color{red}M^2_{\pi}\, ,$\\
\color{red}$M_{B~ {\rm lat}}$ &\color{red}= $\color{red}938 + \color{red}0.0015 ~ \color{red}M^2_{\pi}\, ,$ \\
\color{orange}$M_{S}$ &\color{orange}= $\color{orange}973.4 + \color{orange}0.00023 ~ \color{orange}M^2_{\pi}\, ,$ \\
\color{orange}$M_{S ~{\rm lat}}$ &\color{orange}= $\color{orange}570 + \color{orange}0.0011 ~ \color{orange}M^2_{\pi}\, . $ 
\end{tabular} \vspace{2.cm}

\caption{\label{fig: specta_deformed_vaccum_qcd_pions_analytics}  The top figures show the pion mass squared aginst the UV quark mass. The dashed black lines show the fit we obtained which is given by $M^2_{\pi} = 3871 m_{q} + 13.45 m^2_{q}$. In this formula the pion mass is given in MeV. The bottom plot shows the other meson and baryon masses against $M_\pi$ and the best fits obtained in our model. In these formulae the pion mass is also 
given in MeV.
 }
\end{figure}

We can compute the quark mass dependence of the meson masses also. We display the results in 
\cref{fig: specta_deformed_vaccum_qcd_pions_analytics} including fits and comparisons to lattice data. The top two plots show that at low quark mass the pion mass squared is linear in $m_q$ as required by the Gell-Mann-Oakes-Renner relation whilst at larger $m_q$ the behaviour reverts to depending on $m_q^2$ as for the other mesons.  In the lower plot we show the other meson masses as a function of $M_\pi^2$.  The lattice

\begin{table}[H]   \centering
{\setlength
\doublerulesep{1pt}   
 \begin{tabular}{cccc}
  \toprule[1pt]\midrule[1pt]
    Observables & QCD & AdS/SU(3) & Deviation\\ 
       (MeV)  			& \color{white}{ here's the hidden text } 	& 2 F  2 $\bar{F}$\\ \midrule
    $M_{\rho}$ 			& 775 													& 775$^*$  						&  fitted \\
    $M_{A}$ 			& 1230 													& 1183					& - 4\%\\
    $M_{S}$ 			& 500/990 												& 973  					& +64\%/-2\% \\
    $M_{B}$ 			& 938 													& 1451				& +43\% \\
    $f_{\pi}$  			& 93 														& 55.6						& -50\% \\
    $f_{\rho}$ 				& 345 													& 321 					& - 7\% \\
    $f_{A}$ 				& 433 													& 368						& -16\% \\ &&\\
    $M_{\rho,n=1}$ 		& 1465 													& 1678					&  +14\%\\
    $M_{A,n=1}$ 		& 1655 													& 1922					& +19\% \\
    $M_{S,n=1}$ 		& 990 /1200-1500						& 2009 					&+64\%/+35\% \\
    $M_{B,n=1}$ 		& 1440 													& 2406				& +50\% \\
    \midrule[1pt]\bottomrule[1pt]
  \end{tabular} 
  }   
  \caption{\label{tab: results_qcd_coupling_g5_perturbative} The predictions for masses and decay constants (in MeV) for $N_f=2$ massless QCD. The $\rho$-meson mass has been used to set the scale (indicated by the *). }
\end{table}

  \noindent data is extracted by eye from the plots in for example \cite{Lin:2008pr,Briceno:2016mjc,Werner:2019hxc}  so we don't give errors - they provide a guide to the expected order of magnitude. Note the coefficients are dimensionful so depend on the choice for the setting of the scale. The comparison is reasonable at the level of a factor of two except for the $\sigma$ where our estimate of the mass is high and the gradient low, perhaps reflecting the difficulty with identifying the state we have already encountered.  

Finally it is also interesting to look at the masses of higher excited states of the mesons. We are wary of this comparison - at infinite $N_c$ the AdS/CFT description of excited states remains a point-like supergravity description whilst in QCD, at lower $N_c$, we expect, as the quarks separate, the confining strings between them to become apparent \cite{Shifman:2005zn}. One might therefore only expect the lowest excited state(s) to be well described by the methods we are using.   It has been argued that the excited state masses should scale with the excitation number $n$ as $\sqrt{n}$ \cite{Shifman:2005zn} whilst in standard AdS/QCD models they scale as $n$. In \cite{Karch:2006pv,Evans:2015qaa} it was 

\begin{figure}[H]
\centering
\includegraphics[scale=0.6]{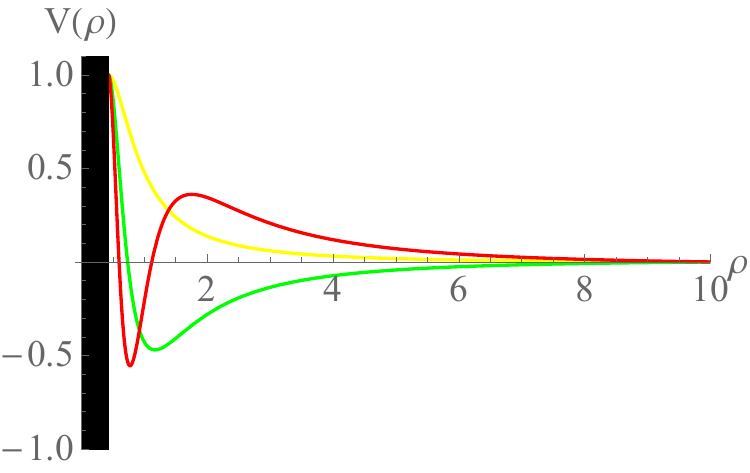} 
\centering
\includegraphics[scale=0.6]{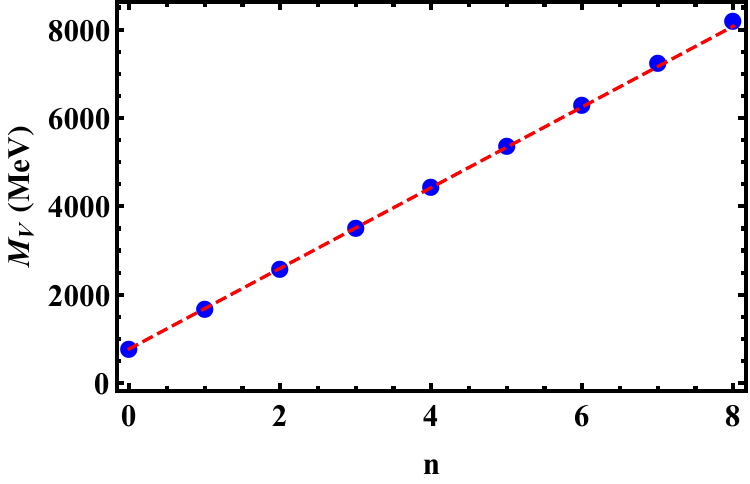} 
\caption{\label{fig: vector_meson_qcd} On the left, we show the normalizable solutions to the equations of motion for the vector meson (the black rectangle covers the region below the IR cut off). They are obtained for $M_{V}=1.337 (775$ MeV$), 2.895 (1677.9$ MeV$), 4.45 (2578.9$ MeV$)$. On the right we show the numerical masses - blue dots - and the spectral curve that we obtained by fitting to the first six states and verified against the next three. Here, $n$ is the number of nodes of the wave-functions.
}
\end{figure}

\noindent argued that rather dramatic changes to the deep IR would be needed to make highly excited states scale as $\sqrt{n}$ - this approach is not obviously reintroducing string like behaviour though. So it is interesting to look at the low lying $n$ masses in our description. In \cref{fig: vector_meson_qcd} we show the wave functions for the first few excited $\rho$-meson states and plot the masses against $n$. In fact they are rather linear in $n$ and the model, unsurprisingly, does not capture the string like behaviour. We display the values of the first excited states in \Cref{tab: results_qcd_coupling_g5_perturbative} where they come out high.  Below we will take a different approach to adding string like structure back into the model by including HDOs which does seem to improve the predictions for at least the $n=1$ states as we will see.

\subsection{The nucleon-$\sigma$ Yukawa coupling} \label{sec: yukawa_qcd}

A further important quantity is the nucleon $\sigma$ Yukawa coupling strength, which we estimate here. 
We must normalize the  kinetic term of the scalar and the baryon using
\begin{equation}\label{Bnorm}
\mathcal{N}_{S} ~ \int d\rho ~ \frac{\rho^3}{ (\rho^2 + L_0^2)^2} ~ S^2(\rho) = 1, \hspace{1cm}
\mathcal{N}_{B} ~ \int d\rho ~ \frac{1}{ \sqrt{ \rho^2 + L_0^2 }} ~ \psi^2(\rho) = 1\, . 
\end{equation}

The precise expression for the dynamically determined Yukawa coupling would depend on the action mixing the $L$ and $\psi$ fields beyond quadratic order and there are a number of terms one could write on dimensional grounds with free couplings. An example term that will contribute is
\begin{equation} \label{eq: yukawa_qcd}
 y_{NN\sigma} = \left| \int d\rho ~ \rho^3  {\partial_\rho S ~\partial_\rho L_0 ~\psi^2 \over (\rho^2 + L_0^2)^2} \right|\, .
 \end{equation}
For this case, we find $y_{NN\sigma}=1.47$,  which is of order one as one might expect. We stress again though that while this is indicative of the expectation that the coupling will be of order one, it is not a prediction because we can multiply by an arbitrary coupling in our holographic model. 

\subsection{Higher dimensional operators}
\label{sec:higherdimQCD}

We now turn to demonstrating the effects of the addition of higher dimensional operators to the Dynamic AdS/YM description of two-flavour QCD. The philosophy is to include a UV cut off at a scale corresponding to the transition region from strong to weak coupling - at higher scales the gravity description is expected to break down (become strongly coupled).  There is an expectation that QCD will have generated HDOs at this matching scale. In addition one can consider the HDOs as potentially including stringy effects into the gravity description as well. We enact this in the holographic model by putting a boundary at $\rho=10$ roughly 10 times the scale of chiral symmetry breaking. Using the $\rho$-meson mass to set the scale this corresponds to a scale of about 6 GeV.

\begin{figure}[H]
\centering
\includegraphics[scale=0.6]{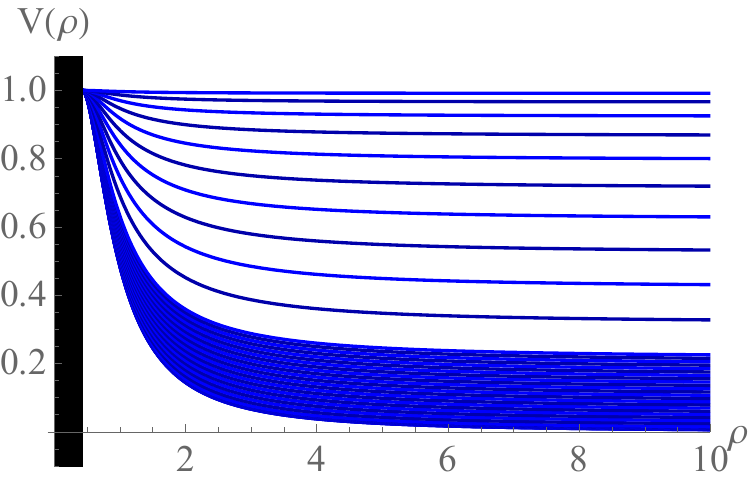} 
\centering
\includegraphics[scale=0.6]{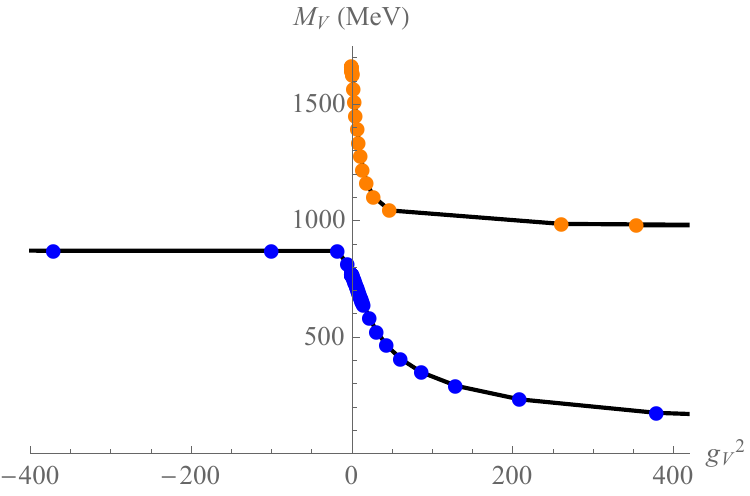} 
\caption{\label{fig: qcd_vector_all_masses} On the left the holographic wave functions of the vector meson ground state for various $g_V^2$ - the ground state at $g_V^2=0$ is the lowest curve; as $M^2$ decreases, $g_V^2$ increases and these are the higher curves. The black region represents the region below the IR cut off $\rho_{IR}$ below which the quarks have become very massive and need to be integrated out. On the right we plot the associated masses against coupling $g_V^2$ extracted from those solutions - blue points are the ground state (corresponding to the left hand points), the orange points are the first excited states. 
}
\end{figure}

Let us start, as an example,  with the analysis of the vector mesons. Previously we solved   \cref{eq: eqn of motion_vector} which has UV asymptotics of the form ${\mathcal J} + \langle {\mathcal O} \rangle /\rho^2$. We only accepted solutions for values of $M_V^2$ where ${\mathcal J}=0$ (see \cref{fig: vector_meson_qcd}). Now though we will enlarge the set of available solutions to those with all values of ${\mathcal J}$ as shown on the left in \cref{fig: qcd_vector_all_masses}. We now interpret these solutions as having ${\mathcal J}=0$ but the higher dimension operator $g_V^2/\Lambda_{UV}^2 |\bar{q} \gamma^\mu q|^2$ present. We extract ${\mathcal J}$ and $\langle {\mathcal O} \rangle$ from the asymptotics and compute the four fermion operator coupling $g_V^2$ using \cref{source}. We can then plot the vector meson mass as a function of $g_V$. This is displayed on the right in \cref{fig: qcd_vector_all_masses}. We see here that the mass of the bound state and the first 

\newpage

\begin{figure}[H]
\centering
\includegraphics[scale=0.6]{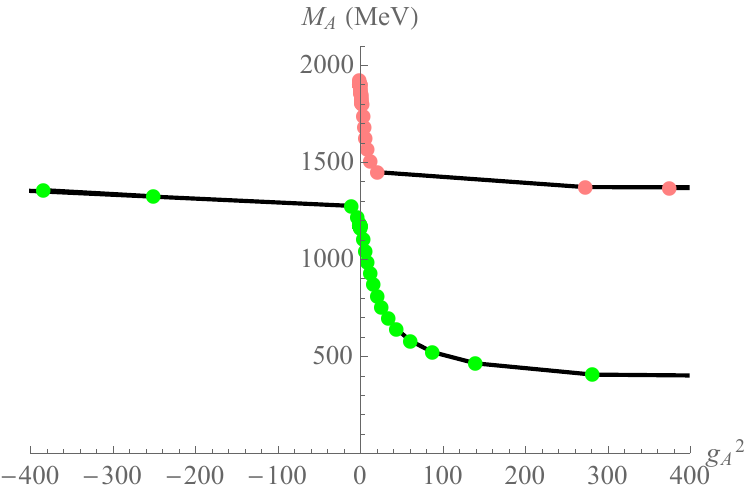} 
\caption{\label{fig: qcd_axial_double_trace} The masses of the  pseudovector meson ground state and first excited state as a function of $g_A^2$.}
\end{figure}

\noindent excited state  fill out the available mass values between the ground state mass and the first excited state masses at $g_V^2=0$ with a discontinuity between $g_V^2=\pm \infty$.  In addition positive $g_V^2$ drives the ground state mass below its value at $g_V^2=0$ and to zero as $g_V^2 \rightarrow \infty$. There is never a
 tachyonic state here. Note that the first excited state's mass does not fall below the mass of the ground state at $g_V^2=0$.

We repeat this computation for the axial vector meson and show the results in \cref{fig: qcd_axial_double_trace}. The behaviour of the mass with $g_A$ is very similar to that of the vector mass with $g_V$ except that it appears to asymptote to a fixed non-zero value as $g_A^2 \rightarrow \infty$.

\begin{figure}[H]
\centering
\includegraphics[scale=0.6]{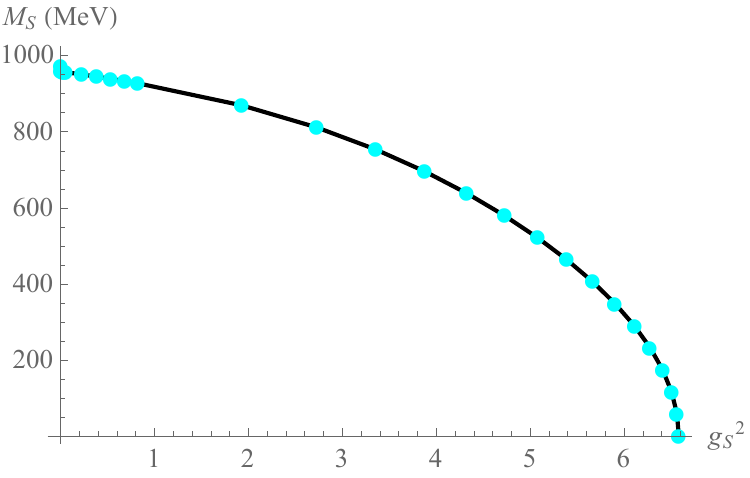} 
\centering
\includegraphics[scale=0.6]{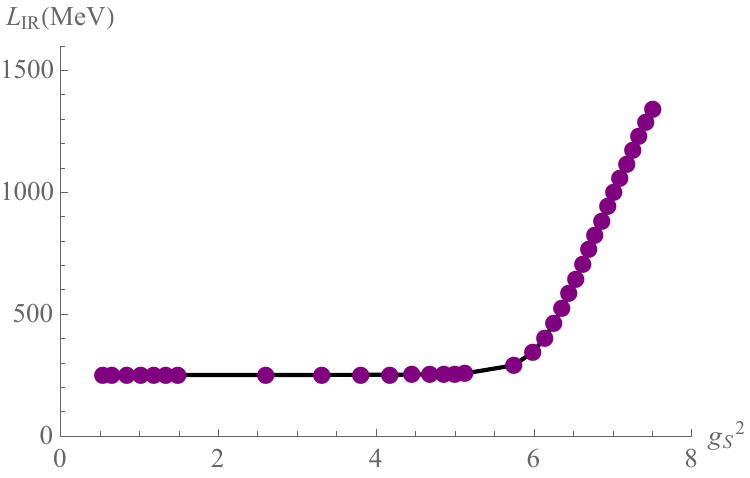} 
\caption{\label{fig: scalar_double_trace_qcd} The instability of the massless embedding in the presence of an NJL interaction: on the left we show the mass of the $\sigma$ scalar in the massless background of 
\cref{fig: QCD_embedding} (shown in red there) - it becomes tachyonic beyond a critical value of $g_s^2$. On the right we show the IR quark mass $L_{IR}$ against the NJL coupling  as interpreted from the embeddings with a source in \cref{fig: QCD_embedding}. We see that the tachyon instability is related to the NJL interaction changing the vacuum by enhancing chiral symmetry breaking.  }
\end{figure}

Next we consider the scalar meson of the theory where there is a new phenomenon. We begin by solving \cref{eq: eqn of motion_scalar} for the scalar meson fluctuations in the background embedding with zero UV quark mass, allowing all $M_S^2$ values and extracting the $g_s^2$ coupling of the higher dimension operator 
\begin{equation}g_S^2/\Lambda_{UV}^2 |\bar{q} q|^2. \end{equation}
We refer to this operator as a Nambu-Jona-Lasinio (NJL) operator. In \cref{fig: scalar_double_trace_qcd} we show that the scalar becomes tachyonic at a finite value of $g_S^2$. This indicates that the vacuum has become unstable at larger $g_S^2$. Here though we understand this instability. Consider again the solutions of the background embedding shown in \cref{fig: QCD_embedding}, including now the solutions with non-zero mass in the UV.  We include all these solutions with non-zero sources as solutions of the theory with the HDO present and at the level of the background determine $g_s^2$. In the  right hand plot in \cref{fig: scalar_double_trace_qcd} we show the IR quark mass $L_{IR}$ against $g_S^2$. Here we interpret the $\rho=0$ behaviour of the function $L_0(\rho)$ as the constituent quark mass.
It shows that around the same  critical value of $g_S^2$, where the scalar became tachyonic, the more massive vacua of the theory with a non-zero UV source emerge. This is the well known dynamics of the Nambu-Jona-Lasinio model \cite{Nambu:1961tp}- this has been investigated before in a holographic context in \cite{Clemens:2017udk}.  Note it is not a pure second order transition with the IR mass rising from zero because the base QCD theory already contains chiral symmetry breaking - the NJL interaction just

\noindent enhances this mass generation. If the $\sigma$ mass is computed in the true vacuum, where $L_0(\rho)$ includes the effect of $g_s^2$, then at any $g_S^2$ there is no tachyonic behaviour.

It is important to note that the vacuum embeddings in \cref{fig: QCD_embedding} have two interpretations - either there is an explicit UV mass for the quarks or a UV HDO is present. At the level of the solutions in \cref{fig: QCD_embedding} there is no distinction but there is at the level of the fluctuations. If  there is a UV quark mass only present, then  in the fluctuation calculation we must require that asymptotically in the UV there is only a vev for the operator and no ${\mathcal J}$. On the other hand, if we interpret all of the UV source in the embedding as being due to the NJL operator  then we must determine the value of $g_S$ from the background. Then we have to enforce that same value at the level of the fluctuations.  Of course, most generally there can also be a mixture of quark mass and NJL operator  in which case one needs to be careful to apply the appropriate  $g_S^2$ for the fluctuation calculation.

\begin{figure}[H]
\centering
\includegraphics[scale=0.6]{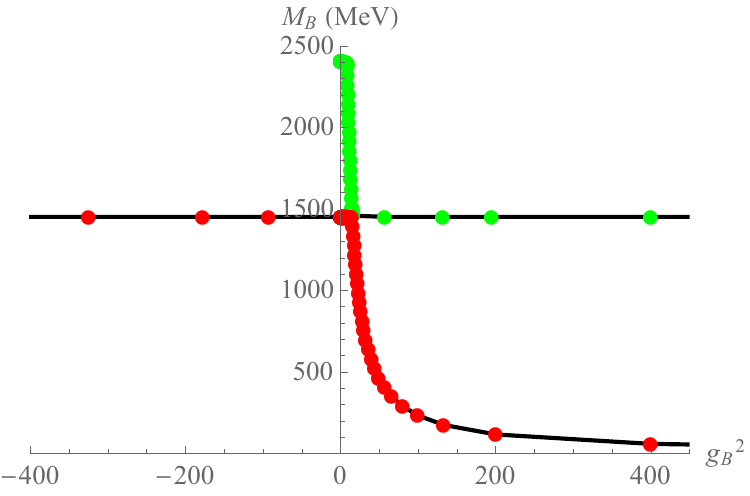} 
\caption{\label{fig: double_trace_qcd} The effect of adding higher-dimensional operators on the mass spectrum of the baryon. The red dots are the results when we drive the ground state lighter and the green ones depict the first excitation.}
\end{figure}

Finally we can introduce a baryon squared HDO, ${g^2_{\rm B} \over \Lambda_{UV}^5} |q q q|^2$, to change the baryon mass. The results are shown in \cref{fig: double_trace_qcd}. In fact this plot was our initial motivation for this work 
since we were interested in bringing the proton mass down relative to the $\rho$-meson mass in AdS/QCD. As we will see later, they may be similarly used  to generate light baryonic top partners in BSM models. Fig.\ref{fig: double_trace_qcd} shows similar features to the ones for the masses of the vector meson
and axial-vector meson.

\subsection{Perfecting two flavour QCD}
\label{sec:prefectingQCD}
Finally for two flavour QCD we will consider perfecting the holographic description \cite{Evans:2005ip,Evans:2006ea}: that is using HDOs to correct for the presence of a finite cut off. We will consider the description to only exist below $\rho=10$ (approximately 6 GeV) and include HDOs to improve the description. These HDOs are intended to represent the physics of the perturbative regime and of the regime where the theory  transitions from weak to strong coupling, which have been integrated out above the cut off. In principle one would like to explicitly match but presumably the intermediate, somewhat strongly coupled regime between perturbative QCD and where the holographic description is sensibly weakly coupled will make this matching hard. Thus we simply tune the HDOs couplings at our somewhat adhoc choice of UV cut off to match the observed mass spectrum.

To bring the decay constant $f_\pi$ to its measured value we allow ourselves to move away from the $L_0$ corresponding vanishing quark mass. This can be interpreted either as including a small bare quark mass or  a four fermion operator for $\bar{q} q$ - we find $m_{q}|_{UV} = 0.06576$ or equivalently  $g_S^2 = 4.59$. Since we use the $\rho$ mass to fix the scale, we can use the $g_V^2$ coupling to tune the ratio of $F_V/M_V^2$  to the observed value. We then use $g_A^2, g_B^2$ to arrange the masses of the axial vector, and baryon to their observed masses.
The resulting spectrum is shown in \Cref{tableperfect}.

\begin{table}[H]
{\setlength
\doublerulesep{1pt}   
 \begin{tabular}{cccc}
  \toprule[1pt]\midrule[1pt]
    Observables & QCD & Dynamic AdS/QCD & HDO coupling \\ \cmidrule{2-4}
       (MeV)  & \color{white}{ here's the hidden text } & \color{white}{ here's the hidden text }  \\ \midrule
    $M_{V}$ & 775 & 775  &  sets scale \\
    $M_{A}$ & 1230 & 1230 & fitted by $g^2_A=5.76149$\\
    $M_{S}$ & 500/990& 597  & prediction $+ 20\%/-40\%$ \\
    $M_{B}$ & 938 & 938 & fitted by $g^2_B=25.1558$ \\
    $f_{\pi}$  & 93 & 93 & fitted by $g^2_S= 4.58981$  \\
    $f_{V}$ & 345 & 345  &  fitted by $g^2_V=4.64807$\\
    $f_{A}$ &433 & 444 & prediction $+2.5\%$\\
    $M_{V,n=1}$ &  1465  & 1532  &  prediction $+4.5\%$\\
    $M_{A,n=1}$ & 1655 & 1789 & prediction $+8\%$\\
    $M_{S,n=1}$ &990/1200-1500 & 1449  & prediction $+46\%/0\%$ \\
    $M_{B,n=1}$ & 1440& 1529 & prediction $+6\%$ \\
    \midrule[1pt]\bottomrule[1pt]
  \end{tabular} 
  }  
  \caption{\label{tab: results_qcd_perfect_v2} The spectum and the decay constants for two-flavour QCD with HDOs from \cref{fig: double_trace_qcd} used to improve the spectrum.}
  \label{tableperfect}
\end{table}

Clearly this is a much better description of the ground state QCD spectrum than in \Cref{tab: results_qcd_coupling_g5_perturbative}  if only because we have tuned most of the parameters! $f_A$ is a prediction and lies closer to the data than before. The scalar mass is also a prediction and here, where we have interpreted the UV quark mass as the presence of  $g_S^2$, the result has dropped closer to the mass of the $f_0(500)$ resonance. The predictions for the first excited states' masses, the final four entries in the table, have all moved closer to the experimental values too - possibly this means that the HDOs are including some of the stringy effects the supergravity approximation excludes. The mass of the  first excited state of the scalar is quite far off again, as in \cref{asection}, suggesting that interpreting these states is difficult. Overall though we conclude that the improvement method used is sensible.
In principle one could go further and allow corrections to the UV matchings of the coupling $g_5^2$ and the normalization of the correlators in \cref{eq: match} but then we would lose essentially all predictivity.

\newpage

\section{Composite Higgs Models} \label{sec: composite_higgs_models}
The holographic model we have used above to describe QCD with higher dimension operators can naturally be extended to other non-abelian gauge theories in which a dimension three, gauge invariant quark bilinear condenses. The key idea is to simply change the running of the anomalous dimension for the quark bilinear. The bound states are then those associated with that operator with inserted gamma matrix structure. It is natural to apply this modelling  to proposed strongly coupled models of  physics beyond the Standard Model (BSM).  In \cite{BitaghsirFadafan:2018efw,Belyaev:2019ybr}, one 
of the authors has already studied predictions of such models for technicolour theories, including examples where the dynamics is enhanced by Nambu-Jona-Lasinio operators \cite{Clemens:2017udk} and where extended technicolour interactions are included as HDOs for the generation of the top mass \cite{Clemens:2017luw}. 
In this section we will apply these techniques to a further class of BSM models, the Composite Higgs Models. 

\subsection{Setting the scene}

\subsubsection{Review of composite Higgs models \label{review}}

The crucial ingredient in composite Higgs models is a strongly coupled sector that breaks a global symmetry generating Nambu-Goldstone bosons. By weakly gauging part of the global symmetries the Standard Model (SM) gauge groups are introduced and 4 of the then pseudo Nambu-Goldstone bosons (pNGB) are identified with the SM Higgs. 
Realistic models have to contain the Higgs fields as a $(2,2)$ representation
of the custodial symmetry group. Gauging the SM $SU(2)_L\times U(1)$ leads to an explicit
breaking of the global group which in turn implies that a potential for the pNGBs is generated
at loop-level.
Moreover, in the low-energy theory one assumes that HDOs have also generated the top Yukawa coupling. The effective cut off on these loops is given by the strong coupling scale $\Lambda_S \simeq 4 \pi f_\pi$ where here $f_\pi$ is the pion decay constant of the SU(4) gauge theory.  Typically, $\Lambda_S$ is assumed to be at a scale of order 1-5 TeV. The potential is of the form \cite{Golterman:2015zwa}
\begin{align}
V_h = - C_{LR} (3 g^2_1 + g_Y^2) \cos^2\left(\frac{h}{f}\right) 
+ \frac{y^2_t}{2} C_t \sin^2\left(\frac{2 h}{f}\right) \,.
\end{align}
Here $C_{LR}$ and $C_t$ are low-energy couplings of the effective theory
below the strongly coupled group's scale, which can be expressed in terms of correlation functions
within the theory (see e.g.~\cite{Golterman:2015zwa} for details in the case of
an explicit $SU(4)$ model). We will not revisit these low energy computations further here, but instead concentrate on the strong dynamics sector at the higher scale that generates the pNGB fields.

Explicit models of the top quark Yukawa coupling require more elaborate models.  In the spirit of extended technicolour \cite{Eichten:1979ah}, one can simply include HDOs of the form
\begin{equation}   
 {1 \over \Lambda_{UV}^2}    \bar{t}_L \bar{\cal F} {\cal F}  t_R  \, ,
\end{equation}

\begin{figure}[H]
\centering
\includegraphics[scale=0.4]{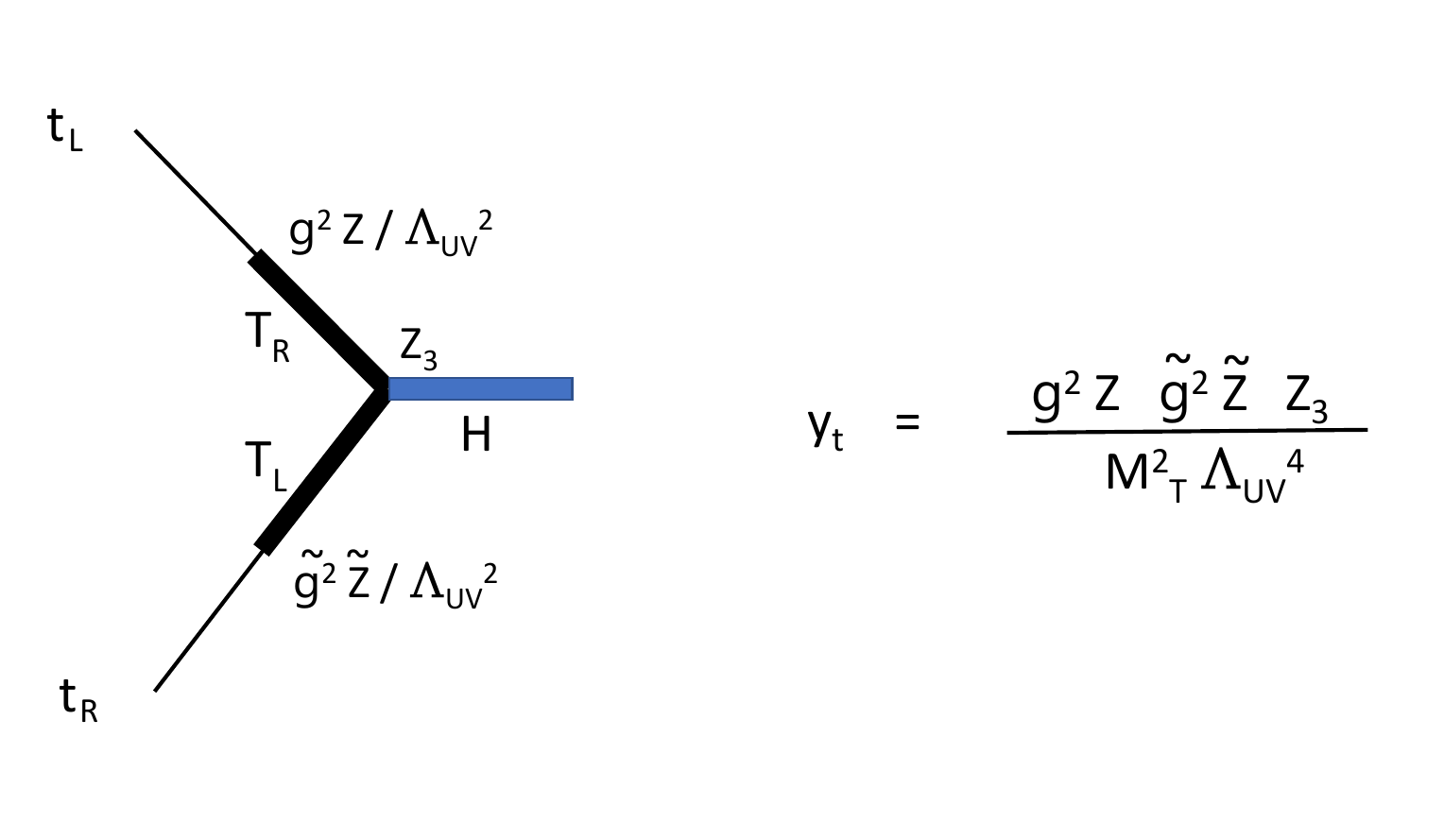} 
\caption{\label{topdiagram} The diagram responsible for the generation of the top Yukawa coupling. $t_L, t_R$ are the standard model top quarks, $T_L$, $T_R$ the top partners - they mix via the HDOs with couplings $g$ and $\tilde{g}$ - there are $Z$ form factors associated with the formation of the top partner baryons. $H$ is the pNGB that becomes the Higgs which has an order one Yukawa coupling to the top partners $Z_3$.}
\end{figure}

\noindent
where ${\cal F}$ are generically the composite fields that make up the Higgs.  $\Lambda_{UV}$  must probably be at least 5 TeV, making it hard to generate the large top mass. Such couplings also potentially suffer from Flavour Changing Neutral Currents.

Another possibility for generating the top mass, often referred to as partial compositeness, is that the left and right-handed top particles $t_L$ and $t_R$   mix with baryon-like spin 1/2 states in the gauge theory  $T_L, T_R$ with the same quantum numbers
\cite{Kaplan:1991dc}. These baryons are frequently called top partners.  They will  be involved in the strong dynamics and so have an order 
 one Yukawa coupling to the Higgs. The diagram in \cref{topdiagram} then generates a contribution to
the top Yukawa coupling as shown.

Here the $Z$ factors are three structure functions that depend on the strong dynamics. The top-top partner mixing factors result from the couplings of HDOs such as 
\begin{equation}  
 {g^2 \over \Lambda_{UV}^2} \bar{t_L} {\cal F}  {\cal F}  {\cal F}  \, ,
\label{eq:mixing} 
\end{equation}
where the ${\cal F}$ are again generically representing the fermions that $T_R$ is made from. 
We expect $Z_3$ to be of order one since it is generated by the strong dynamics - it is analogous to the nucleon-$\sigma$ or $\pi$ coupling in QCD.  The $Z$ and $\tilde{Z}$ factors  (setting $g=\tilde{g}=1$) will take the form $\Lambda_S^3 / \Lambda_{UV}^2$ where $\Lambda_S$ is the strong coupling scale. If the top partner's masses are of order $\Lambda_S$, then the Yukawa is given by 
\begin{equation} y_t \simeq  \Lambda_S^4 / \Lambda_{UV}^4 \end{equation}
which, assuming a separation of at least a factor of 3 between the flavour scale and the strong coupling scale, makes the top mass a factor of 100 too light. We will compute the $Z$  factors and $M_T$ holographically below where we indeed find that  a large top Yukawa cannot be achieved in this way.  

To combat the small Yukawa coupling size one could try to  lower $M_T$ or reduce the power of $\Lambda_{UV}$ in the denominator. One proposed solution  is walking dynamics \cite{Holdom:1981rm}. In a walking theory  the dimension of the fermions in \cref{eq:mixing} are lower at $\Lambda_{UV}$ and then the powers of $\Lambda_{UV}$ reduce (see for example \cite{Barnard:2013zea,Kim:2020yvr} for discussion).  
Here, though,  we will provide a new mechanism that allows an order one top Yukawa coupling as needed for the top mass. To generate the large top mass one could hope the top partners are anomalously light relative to the strong scale  $\Lambda_S$ by a factor of  3 or more, but generically there is no reason to expect this. However, here we will realize such a mechanism: in particular we will include a new HDO that  reduces to a shift in the top partners’ mass at low energies, using the holographic HDO implementation introduced in sections \ref{sec:higherdim} and \ref{sec:higherdimQCD}.  We show that the top Yukawa coupling can be made of order one by lowering the top partners’ mass to roughly half the vector meson mass in the strongly coupled sector. This appears to be consistent with experimental constraints and
provides a mechanism for generating an order one top Yukawa coupling.

A comprehensive analysis of the group theoretic possibilities for the strong sector underlying composite Higgs models with top partners was performed in \cite{Ferretti:2013kya}. We will analyze all 26 models using our holographic techniques. However, we also show that some of these models lie, at least based on the ansatz of the two loop running of the coupling, in the conformal window \cite{Appelquist:1996dq,Dietrich:2006cm} with a infra-red (IR) fixed point that is too small to break chiral symmetries. 
In the theories that do break symmetries dynamically we derive the values of the masses of the vector, scalar, axial mesons, and spin-$1/2$ baryon as well as the decay constants.

There are three scenarios that we will consider in considerable detail here, since they were already studied within lattice gauge theory \cite{Arthur:2016dir,Arthur:2016ozw,Bennett:2019cxd,Bennett:2019jzz,Ayyar:2018zuk}.  We will  start with a simple $SU(2)$ gauge theory with quarks in the fundamental representation (which in the classification of \cite{Ferretti:2013kya}
is among the $Sp(2 N)$ models). We will  then discuss two models, one based on the gauge group
$Sp(4)$, originally proposed in \cite{Barnard:2013zea}, and one based on the gauge group
$SU(4)$ proposed in \cite{Ferretti:2014qta}. These models contain additional pNGBs beyond the Higgs. We will not address their mass generation in the low-energy theory, though. Instead, we will concentrate on the bound states at the higher, strongly coupled scale.

\subsubsection{Model classification \label{classification}}

Since we will be discussing many different models, it is important to be able to clearly but succinctly identify them.
We will label models by their gauge group and the matter content of the model. We give the number of Weyl fermions in the representations $F$  for the fundamental, $A_n$ for the $n$ index antisymmetric representation, $S_n$ for the $n$ index symmetric representation, $G$ for the adjoint and $s$ for the spinor representation. We use a bar for the anti-representation.
Thus, for example, we can fully specify a model as $Sp(2N_c)$ $a$G$,$~$b$F,  which means an $Sp(2N_c)$ gauge group with $a$ Weyl  flavours  in the adjoint and $b$ in the fundamental of the group.
We will refer to the holographic description of such a  model as  $AdS/Sp(2N)$  $a$G$,$~$b$F.

We also note that we will refer to all fields in representations of the flavour group as `quarks' in analogy to QCD.

\subsubsection{Lattice data in a normalization adapted to holography}

In the sections below, we will present data from a variety of lattice collaborations \cite{Arthur:2016dir,Arthur:2016ozw,Bennett:2019cxd,Bennett:2019jzz,Ayyar:2018zuk}. In order to present them in a uniform manner we have manipulated the data from some of the original papers.  In particular, we choose to present all quantities as  dimension one quantities (mass or decay constant) using one of the representation's vector meson mass to set the scale.  Wherever possible, we give errors on the quantities we have extracted from lattice papers. We propagate them using simple differential formulae. Thus for example if
\begin{equation}  C = \sqrt{A \over B}, \hspace{1cm}{\rm  then} \hspace{1cm} dC = {1 \over 2} \left( {dA\over \sqrt{AB}} + {\sqrt{A} ~ dB \over \sqrt{B^3}} \right).  \end{equation}

We note again that we are using the notation common in the AdS/QCD literature that the dimension two coupling between the meson and its source is called $F_V^2$. It is common in the phenomenology and lattice literature to call this quantity $\tilde{F}_V M_V$ (see for example \cite{Bennett:2017kga}). We have moved any lattice results we quote below to our definition of $F_V$ as discussed in \cref{bsection}.

\newpage


\subsection{$SU(2)$ gauge theory with 2 Dirac fundamental quarks - $SU(2) ~4 F$ \label{sec: cfSU(2)}}

One of the simplest gauge theories that can underlie composite Higgs models is an $SU(2)$ gauge theory with two Dirac quarks in the fundamental representation \cite{Appelquist:1999dq} (or two Weyl fermions in each of the $F$ and $\bar{F}$). The pseudo-real nature of the fundamental of $SU(2)$ means that the naive $SU(2)_L \times SU(2)_R$ symmetry of the quarks is enhanced to an $SU(4)$ flavour symmetry \cite{Lewis:2011zb} (the 2 and $\bar{2}$ are identical). The condensation pattern is of similar structure as in QCD ($\langle \bar{u}_L u_R + \bar{d}_L d_R + h.c. \rangle$), which then breaks the $SU(4)$ flavour symmetry to $Sp(4)$. Five generators are broken so there are 5 pNGBs. 

It is straightfoward to describe the model using our AdS/YM description - we simply dial $N_c=2, N_f=2$ in the running of $\alpha$ in \cref{running} and $\gamma$ \cref{grun}. These then feed into $\Delta m^2$ in \cref{dm}. With these values, we repeat our computations as in holographic QCD. We have again
\begin{equation} \label{eq: renormalization_su2}
\begin{aligned}
b_0 &= \frac{1}{6 \pi} \left(11 N_c - (N_f+\bar{N}_f) \right)\, , \\ 
b_1 &= \frac{1}{24 \pi^2} \left(34 {N_c}^2 -5 N_c (N_f+\bar{N}_f) - {3 \over 2} \frac{{N_c}^2-1}{N_c} (N_f+\bar{N}_f)\right)\, , \\
\gamma &= \frac{3({N_c}^2-1)}{4 N_c \pi} \alpha\, .
\end{aligned}
\end{equation}
Note that the $Sp(4)$ multiplets of mesons include the usual $SU(2)_V$ multiplets, so we compute as in QCD to find masses and decay constants. Our results for the massless theory are shown in \cref{tab: results_su2_lattice} normalized to the $\rho/V$ mass.

There is lattice work on this model in \cite{Arthur:2016dir,Arthur:2016ozw}, where unquenched Wilson fermions are used, i.e. the determinant of the Dirac operator is calculated instead of setting it to one, as in quenched theories.
In the holographic approach this corresponds to including quark loop contributions to the gauge propagator.

 We show these results in the massless limit for the $V,A$ and $\sigma$ masses also in  \Cref{tab: results_su2_lattice}. Comparing to our holographic results, we see sensible agreement, as we found in the QCD 
\begin{table}[H]
{\setlength
\doublerulesep{1pt}   
 \begin{tabular}{cccc}
  \toprule[1pt]\midrule[1pt]
    Observables & Lattice & AdS/$SU(2)$ & \color{white}{ here's the hidden text } \\ 
       \color{white}{ here's the hidden text }  & \color{white}{ here's the hidden text } & 2 F, $2 \bar{F}$  \\ \midrule
    $M_{V}$ 	& 1.00(3) 			& 1*  				    			& sets scale \\
    $M_{A}$ 	& 1.11(46) 		& 1.66  		 			& \\
    $M_{S}$ 	& 1.5(1.1)		& 1.27   		 			& \\
    $f_{\pi}$  	& 0.076(13) 	& 0.0609  		 			& \\
    $f_{V}$ 		&   					& 0.376  		 			& \\
    $f_{A}$ 		&  					& 0.474  		 			& \\
    \midrule[1pt]\bottomrule[1pt]
  \end{tabular} 
  }  
  \caption{\label{tab: results_su2_lattice} Comparison of the lattice studies \cite{Arthur:2016dir,Arthur:2016ozw} of the massless $SU(2)$ gauge theory to our holographic model's predictions for meson masses and decay constants in units of the vector meson mass.}
\end{table}

\begin{figure}[H]
\centering
\includegraphics[scale=0.6]{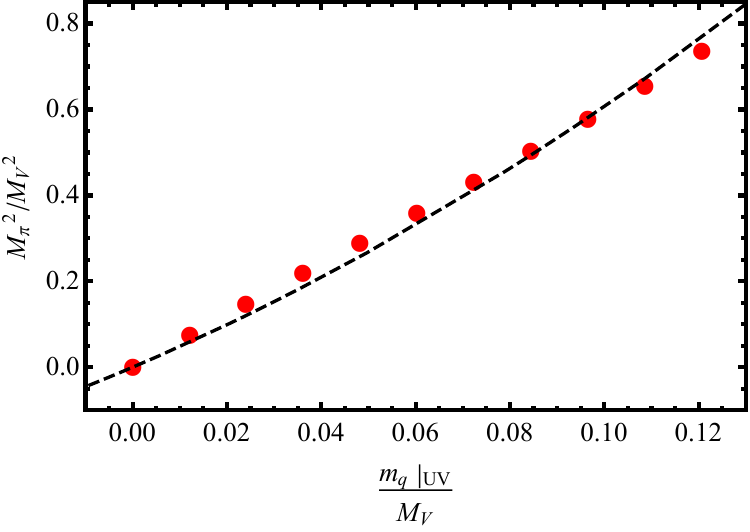} 
\centering
\includegraphics[scale=0.6]{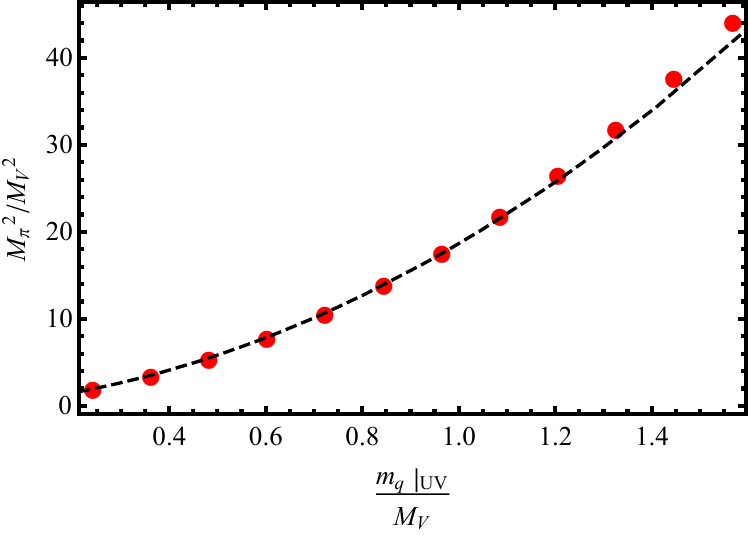} 
\caption{We plot the pNGB mass against the UV quark mass in the small and intermediate quark mass regions for the $SU(2)$ gauge theory (in units of the vector meson mass at $m_{q}|_{UV}=0$). The red points are the numerical results. The dashed black lines are obtained as a simple analytic fit: $M^2_{\pi} = 4.67 ~ m_{q}|_{UV} + 13.97~ m^2_{q}|_{UV}$.
 \label{su2_pionfit}
 }
\end{figure}

\begin{figure}[H]
\includegraphics[scale=0.6]{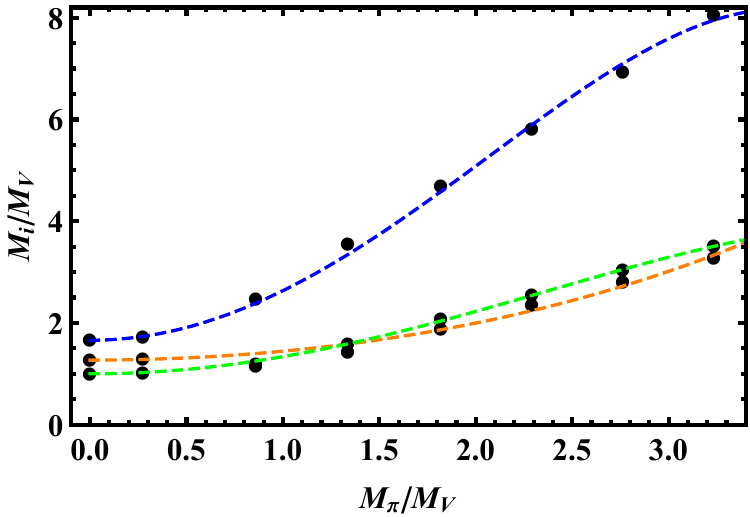} $\left. \right.$ 
\hspace{8cm}  \vspace{-4.7cm}

$\left. \right.$ \hspace{7.5cm} 
\begin{tabular}{cl} 
\color{orange}$M_{S}$ &\color{orange}= $\color{orange}1.27 + \color{orange}0.172 ~ \color{orange}M_{\pi}^2  + 0.00239 M_\pi^4\, ,$ \\
\color{orange}$M_{S {\rm lat} }$ &\color{orange}= $\color{orange}1.5(6)+ \color{orange}0.34(7) ~ \color{orange}M_{\pi}^2 -0.016(8) M_\pi^4\, ,$ \\
\color{green}$M_{V}$ &\color{green}= $\color{green}1 + \color{green}0.348 ~ \color{green}M_{\pi}^2  - 0.0104 M_\pi^4\, ,$ \\
\color{green}$M_{V{\rm lat} }$ &\color{green}= $\color{green}1.01(3)+ \color{green}0.47(3) ~ \color{green}M_{\pi}^2 -0.039(6) M_\pi^4\, ,$ \\
\color{blue}$M_{A}$ &\color{blue}= $\color{blue}1.66 + \color{blue}1.01 ~ \color{blue}M_{\pi}^2 -0.0394 M_\pi^4\, ,$ \\
\color{blue}$M_{A {\rm lat}}$ &\color{blue}= $\color{blue}1.1(1) + \color{blue}0.8(1) ~ \color{blue}M_{\pi}^2 -0.09(3)M_\pi^4\, .$
\end{tabular} 
\vspace{1.6cm}

\caption{\label{fig: specta_deformed_vaccum_su2_analytics} The growth of the spectra in the $SU(2)$ theory as we increase the quark mass in the UV. The masses are rescaled with respect to the vector meson mass at $m_{q}|_{UV}=0$. In our analytic formulae quantities are again normalized to the vector meson mass at $m_{q}|_{UV}=0$.
 }
\end{figure}

\noindent analysis above. The holographic $A$ mass  is perhaps a little high.  The lattice errors on the scalar mass are sufficiently large to incorporate our result.

Since the lattice studies also provide fits to the quark mass dependence in the model, we make that comparison as well. In \cref{su2_pionfit} we show the small (linear) and larger (quadratic) $m_q$ dependence of the pNGB mass squared. At larger pNGB masses, higher order terms in the expansion in $m_q$ would be needed. We then plot the meson masses as a function of $M_\pi^2$ in 
\cref{fig: specta_deformed_vaccum_su2_analytics} and present our fits and those from the lattice. The holographic model agrees rather well with the lattice fit and certainly lends strength to the view that the holographic model provides a credible and useful description of the dynamics. 
 
\newpage
\subsection{$Sp(4)$ gauge theory with top partners - $Sp(4) ~4F,6A_2$} \label{sec: sp4 model}
The $SU(2)$ model of the previous subsection can realize a composite pseudo-Goldstone Higgs but can not contain top partners since there are no baryons in an $SU(2)$ gauge theory. The same global symmetry breaking pattern ($SU(4) \rightarrow Sp(4)$) can be achieved with any $Sp(2N)$ gauge theory with again two Dirac fermions in the fundamental representation (4 Weyl fermions in the $F$). It is natural to concentrate on the next most minimal $Sp(4)$ case, as $Sp(2) \simeq SU(2)$.

Top partners can be introduced \cite{Barnard:2013zea} into the $Sp(4)$ model by the inclusion of three additional Dirac fermion in the sextet, two index anti-symmetric representation of the gauge group (we will refer to them as $A_2$s)  (in the nomenclature of \cite{Belyaev:2016ftv} this is model M8). The three copies are the three QCD colours although we drop the colour interactions since they are only weakly coupled at the energy scales we consider. The top partners are $F A_2 F$ bound states. From the point of view of the $Sp(4)$ dynamics there is an $SU(6)$ symmetry on the six Weyl fermion $A_2$s which are in a real representation. When the $A_2$ condensate forms this symmetry is broken to $SO(6)$.
The full symmetry breaking pattern is characterized by
\begin{align}
SU(4) \times SU(6) \times U(1) & 
  \to \underbrace{Sp(4)}_{SU(2)_L  \times U(1)} \times \underbrace{SO(6)}_{SU(3)\times U(1)} \times U(1)
\end{align}
where the $U(1)$ factors give eventually the hypercharge.

For the holographic model we need  the running of the coupling  \cref{running} and $\gamma$ \cref{grun}. These then feed into $\Delta m^2$ in \cref{dm} to define the model.The beta function coefficients for the running of $\alpha$ and $\gamma$ in the UV are
\begin{equation}  \label{eq: renormalization_gherghetta_1}
\begin{aligned}
b_0 &= \frac{1}{6 \pi} \left(\vphantom{\frac{1}{2}} 11(N+1) -  N_{f_1} - 2(N-1) N_{f_2}   \vphantom{\frac{1}{2}} \right)  \\
b_1 &=  \frac{1}{24 \pi^2} \left( \vphantom{\frac{1}{2}} 34(N +1)^2  -5(N+1) N_{f_1} - \frac{3}{4} (2N+1)N_{f_1} \right. \\
    &\left.- 10(N+1)(N-1) N_{f_2} -6 N (N-1) N_{f_2}  \vphantom{\frac{1}{2}} \right)  
\end{aligned}
\end{equation}
and the one-loop anomalous dimensions for the different representations are 
\begin{equation}  \label{eq: renormalization_gherghetta_2}
\begin{split}
\gamma_{A_2}&= \frac{3}{2 \pi} N \alpha\, , \\
\gamma_{F} &=\frac{3}{2 \pi} \frac{2N+1}{4}   \alpha\, ,
\end{split}
\end{equation}
In the above $N_{f_1}=4$ denotes the flavours in the fundamental and $N_{f_2}=6$ in the two-index antisymmetric.  $N=2$ for $Sp(4)$. 

Generically one would expect the $A_2$ fermions to condense ahead of the fundamental fields since the critical value for $\alpha$ where $\gamma=1/2$ (the criteria discussed below \cref{dm}) is smaller. If we extend the perturbative results into the non-perturbative regime we find
\begin{equation} 
\alpha_c^{A_2} =  {\pi \over 6} = 0.53\, , \hspace{1cm} \alpha_c^{F} = {4 \pi \over 15} = 0.84\, .
\end{equation}
When the $A_2$s  condense  their condensate breaks their flavour $SU(6)$ to $SO(6)$. At this point the $A_2$s become massive but it is unclear how quickly they decouple from the running of $\alpha$ - we will investigate this point below. The usual assumption is that  both species of fermion condense close to the same scale.

\subsubsection{The holographic vacuum of the theory}
\begin{figure}[H]
\centering
\includegraphics[scale=0.6]{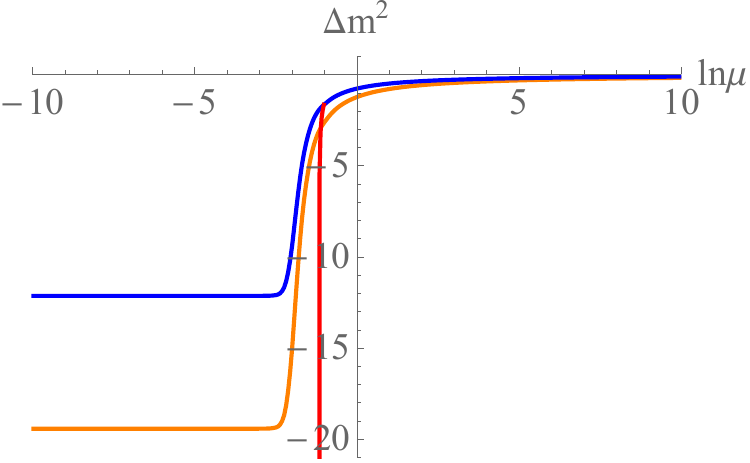} 
\centering
\includegraphics[scale=0.6]{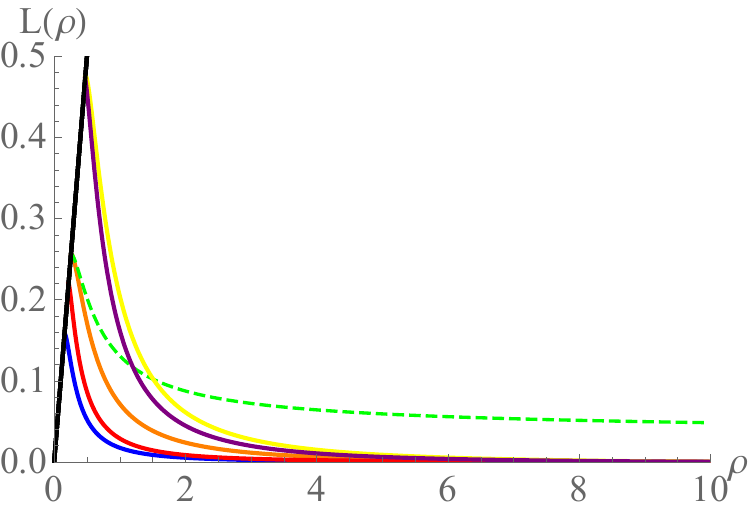} 
\caption{\label{fig: gherghetta_embedding} AdS/$Sp(4) ~4F,6A_2$. Left panel: The running of $\Delta m^2$ against RG scale for the fundamental (blue line), $A_2$ (orange) and in red the running of  the fundamental representation after $A_2$ have been integrated out. Right panel: The vacuum solution $L(\rho)$: the orange line for the $A_2$ representation and blue the fundamental without decoupling. The red solution is when we consider the decoupling of the $A_2$ which condenses before the fundamental. The dashed green line is the fundamental when we consider additional NJL-terms such that it matches in the IR the $A_2$ representation. Finally, the yellow and purple vacuum solution correspond to the quenched models for the $A_2$ and fundamental representations respectively.
Here units are set by $\alpha(\rho = 1) = 0.65$.}
\end{figure}

Let us begin by investigating the question of the scale of the condensates in the vacuum of the theory using our holographic model. As a first run we use the AdS/YM theory with the running of $\alpha$ including both fermion species - that is we use \cref{eq: renormalization_gherghetta_1} at all energy scales. We then track the running of the anomalous dimension $\gamma$ for the two representations using \cref{eq: renormalization_gherghetta_2}. Note the scale where the BF bound is violated is similar for the two representations because the coupling is running quickly near the BF bound violation point.  These give us two $\Delta m^2$ in  \cref{dm}, one for each representation, which are shown in blue (F) and orange ($A_2$) on the left in \cref{fig: gherghetta_embedding}. Each of the condensates is a distinct operator which we represent by a distinct field $L$ - in other words we run two copies of the AdS/YM equations for the vacuum expectation values of the two condensates. The results for the two resulting $L$ functions are shown in \cref{fig: gherghetta_embedding} on the right - again blue (F) and orange ($A_2$). 
The $A_2$ fields condense at a higher scale than the $F$ because its $\Delta m^2$ passes through the BF bound first.

There is though a tricky and interesting decoupling problem here. When the $A_2$ fields condense and become massive should we integrate them out of the running of $\alpha$? At weak 

\noindent  coupling massive quarks do decouple from the running but it is less clear what is appropriate at strong coupling. We have computed an example of such a possible decoupling. Here as soon as the scale $L_{IR}$ for the $A_2$ fermions is reached we remove them from the running of $\alpha$ at lower $\rho$ - the running of the $F$ fields $\Delta m^2$ then deviates from the blue to the red curve on the left in  \cref{fig: gherghetta_embedding}. It runs faster than before and the condensation scale for the fundamental
 fields moves closer to that of the $A_2$s.   The resulting $L(\rho)$ function for the $F$ is shown by the red line in the right hand plot of \cref{fig: gherghetta_embedding}.

The lattice should be able to shed light on the rate of decoupling of massive flavours but to date only quenched calculations have been performed for this model as we will review below. We therefore also show results for the embeddings that result from the fully quenched (ie setting all $N_f=0$ in \cref{{eq: renormalization_gherghetta_1}}) running in \cref{fig: gherghetta_embedding} on the right - the yellow ($F$) and purple ($A_2$) curves. The coupling now runs faster at all scales and the condensation scale for both fermion species rises, with again the $A_2$s condensing first. The gap between the F and $A_2$ condensation scales is yet smaller due to the very fast running of the pure glue theory.

If the IR separation in the condensation scales for the two fermion species is undesirable then they can be brought together by including an NJL term for the $F$ fields ($g_s^2/\Lambda_{UV}^2 |\bar{F} F|^2$) that enhances the fundamental condensation scale. We have also looked at this case, adding a NJL four fermion term to make the values of $L_{IR}$ equal for the two representations. The $A_2$ embedding function is our original orange curve but the  embedding for the $F$ representation becomes the green dotted curve in \cref{fig: gherghetta_embedding}.

It is worth commenting on the size of the IR mass, $L_{IR}$ in physical units. We will compute the spectrum in the next section but borrowing ahead we can write $L_{IR}$  in units of the vector meson's mass in the $A_2$ representation for the  case discussed. For the model where we do not integrate out the $A_2$ fields we have $L_{IR}^{A_2}=0.304 m_V$ and $L_{IR}^{F}=0.187 m_V$. When we integrate out the $A_2$s on mass shell we have $L_{IR}^{A_2}=0.304 m_V$ and $L_{IR}^{F}=0.26 m_V$. For the model with the NJL interaction for the fundamental we have  $L_{IR}^{A_2}=L_{IR}^{F}=0.304 m_V$. For the quenched model we have $L_{IR}^{A_2}=0.317 m_V$ and $L_{IR}^{F}=0.314 m_V$.

\subsubsection{Holographic spectrum}

We  now  compute the spectrum of the theory holographically. We will do this for each of the scenarios we have outlined - the quenched theory; the theory where the $A_2$s are integrated out at their IR mass scale; the theory where $A_2$s are not integrated out; and the theory with a NJL term to enforce an equal scale of condensation.

We assume that there is only a small mixing between bound states made of the two fermion species so that we do not have to mix the states associated with fluctuations of each $L_0$ embedding (indeed to include that mixing would be hard requiring the fluctuations to know of both embeddings in some sort of non-abelian DBI action). Now we simply fluctuate around each vacuum solution separately from \cref{eq: general_action} with 
\begin{equation}
g^2_5|_{F} = \frac{48 \pi^2}{2 N_{f_{1}} N_c}, \qquad g^2_5|_{A_2} = \frac{48 \pi^2}{N_{f_{2}}(N_c(N_c-1)-1)}\, .
\end{equation}

\begin{table}[H]  
{\setlength
\doublerulesep{1pt}   
 \begin{tabular}{cccccccc}  
  \toprule[1pt]\midrule[1pt]
     								& AdS/$Sp(4)$  	& AdS/Sp(4) 		& AdS/Sp(4) 	& lattice \cite{Bennett:2019cxd}  	& lattice \cite{Bennett:2019jzz} 	&AdS/Sp(4) \\ 
     								& no decouple 	& A2 decouple 	& quench  		& quench  									& unquench 								& + NJL  \\    \midrule
    $f_{\pi A_2}$  				& 0.120  			& 0.120 			& 0.103			& 0.1453(12) 								&  											& 0.120\\
    $f_{\pi F}$  				&  0.0569 			&0.0701 			&  0.0756 		&0.1079(52) 								& 0.1018(83) 								& 0.160\\
    $M_{V A_2}$ 				& 1* 					& 1* 					& 1*				& 1.000(32) 								& 												& 1*\\
    $f_{V A_2}$ 				& 0.517 			& 0.517 			& 0.518 		& 0.508(18)  								& 												& 0.517\\
    $M_{V F}$ 					& 0.61   			& 0.814 			& 0.962 		& 0.83(19)									& 0.83(27) 								& 1.03\\
    $f_{V F}$ 					& 0.271   			& 0.364 			& 0.428 		&  0.411(58)								& 0.430(86)  								& 0.449\\
    $M_{A A_2}$ 				& 1.35  				& 1.35 				& 1.28 			& 1.75 (13)								& 												& 1.35\\
    $f_{A A_2}$ 				& 0.520 			& 0.520 			& 0.524			& 0.794(70) 								& 												& 0.520 \\
    $M_{A F}$ 					&  0.938 			&1.19 				& 1.36 			&  1.32(18) 								& 1.34(14) 								& 1.70\\
    $f_{A F}$ 					&  0.303 			&0.399 				& 0.462 		& 0.54(11) 									& 0.559(76) 								& 0.449\\
    $M_{S A_2}$ 				& 0.375 			& 0.375 			& 1.14			& 1.65(15)	$^\dagger$								& 												& 0.375\\
    $M_{S F}$ 					& 0.325  			& 0.902 			& 1.25 			& 1.52 (11)$^\dagger$									& 1.40(19)	$^\dagger$								& 0.375\\
    $M_{B A_2}$ 				& 1.85 				& 1.85 				& 1.86 			&												& 												& 1.85\\
    $M_{B F}$ 					& 1.13 				&1.53 				& 1.79 			& 												& 												& 1.88\\
    \midrule[1pt]\bottomrule[1pt]
  \end{tabular} 
  }  
  \caption{ \label{Sp4table} AdS/$Sp(4) ~4F,6A_2$. Ground state spectra and decay constants for our 
  various holographic models and comparison to lattice results - we use the subscript $A_2$ and $F$ for the quantity in each of the two different representation sectors. Note the lattice scalar is the $a_0$ not the isospin singlet $\sigma$ which we compute holographically - we present the results as a guide to lattice expectations of quark anti-quark meson masses though. Note here for the unquenched lattice results, which do not include the $A_2$ fields, we have normalized the $F$ vector meson mass to that of the quenched computation.}
\end{table}

\noindent Similarly we split the normalizations for the external currents in \cref{eq: match}. 

We show the resulting spectrum for each of the cases we consider in \Cref{Sp4table} for the case where all fermion representations are massless. 

In each case, without a NJL term, the bound states of the $A_2$ fields are heavier and have higher decay constants than those made of the fundamental fields $F$, reflecting the $A_2$s' higher condensation scale. The separation in scale between the two sectors does depend quite strongly on the decoupling assumptions. If the $A_2$s  are not decoupled at all, the separation, as measured by the vector meson masses, is almost a factor of two whilst in the quenched limit it barely exists. The slowing of the running of the gauge coupling with the inclusion of flavours is important. The case where the $A_2$s are integrated out at their IR mass scale lies between these two extremes. 

The greatest impact in the spectrum shows up in the scalar meson ($S$) masses. The rate of running measures the departure from conformality which shows up in the flatness of the effective potential for the quark condensates. The slower the running the lighter the resultant scalar - here there is as much as a factor of four in the prediction.

When the NJL term is used to enforce equal IR mass scales for the two fermion species the bound states of the fundamental fields become just slightly heavier than those with $A_2$ constituents, reflecting the higher UV mass. 

Finally in \Cref{Sp4table} we also show results for the baryon top partner. This state is a bound state of two $F$ and an $A_2$ so should know about both vacuum solutions $L_{0F}$ and $L_{0A_2}$. The present holographic framework does not allow us to include two $L_0$ at the same time so instead we compute the mass of the baryon using each of the two embedding functions - this is as if each constituent had the same constituent mass, either that of the $F$ or that of the $A_2$. We expect that the mixed state's mass will be between these two values.

\subsubsection{Comparison to lattice results}

Lattice studies of this model, in the quenched approximation, have been made in \cite{Bennett:2019cxd}. In \cite{Bennett:2019jzz} the group followed up that work by unquenching the fundamental quark sector using Wilson fermions. We show the results of these studies in \Cref{Sp4table} for direct comparison to the holographic results.  We have normalized the quenched results to the vector meson mass from the $A_2$ sector. For the unquenched calculation, which does not include the $A_2$ fields, we align the vector meson mass in the F sector to the quenched theory to allow the changes to be seen in the $F$ sector. One notes that the variation from quenched to unquenched lattice simulations are not large. Note the lattice results for scalar masses are for the $a_0$ like states rather than the $\sigma$ state we compute with holography - they provide a guide to the lattice expectation for scalar states though.

An initial view of the quenched results from both the lattice and the holographic model is that they show considerable correlation. As in QCD, the holographic approach appears to be a decent stab at the spectrum! This lends confidence that trends as the fields are unquenched may be trustworthy. Thus as discussed above we would expect that if the $A_2$ fields were included as unquenched fields the $F$ sector would decrease in mass by 20-40$\%$. We also expect the scalar meson masses to be considerably lower than predicted by the quenched lattice computation.

Here the lattice computations to date don't provide guidance on a prescription for decoupling the $A_2$ fields since they have always been quenched.

\subsubsection{Quark mass dependence}

\begin{figure}[H]
\centering
$A_2$ Sector - quenched \vspace{0.25cm}

\includegraphics[scale=0.5]{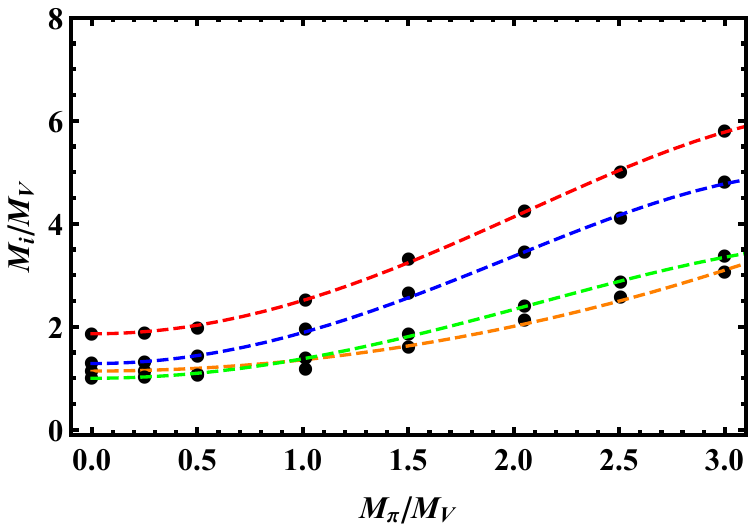} $\left. \right.$ \hspace{8cm}  \vspace{-4.2cm}

$\left. \right.$ \hspace{7.5cm} 
\begin{tabular}{cl} 
\color{orange}$M_{S}$ &\color{orange}= $\color{orange}1.14 + \color{orange}0.217 ~ \color{orange}M^2_{\pi}\, ,$ \\
\color{orange}$M_{S {\rm lat}}$ &\color{orange}= $\color{orange}1.65(15) + \color{orange}0.17 (5) M_\pi^2\, ,$  \\
\color{green}$M_{V}$ &\color{green}= $\color{green}1 + \color{green}0.392 ~ \color{green}M^2_{\pi}\, ,$  \\
\color{green}$M_{V {\rm lat}}$ &\color{green}= $\color{green}1.000(32) + \color{green}0.45(26) ~ \color{green}M_{\pi}^2\, ,$  \\
\color{blue}$M_{A}$ &\color{blue}= $\color{blue}1.28 + \color{blue}0.627 ~ \color{blue}M^2_{\pi}\, ,$ \\
\color{blue}$M_{A {\rm lat}}$ &\color{blue}= $\color{blue}1.75(13) + \color{blue}0.40(12) ~ \color{blue}M^2_{\pi}\, ,$ \\
\color{red}$M_{B}$ &\color{red}= $\color{red}1.86 + \color{red}0.673 ~ \color{red}M^2_{\pi}\, .$
\end{tabular}  \vspace{0.75cm}

$A_2$ Sector - no decoupling \vspace{0.25cm}

\includegraphics[scale=0.5]{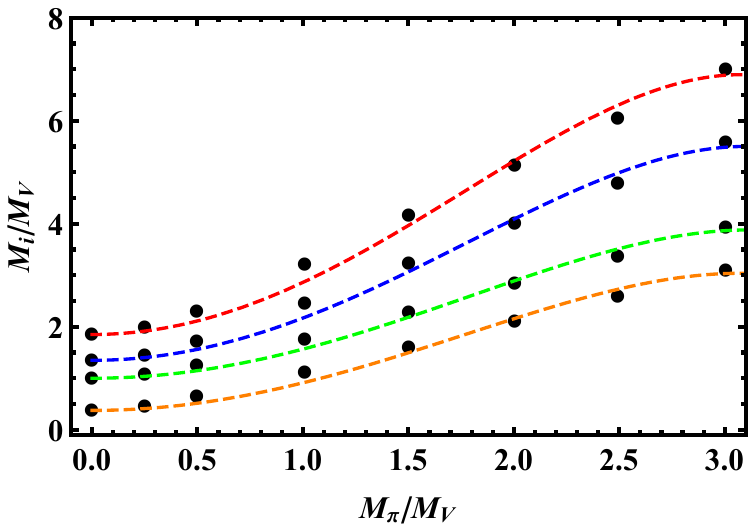} $\left. \right.$ \hspace{8cm}  \vspace{-3.6cm}

$\left. \right.$ \hspace{7.5cm} 
\begin{tabular}{cl} 
\color{orange}$M_{S}$ &\color{orange}= $\color{orange}0.375 + \color{orange}0.564 ~ \color{orange}M^2_{\pi} - \color{orange}0.0299 ~ \color{orange}M^4_{\pi}\, ,$ \\
\color{green}$M_{V}$ &\color{green}= $\color{green}1 + \color{green}0.595 ~ \color{green}M^2_{\pi} - \color{green}0.0307 ~ \color{green}M^4_{\pi}\, ,$  \\
\color{blue}$M_{A}$ &\color{blue}= $\color{blue}1.35 + \color{blue}0.865 ~ \color{blue}M^2_{\pi} -  \color{blue}0.0450 ~ \color{blue}M^4_{\pi}\, ,$ \\
\color{red}$M_{B}$ &\color{red}= $\color{red}1.85 + \color{red}1.07 ~ \color{red}M^2_{\pi} - \color{red}0.0564 ~ \color{red}M^4_{\pi}\, .$
\end{tabular}   \vspace{1.75cm}

$F$ Sector - quenched \vspace{0.25cm}

\includegraphics[scale=0.5]{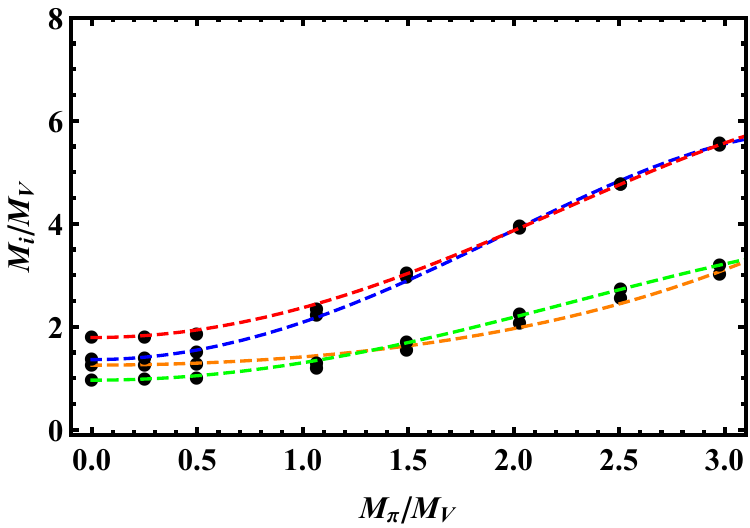} $\left. \right.$ \hspace{8cm}  \vspace{-4.2cm}

$\left. \right.$ \hspace{8cm} 
\begin{tabular}{cl} 
\color{orange}$M_{S}$ &\color{orange}= $\color{orange}1.25 + \color{orange}0.154 ~ \color{orange}M^2_{\pi}\, ,$ \\
\color{orange}$M_{S {\rm lat}}$ &\color{orange}= $\color{orange}1.52(1)  + \color{orange}0.09(10) ~ \color{orange}M^2_{\pi}\, ,$ \\
\color{green}$M_{V}$ &\color{green}= $\color{green}0.962 + \color{green}0.349 ~ \color{green}M^2_{\pi}\, ,$  \\
\color{green}$M_{V {\rm lat}}$ &\color{green}= $\color{green}0.83(19) + \color{green}0.50(16) ~ \color{green}M^2_{\pi}\, ,$  \\
\color{blue}$M_{A}$ &\color{blue}= $\color{blue}1.36 + \color{blue}0.758 ~ \color{blue}M^2_{\pi}\, ,$ \\
\color{blue}$M_{A{\rm lat}}$ &\color{blue}= $\color{blue}1.32(18) + \color{blue}0.42(0.20) ~ \color{blue}M^2_{\pi}\, ,$ \\
\color{red}$M_{B}$ &\color{red}= $\color{red}1.79 + \color{red}0.599 ~ \color{red}M^2_{\pi}\, .$
\end{tabular}   \vspace{0.75cm}

$F$ Sector - no decoupling  \vspace{0.25cm}

\includegraphics[scale=0.5]{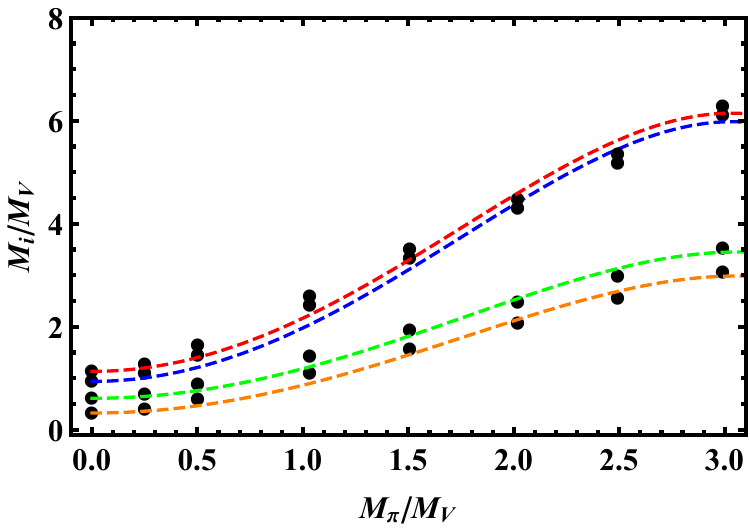}  $\left. \right.$ \hspace{8cm}  \vspace{-3.6cm}

$\left. \right.$ \hspace{8cm} 
\begin{tabular}{cl} 
\color{orange}$M_{S}$ &\color{orange}= $\color{orange}0.325 + \color{orange}0.57 ~ \color{orange}M^2_{\pi} -  \color{orange}0.031 ~ \color{orange}M^4_{\pi}\, ,$ \\
\color{green}$M_{V}$ &\color{green}= $\color{green}0.61 + \color{green}0.61 ~ \color{green}M^2_{\pi} - \color{green}0.033 ~ \color{green}M^4_{\pi}\, ,$  \\
\color{blue}$M_{A}$ &\color{blue}= $\color{blue}0.938 + \color{blue}1.1 ~ \color{blue}M^2_{\pi} -  \color{blue}0.06 ~ \color{blue}M^4_{\pi}\, ,$ \\
\color{red}$M_{B}$ &\color{red}= $\color{red}1.13 + \color{red}1.09 ~ \color{red}M^2_{\pi} -  \color{red}0.059 ~ \color{red}M^4_{\pi}\, .$
\end{tabular} 
\vspace{1.25cm}

\caption{\label{fig: pion_masses_sp4} AdS/$Sp(4) ~4F,6A_2$ -  results for the spectrum as a function of the pNGB mass in the quenched theory and the case with no decoupling of the $A_2$ - lattice results from \cite{Bennett:2019cxd} are included for comparison. In our analytic formulae we use units of the vector meson mass at $m_{q}|_{UV}=0$.
}
\end{figure}

The quenched lattice study of \cite{Bennett:2019cxd} provides fits to the mass dependence of the spectrum so for comparison we reproduce the same fits in \cref{fig: pion_masses_sp4} .  We also display the same plots and fits for the fully undecoupled model (the furthest extreme from the quenched version of our models). The fits for the quenched theory are reasonably close with gradients matching better than a factor of two in most cases. We note that our holographic model predicts that the slower the running of the coupling (the less quenched the quarks are) the sharper the slopes with $M_\pi$  - this effect was previously seen for walking theories in \cite{Erdmenger:2014fxa}. It would be interesting to see if this result was reproduced in unquenched lattice computations.

\subsubsection{Holography of the top partners \label{sec: yukawa_sp4}  }

The top partners are $F A_2 F$ spin 1/2 baryons of the strongly coupled dynamics that play a key role in the generation of the top quark mass as described in the \cref{review}. We have computed their masses in the Sp(4) model which are shown in \Cref{Sp4table} - we remind that we have computed the masses as if all constituents have a dynamical mass given by first the fundamental and secondly the $A_2$ representations. The true mass is likely to lie between these values.

For the top mass there are two key contributions as we can see in \cref{topdiagram}. The top Yukawa coupling, 
\begin{equation} \label{topyuk} y_t = { g^2 Z ~ \tilde{g}^2 \tilde{Z} ~  Z_3 \over M_T^2~ \Lambda_{UV}^4}  \, ,\end{equation}
is inversely proportional to the top partner mass squared. It is proportional to the $Z_3$ and $Z/\tilde{Z}$ factors which we will set equal.  The $Z$ factors, like the baryon-$\sigma$ vertex in QCD,  are not direct predictions of the holographic framework since they must be generated by couplings beyond the basic quadratic  terms of the holographic action \cref{eq: general_action} and so in principle one can add new couplings. We can though write down holographic terms that are likely to be the dominant contributions and look at their order of magnitude behaviour. In particular we have
\begin{equation} \label{eq: top_mass_1}  
Z_3 \simeq \int d \rho ~ \rho^3~ { \partial_\rho \pi(\rho) ~ \psi_B(\rho)^2 \over ( \rho^2 +L^2)^2} \, ,
\end{equation}
\begin{equation} \label{eq: top_mass_2}
Z = \tilde{Z}  \simeq \int d \rho ~\rho^3 \partial_\rho\psi_B(\rho) \, .
\end{equation}
Here $\pi(\rho)$ and $\psi_B(\rho)$ are the holographic wavefunctions for the pNGB and the baryon repectively.
They are normalized to give canonical kinetic terms for these states as in \cref{Bnorm}.

If we naively compute the top Yukawa coupling, from the full set of factors in \cref{topyuk} (with $g=\tilde{g}=1$),  in the $Sp(4)$ model,  with a cut off on the HDOs of roughly 6 times the vector meson mass, we find the top Yukawa coupling is only of order 0.01 which is far below the value of one needed. 

\begin{figure}[H]
\centering
\includegraphics[scale=0.6]{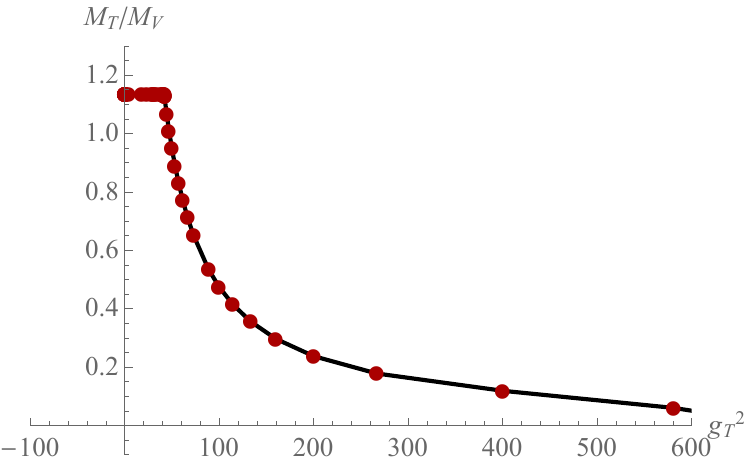} 
\centering
\includegraphics[scale=0.6]{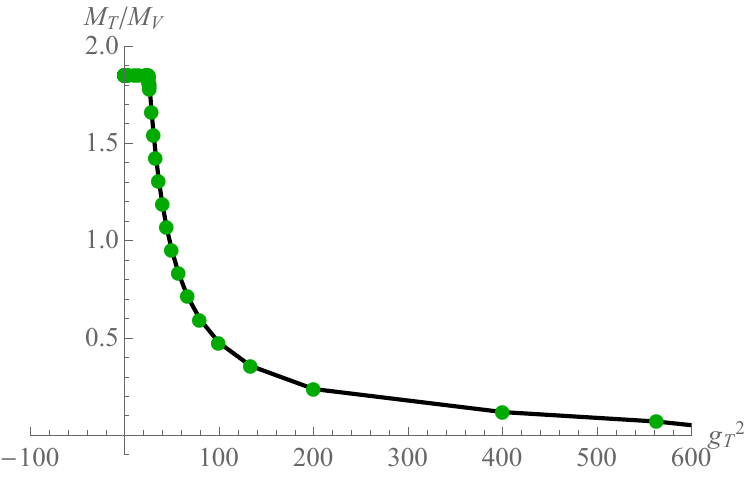} 
\caption{\label{fig: spin_1_2_dt_sp4} AdS/$Sp(4) ~4F,6A_2$ - We show the effect of adding the double-trace operator \cref{Tphdo} to the spin-$1/2$ baryon's mass. On the left we use the $L_0(\rho)$ from the  F representation  and on the right $L_0(\rho)$ from the the $A_2$ representation. Note the initial linear behaviour when $g_T^2$ is perturbative but then as it passes a critical value the effect on the mass is much larger.
}
\end{figure}

The top Yukawa would be enhanced if the top partners were anomalously light relative to the strong coupling scale (roughly the scale 1 in our \Cref{Sp4table}). As we have described in QCD, it is possible to drive the baryons light by including a HDO - see \cref{fig: double_trace_qcd} for example. In the Sp(4) theory we can also look to include a HDO of the form
\begin{equation}  \label{Tphdo}
{\cal L}_{HDO} = {g_T^2 \over \Lambda_{UV}^5}  |F A_2 F|^2 \, .
\end{equation}
As the operator $F A_2 F$ becomes the top partner field, this is directly a shift in the top partner mass.\footnote{A similar effective operator was mentioned in \cite{Kaplan:1991dc}, but there it is  not included in the dynamical calculations.} In \cref{fig: spin_1_2_dt_sp4} we show the dependence of the top partner mass on $g_T^2$ - we show the effect using both the $F$ and $A_2$ embeddings as $L_0(\rho)$ in \cref{eq: eqn of motion_fermions}. The HDO can indeed be used to reduce

\noindent the top partner mass - for small $g_T^2$ the effect is linear and small but after a critical value the effect is much larger, as shown.

\begin{figure}[H]
\centering
\includegraphics[scale=0.6]{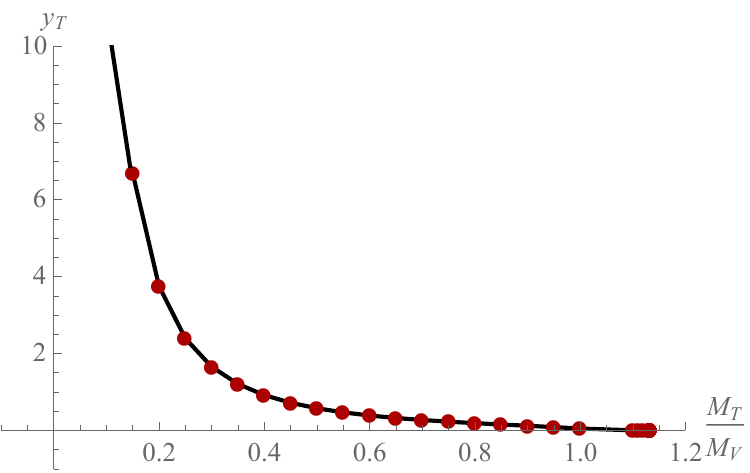} 
\centering
\includegraphics[scale=0.6]{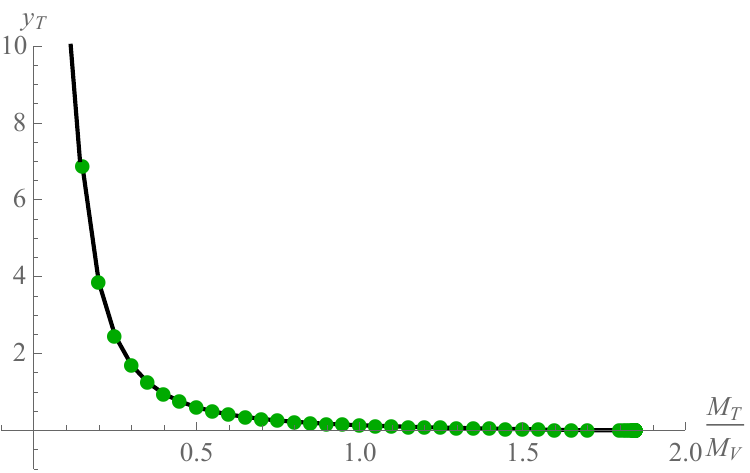} 
\caption{\label{fig: spin_1_2_dt_sp4_Z} AdS/$Sp(4) ~4F,6A_2$ -   the top Yukawa coupling, as given by \ref{topyuk}, is  plotted against the top partner mass in units of the vector meson mass. $M_T$ is controlled by adding a HDO as in \cref{fig: spin_1_2_dt_sp4}. We compute on the left with $L_0(\rho)$ for the fundamental quark and on the right we use the $A_2$ $L_0(\rho)$. 
}
\end{figure}

One must be careful though because as the top partners' mass changes so also do the $Z$ factors in \cref{eq: top_mass_1}   and \cref{eq: top_mass_2}. In particular as the HDO in \cref{Tphdo} plays a large role it induces a sizeable non-normalizable piece in the UV holographic wave function of the top partner. This means that the integrals in the equivalent of the normalization factors in \cref{Bnorm} and directly in the expressions for the $Z$  factors are more dominated by the UV part of the integral. The overlap between different states can change substantially. We therefore plot the full expression for the Yukawa coupling from \cref{topyuk} against the top partner  mass (which changes as we dial $g_T^2$) in \cref{fig: spin_1_2_dt_sp4_Z}. We see that the top Yukawa does indeed grow as
 the

\noindent  top partner's mass falls and can become of order one as the top partners mass falls to about half of the vector meson mass. This suggests, that after fixing the strong coupling scale to a sensible large value in the 1-5 TeV range,  we should be able to realize a top partner mass of about 1 TeV  and the required top mass.

\newpage
\subsection{$SU(4)$ gauge theory with top partners - $SU(4) ~ 3F, 3 \bar{F}, 5 A_2$ } \label{sec: su4_model}

The next model we choose to study is one taken from \cite{Ferretti:2014qta,Ferretti:2016upr} for which there has been related lattice work \cite{Ayyar:2018zuk,Ayyar:2017qdf}. The gauge group is $SU(4)$. There are five Weyl fields in the sextet $A_2$ representation. When these $A_2$ condense they break their $SU(5)$ symmetry to $SO(5)$ - the pNGBs inlcude the Higgs.

To include top partner baryons, fermions in the fundamental representation $F$ are added allowing $F A_2 F$ states. To make these states QCD coloured we need three Dirac spinors in the fundamental. When these fields condense the chiral $SU(3)_L \times SU(3)_R$ symmetry is broken to the vector $SU(3)$ subgroup - the $SU(3)$ sub-group is identified with weakly coupled QCD (which we will neglect since it is weak at the scales in question).

The full symmetry breaking pattern and embedding of the SM groups is
\begin{align}
SU(5) \times SU(3)_L \times SU(3)_R \times U(1) & 
  \to \underbrace{SO(5)}_{SU(2)_L  \times U(1)} \times \underbrace{SU(3)}_{SU(3)} \times U(1)
\end{align}

For the holographic model we need the running of the coupling  \cref{running} and $\gamma$ \cref{grun}. These then feed into $\Delta m^2$ in \cref{dm} to define the model. The coefficients of the one and two-loop $\beta$-function read
\begin{align} \label{eq: renormalization_ferretti_1}
\begin{split}
b_0 &= \frac{1}{6 \pi} \Big(11N_c  -  N_{f_{1}} -  (N_c-2) N_{f_{2}} \Big)\, , \\
b_1 &= \frac{1}{24 \pi^2} \left(34 N_c^2 - 5 N_c N_{f_{1}} - {3\over 2} \frac{N_c^2-1}{N_c} N_{f_{1}}  \right. \\
&\left. -5 N_c (N_c-2) N_{f_{2}} - 3 \frac{(N_c+1)(N_c-2)^2}{N_c} N_{f_{2}} \right)\, .
\end{split}
\end{align}
and the one-loop anomalous dimensions for the different representations are 
\begin{equation}  \label{eq: renormalization_ferretti_2}
\begin{split}
\gamma_{A_2} &= \left( \frac{6}{4 \pi} \frac{(N_c+1)(N_c-2)}{N_c}  \right)   \alpha\, , \\
\gamma_{F} &= \left( \frac{3}{4 \pi} \frac{N_c^2-1}{N_c}   \right)   \alpha\, .
\end{split}
\end{equation}

Naively one would expect the $A_2$ fermions to condense ahead of the fundamental fields since the critical value for $\alpha$ where $\gamma=1/2$ (the criteria discussed below \cref{dm})  is smaller. If we extend the perturbative results into the non-perturbative regime we find
\begin{equation} 
\alpha_c^F =  {8 \pi \over 45} = 0.56\, , \hspace{1cm} \alpha_c^{A_2} = {2 \pi \over 15} = 0.42\, .
\end{equation}
As in the $Sp(4)$ model we will ask how quickly the $A_2$ fields decouple from the running of $\gamma$ below their IR mass scale. 

This model is hard to simulate on the lattice because of the fermion doubling problem and the sign problem associated to chiral theories so instead lattice work \cite{Ayyar:2017qdf,Ayyar:2018zuk} has focused on the theory with just two Dirac $A_2$s and 2 Dirac fundamental quarks. In the next subsection we  will switch to the holographic description of that model and the comparison to the lattice data before returning to the full model thereafter. Of course, the ability to simply switch fields in and out is one of the huge benefits of the holographic approach.

\subsubsection{The lattice variant of the model - $SU(4) ~ 2F, 2 \bar{F}, 4 A_2$}
Here we consider a model with an $SU(4)$ gauge theory with two Dirac sextet and two Dirac fundamental quarks. 
We again run two separate holographic models for the $F$ and $A_2$ (though linked through the different representations contributions to the running of $\alpha$) which neglects mixing between the two sectors. 

\begin{figure}[H]
\centering
\begin{minipage}{.5\linewidth}
\subfloat[] {\label{main:a_su4_vacua_lattice} \includegraphics[scale=0.6]{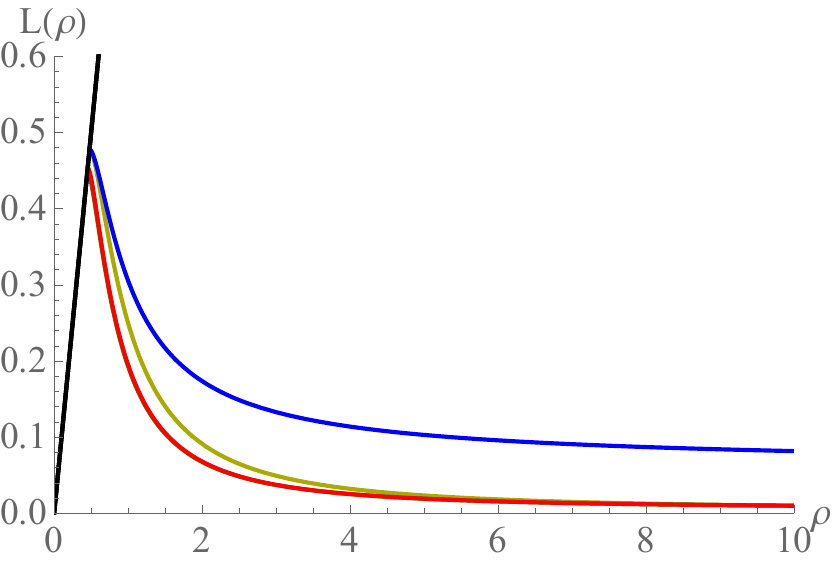}}
\end{minipage}%

\centering
\caption{$SU(4)$  $2F, 2 \bar{F}, 4 A_2$ - We display the vacuum solutions $L(\rho)$: the gold line corresponds to  the $A_2$ representation and the red is the F.  The blue line is the fundamental when we consider an additional NJL-term such that it matches in the IR the $A_2$ representation. }
\label{fig: SU4AllVacua_v2_lattice}
\end{figure}

We set our model parameters, as defined in \cref{sec: QCD},  using
\begin{align}
m_{q}|_{UV} &= 0, \qquad \alpha(0) =0.65, \qquad g_{5}|_{4} = \frac{24 \pi^2}{N_{f_{1}} N_c}, \qquad \qquad g_{5}|_{6} =  \frac{48 \pi^2}{N_{f_{2}} N_c(N_c-1)}\, .
\end{align}
  We run two schemes - one where the $A_2$s contribute to the running of $\alpha$ at all scales and one where we decouple them at their IR mass scale. The vacuum profiles for $L(\rho)$ are shown in \cref{fig: SU4AllVacua_v2_lattice} - here the coupling runs sufficiently quickly with or without the $A_2$ fields that the differences in the $L(\rho)$ function for the F quarks lies within the line width, whether the $A_2$ decouple or not. It would be nice if the lattice could teach us how to enact this decoupling. Here though  the errors on the  lattice data are still too large to distinguish these two scenarios, so again we lack data on precisely how to decouple quarks in the strong coupling regime.  Finally we also compute, and display in  \cref{fig: SU4AllVacua_v2_lattice}, for the theory with an NJL operator  ($g_s^2/\Lambda_{UV}^2 |\bar{F} F|^2$) which allows us to bring the $F$ IR mass equal to the $A_2$ IR mass.

\newpage

Next we compute the spectrum and display the predictions in  \Cref{tab: results_su4_lattice_two_runnings}. We also display lattice data  from  \cite{Ayyar:2018zuk} (also  \cite{Ayyar:2017qdf} and there is a relevant chiral perturbation theory analysis for the model in  \cite{DeGrand:2016pgq}).  The holographic model and the lattice data agree well in describing  the split in mass between the vector mesons of the $F$ and $A_2$ sectors (the differences in decoupling choices lie within the error bars).

The top partner baryon is a mixed $FA_2 F$ state. Again we estimate the possible spread of its mass by using in turn the $L_0(\rho)$ from the $F$ and $A_2$ sectors, essentially assuming the $F$ and $A_2$ have the same constituent masses  at either the lower $F$ or higher $A_2$ scale. The

\begin{figure}[H]
A2 Sector  $\left. \right.$ \hspace{5cm} $\left. \right.$  \bigskip

\includegraphics[scale=0.6]{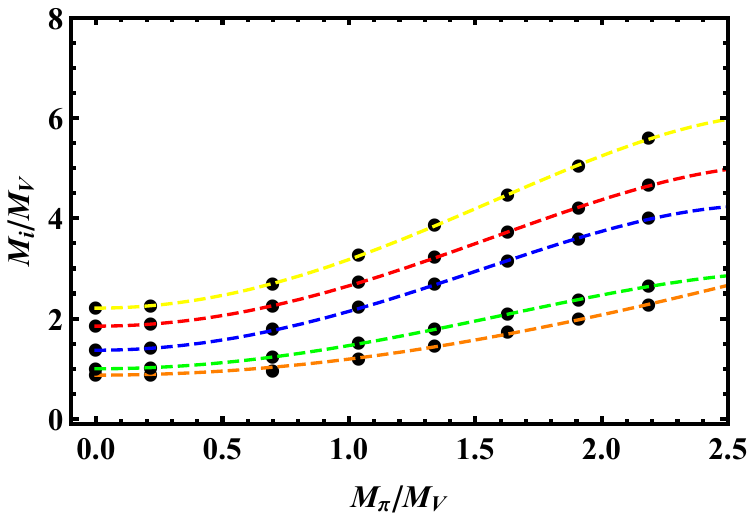} $\left. \right.$ \hspace{8cm}  \vspace{-3.9cm}

$\left. \right.$ \hspace{8cm} 
\begin{tabular}{cl} 
\color{orange}$M_{S}$ &\color{orange}= $\color{orange}0.873 + \color{orange}0.326 ~ \color{orange}M^2_{\pi} - \color{orange}0.00626 ~ \color{orange}M^4_{\pi}\, ,$ \\
\color{green}$M_{V}$ &\color{green}= $\color{green}1 + \color{green}0.494 ~ \color{green}M^2_{\pi} - \color{green}0.0315 ~ \color{green}M^4_{\pi}\, ,$  \\
\color{blue}$M_{A}$ &\color{blue}= $\color{blue}1.37 + \color{blue}0.834 ~ \color{blue}M^2_{\pi} - \color{blue}0.0602 ~ \color{blue}M^4_{\pi}\, ,$ \\
\color{red}$M_{B}$ &\color{red}= $\color{red}1.85 + \color{red}0.861 ~ \color{red}M^2_{\pi} - \color{red}0.0577 ~ \color{red}M^4_{\pi}\, ,$ \\
\color{yellow}$M_{J}$ &\color{yellow}= $\color{yellow}2.21 + \color{yellow}1.04 ~ \color{yellow}M^2_{\pi} - \color{yellow}0.0696 ~ \color{yellow}M^4_{\pi}\, .$
\end{tabular}  \vspace{2cm}

$F$ Sector  $\left. \right.$ \hspace{5cm} $\left. \right.$  \bigskip

\includegraphics[scale=0.6]{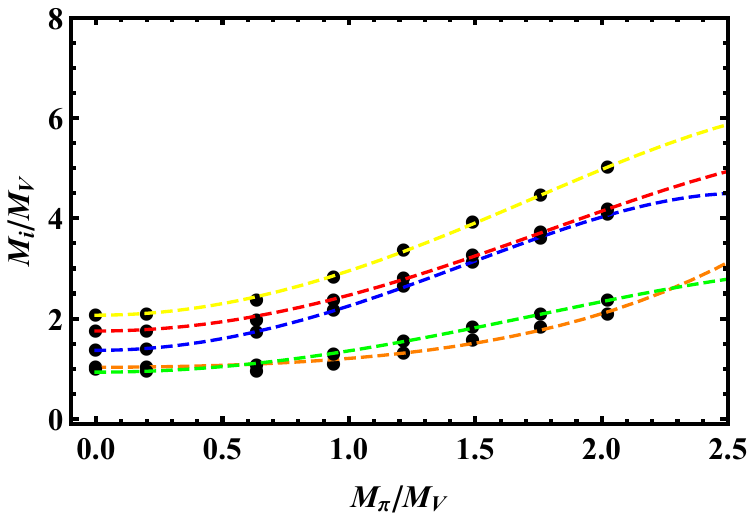} $\left. \right.$ \hspace{8cm}  \vspace{-3.9cm}

$\left. \right.$ \hspace{8cm}  
\begin{tabular}{cl} 
\color{orange}$M_{S}$ &\color{orange}= $\color{orange}1.03 + \color{orange}0.146 ~ \color{orange}M^2_{\pi} + \color{orange}0.0302 ~ \color{orange}M^4_{\pi}\, ,$ \\
\color{green}$M_{V}$ &\color{green}= $\color{green}0.933 + \color{green}0.448 ~ \color{green}M^2_{\pi} - \color{green}0.0241 ~ \color{green}M^4_{\pi}\, ,$  \\
\color{blue}$M_{A}$ &\color{blue}= $\color{blue}1.37 + \color{blue}0.957 ~ \color{blue}M^2_{\pi} - \color{blue}0.0731 ~ \color{blue}M^4_{\pi}\, ,$ \\
\color{red}$M_{B}$ &\color{red}= $\color{red}1.75 + \color{red}0.753 ~ \color{red}M^2_{\pi} - \color{red}0.039 ~ \color{red}M^4_{\pi}\, ,$ \\
\color{yellow}$M_{J}$ &\color{yellow}= $\color{yellow}2.07 + \color{yellow}0.935 ~ \color{yellow}M^2_{\pi} - \color{yellow}0.0522 ~ \color{yellow}M^4_{\pi}\, .$
\end{tabular} 
\vspace{2.2cm}

\caption{\label{fig: specta_deformed_vaccum_su4_lattice_A2_analytics} $SU(4)$  $2F, 2 \bar{F}, 4 A_2$ - The growth of the $A_2$ and $F$ sectors spectra as we increase the quark mass in the UV. The masses are rescaled with respect to the vector meson mass in the $A_2$ representation at $m_{q}|_{UV}=0$ in accord with the presentation in  \Cref{tab: results_su4_lattice_two_runnings}.  Here  $M_{\pi}$ is the pNGB mass in units of the vector meson mass at $m_{q}|_{UV}=0$. 
}
\end{figure}

 \noindent holographic model over estimates the top partner mass by 30\%.

There is lattice data for an additional spin zero state made of four quarks (either all $F$s or all $A_2$s), that we refer to as a tetraquark,  and denote as the $J$ in \cref{tab: results_su4_lattice_two_runnings}. We have computed the mass of such a state using \cref{eq: eqn of motion_spinless} - here the holographic prediction is that the $F$ and $A_2$ tetraquarks' masses lie within 10\%.  In contrast the lattice prediction suggests a factor of two between the masses of the states. It is hard to understand how such a large separation could occur  when the constituent quark masses are very similar for the $F$s and $A_2$ as measured by the vector meson masses. It would be interesting to look into the origin of the splitting in the lattice simulations further.

Finally in \cref{fig: specta_deformed_vaccum_su4_lattice_A2_analytics} we display the $M_\pi$ dependence of the spectrum in the non-decoupling scenario although here we do not have lattice data for comparison. \vspace{2cm}

\begin{table}[H]
{\setlength
\doublerulesep{0.5pt}   
\resizebox{\textwidth}{!}
{
\begin{tabular}{cccccccc}  
  \toprule[0.5pt]\midrule[0.5pt]
     & Lattice \cite{Ayyar:2018zuk} & AdS/$SU(4)$ & AdS/$SU(4)$ & AdS/$SU(4)$ & AdS/$SU(4)$ & AdS/$SU(4)$ & AdS/$SU(4)$  \\
     & $ 4A_2, 2 F, 2 \bar{F} $ & $ 4A_2, 2 F, 2 \bar{F} $ & $ 4A_2, 2 F, 2 \bar{F} $ & $ 5A_2, 3 F, 3 \bar{F} $ & $ 5A_2, 3 F, 3 \bar{F} $ & $ 5A_2, 3 F, 3 \bar{F}$ & $ 5A_2, 3 F, 3 \bar{F}$ \\ 
             					& unquench 	& no decouple	& decouple	 	& no decouple 	& decouple 	& quench 	& + NJL \\ \midrule
$f_{\pi A_2}$				& 0.15(4) 		&  0.0997  			& 0.0997			& 0.111				& 0.111 			& 0.102		& 0.11				\\
$f_{\pi F}$ 				& 0.11(2) 		& 0.0949  			& 0.0953			&0.0844			& 0.109			& 0.892		& 0.139				\\
$M_{V A_2}$ 			& 1.00(4) 		&  1*  				&	1*					& 1*					& 1*				& 1*			& 1* 					\\
$f_{V A_2}$ 			& 0.68(5)		&  0.489  			&	0.489			& 0.516				& 0.516			& 0.517		& 0.516				\\
$M_{V F}$ 			& 0.93(7) 		& 0.933  			&	0.939			& 0.890 			& 0.904 		& 0.976		& 1.02				\\
$f_{V F}$  			& 0.49(7) 		& 0.458  			&	0.461			&0.437  			& 0.491 		& 0.479		& 0.495				 \\
$M_{A A_2}$ 				&   				&  1.37  			&	1.37				&1.32				& 1.32 			& 1.28		& 1.32				 \\
$f_{A A_2}$ 				&   				&  0.505  			&	0.505			& 0.521				& 0.521 		& 0.522		& 0.521				 \\
$M_{A F}$ 				&  				& 1.37  				&	1.37				& 1.21 				& 1.23			& 1.28		& 1.46				 \\
$f_{A F}$  				&  				& 0.501 			&	0.504			& 0.453 			& 0.509			& 0.492		& 0.489				  \\
$M_{S A_2}$ 			&  				& 0.873  			&	0.873			& 0.684 			& 0.684			& 1.18		& 0.684				   \\
$M_{S F}$  			&  				& 1.03 				&	1.02				& 0.811 				& 0.798			& 1.25		& 0.815					\\
$M_{J A_2}$ 				& 3.9(3) 		& 2.21   			&	2.21				& 2.21				& 2.21 			& 2.22		& 2.21						\\
$M_{J F}$ 				& 2.0(2)  		&  2.07  			&	2.08				&  1.97 				& 2.00 			& 2.17		& 2.24					\\
$M_{B A_2}$ 			& 1.4(1) 		&  1.85  			&	1.85				&   1.85 			& 1.85 			& 1.86		& 1.85					\\
$M_{B F}$ 			& 1.4(1) 		&  1.74  			&	1.75				&  1.65				& 1.68 			& 1.81		& 1.88					\\
   \midrule[0.5pt]\bottomrule[0.5pt]
  \end{tabular} 
}
  }  
  \caption{\label{tab: results_su4_lattice_two_runnings} $SU(4)$  theories - the spectrum in a variety of scenarios and lattice data for comparison. 
}
\end{table}

\newpage

\subsubsection{$SU(4) ~ 3F, 3 \bar{F}, 5 A_2$  model - vacuum configuration}
\label{sec:M6}
After this small digression to the lattice variant, we return to the study of the model actually proposed for composite Higgs models - $SU(4)$  $3F, 3 \bar{F}, 5 A_2$. The coefficients of the one and two-loop $\beta$-function are still given by \cref{eq: renormalization_ferretti_1} and the $\gamma$s in \cref{eq: renormalization_ferretti_2} with appropriate choices of numbers of flavours. We choose as previously, see \cref{sec: QCD},
\begin{align}
m_{q}|_{UV} &= 0, \qquad \alpha(0) =0.65, \qquad g_{5}|_{4} = \frac{24 \pi^2}{N_{f_{1}} N_c}, \qquad \qquad g_{5}|_{6} =  \frac{48 \pi^2}{N_{f_{2}} N_c(N_c-1)}\, .
\end{align}

\begin{figure}[H]
\begin{minipage}{.5\linewidth}
\centering
\subfloat[] {\label{main:a_su4_vacua} \includegraphics[scale=0.5]{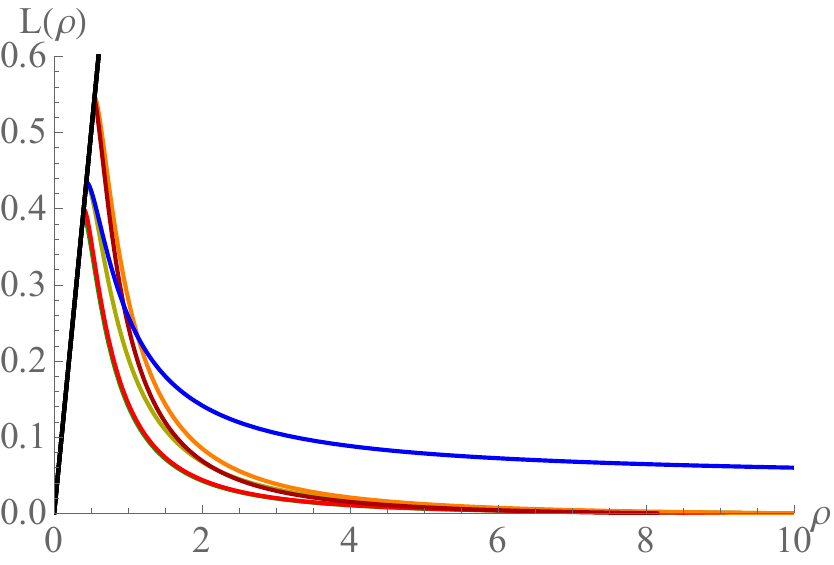}}
\end{minipage}%
\begin{minipage}{.5\linewidth}
\centering
\subfloat[]{ \label{main:b_su4_vacua} \includegraphics[scale=0.55]{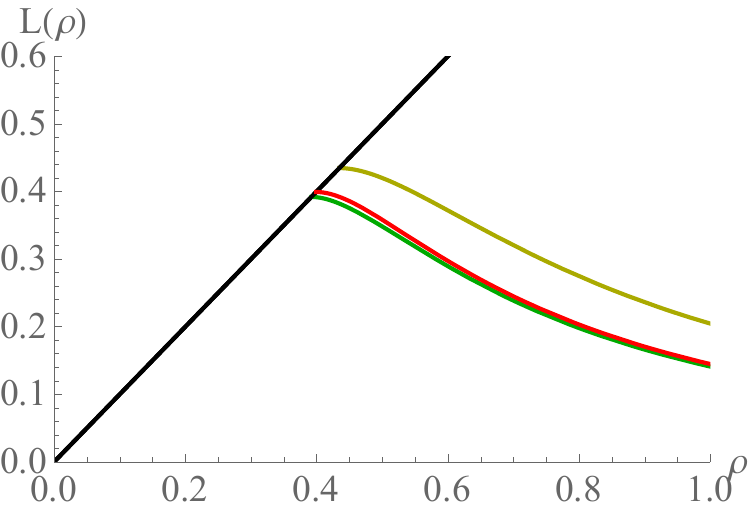}}
\end{minipage}\par\medskip
\centering
\caption{$SU(4)$  $3F, 3 \bar{F}, 5 A_2$ - In the left plot we display the vacuum soluions $L(\rho)$: the gold line corresponds to $AdS/SU(4)$ for the $A_2$ representation and the green is the fundamental. The red vacuum solution is when we consider the decoupling of the $A_2$ which condenses before the fundamental. The blue line is the fundamental when we consider additional NJL-terms such that it matches in the IR the $A_2$ representation. Finally, the orange and purple vacua correspond to the quenched models for the $A_2$ and fundamental representations respectively. The right hand plot is a zoom in when considering the $AdS/SU(4)$ model without decoupling and when we consider the decoupling of the $A_2$ quark fields.}
\label{fig: SU4AllVacua_v2}
\end{figure}

To address the decoupling of the $A_2$ we will present results for the vacuum solution, $L(\rho)$, in a number of different cases in \cref{fig: SU4AllVacua_v2}. Firstly we do not decouple the $A_2$ from the running of $\alpha$ at any scale - the gold line corresponds to  the $A_2$ representation and the green the fundamental. There is a small gap with the fundamentals a little lighter. 
If we decouple the $A_2$ fields at scales below their IR mass $L_{IR}$ then the fundamental $L(\rho)$ becomes the red embedding. Plot \ref{main:b_su4_vacua} is a zoom in showing the difference between the non-decoupled and the decoupling cases - in this model the separation barely changes when the decoupling is implemented. 

It is possible to make the IR mass scales the same for both representations by including an NJL interaction for the fundamental fields ($g_s^2 / \Lambda_{UV}^2 |\bar{F}F|^2$). The blue line in \cref{fig: SU4AllVacua_v2} is for the fundamental representation when we consider an additional NJL-term such that it matches in the IR the $A_2$ representation $L_{IR}$.

Finally, the orange and purple vacua correspond to the quenched models for the $A_2$ and fundamental representations respectively - here we don't include the fermions in the running at all. We include this example because it would be relatively cheap to perform a lattice simulation of the theory in the quenched limit so our results may be of future interest.


It is worth commenting on the size of the IR mass, $L_{IR}$ in physical units. We will compute the spectrum in the next section but borrowing ahead we can write $L_{IR}$  in units of the vector meson's mass in the $A_2$ representation for the  case discussed. For the model where we do not integrate out the $A_2$ fields we have $L_{IR}^{A_2}=0.308 m_V$ and $L_{IR}^{F}=0.278 m_V$. When we integrate

\begin{figure}[H]
A2 Sector  $\left. \right.$ \hspace{5cm} $\left. \right.$  \bigskip

\includegraphics[scale=0.6]{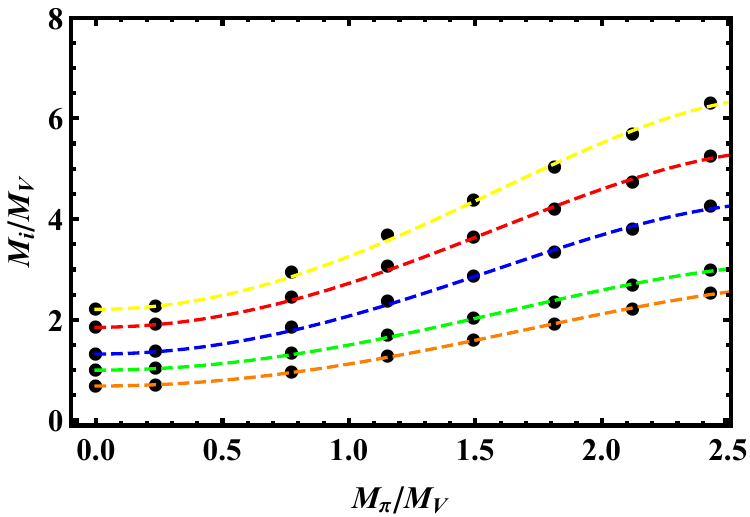} $\left. \right.$ \hspace{8cm}  \vspace{-3.9cm}
 
$\left. \right.$ \hspace{8cm} 
\begin{tabular}{cl} 
\color{orange}$M_{S}$ &\color{orange}= $\color{orange}0.684 + \color{orange}0.463 ~ \color{orange}M^2_{\pi} - \color{orange}0.0263 ~ \color{orange}M^4_{\pi}\, ,$ \\
\color{green}$M_{V}$ &\color{green}= $\color{green}1 + \color{green}0.532 ~ \color{green}M^2_{\pi} - \color{green}0.0338 ~ \color{green}M^4_{\pi}\, ,$  \\
\color{blue}$M_{A}$ &\color{blue}= $\color{blue}1.32 + \color{blue}0.808 ~ \color{blue}M^2_{\pi} - \color{blue}0.0541 ~ \color{blue}M^4_{\pi}\, ,$ \\
\color{red}$M_{B}$ &\color{red}= $\color{red}1.85 + \color{red}0.933 ~ \color{red}M^2_{\pi} - \color{red}0.0619 ~ \color{red}M^4_{\pi}\, ,$ \\
\color{yellow}$M_{J}$ &\color{yellow}= $\color{yellow}2.21 + \color{yellow}1.13 ~ \color{yellow}M^2_{\pi} - \color{yellow}0.0748 ~ \color{yellow}M^4_{\pi}\, ,$
\end{tabular} \vspace{2cm}

F Sector  $\left. \right.$ \hspace{5cm} $\left. \right.$  \bigskip

\includegraphics[scale=0.6]{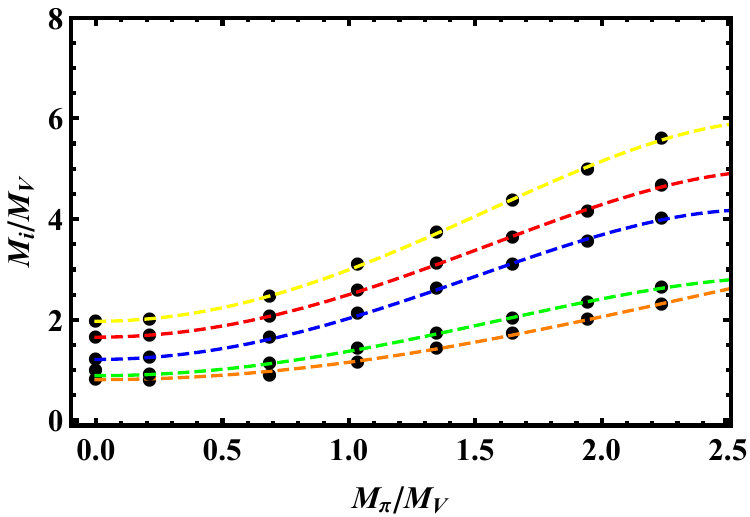} $\left. \right.$ \hspace{8cm}  \vspace{-3.9cm}

$\left. \right.$ \hspace{8cm} 
\begin{tabular}{cl} 
\color{orange}$M_{S}$ &\color{orange}= $\color{orange}0.811 + \color{orange}0.356 ~ \color{orange}M^2_{\pi} - \color{orange}0.0109 ~ \color{orange}M^4_{\pi}\, ,$ \\
\color{green}$M_{V}$ &\color{green}= $\color{green}0.89 + \color{green}0.519 ~ \color{green}M^2_{\pi} - \color{green}0.0344 ~ \color{green}M^4_{\pi}\, ,$  \\
\color{blue}$M_{A}$ &\color{blue}= $\color{blue}1.21 + \color{blue}0.879 ~ \color{blue}M^2_{\pi} - \color{blue}0.065 ~ \color{blue}M^4_{\pi}\, ,$ \\
\color{red}$M_{B}$ &\color{red}= $\color{red}1.65 + \color{red}0.907 ~ \color{red}M^2_{\pi} - \color{red}0.062 ~ \color{red}M^4_{\pi}\, ,$ \\
\color{yellow}$M_{J}$ &\color{yellow}= $\color{yellow}1.97 + \color{yellow}1.1 ~ \color{yellow}M^2_{\pi} - \color{yellow}0.0755 ~ \color{yellow}M^4_{\pi}\, ,$
\end{tabular} 
\vspace{2.2cm}

\caption{\label{fig: specta_deformed_vaccum_su4_full_A2_analytics} $SU(4)$  $3F, 3 \bar{F}, 5 A_2$ - The growth of the spectra as we increase the quark mass in the UV. The masses are rescaled with respect to the vector meson mass in the $A_2$ representation at $m_{q}|_{UV}=0$ in accord with the presentation in  \Cref{tab: results_su4_lattice_two_runnings}.  In our analytic formulae  the scale is again set by the $A_2$ vector meson mass at $m_{q}|_{UV}=0$.}
\end{figure}

\noindent out the $A_2$s on mass shell we have $L_{IR}^{A_2}=0.308 m_V $ and $L_{IR}^{F}=0.283 m_V $.For the model with the NJL interaction for the fundamental we have  $L_{IR}^{A_2}=L_{IR}^{F}=0.308 m_V $. For the quenched model we have $L_{IR}^{A_2}=0.318 m_V $ and $L_{IR}^{F}=0.316 m_V $.

\subsubsection{$SU(4) ~ 3F, 3 \bar{F}, 5 A_2$  model - spectrum}

We can now compute the spectrum of the theory in each of these cases. We display the results in \cref{tab: results_su4_lattice_two_runnings} so it is easy to compare to the lattice variant model. The spectra are fairly similar in all cases but the key changes occur as more fermions are included in the running. Thus increasing the number of fields slows the running which firstly increases the gap between 
the $A_2$ and $F$ sectors and secondly reduces the $\sigma$ scalar mass. 
For completeness in \cref{fig: specta_deformed_vaccum_su4_full_A2_analytics} we show the dependence of the meson and baryon masses on the pNGB mass for of the $F$ and $A_2$ sectors, although there is no lattice data to compare to here. 

\subsubsection{Top partners}

The top partners are $F A_2 F$ spin 1/2 baryons of the strongly coupled dynamics that play a key role in the generation of the top quark mass as described in the \Cref{review}. We have computed their masses in the $SU(4)$ model which are shown in \Cref{tab: results_su4_lattice_two_runnings}  - we remind that we have computed the masses as if all constituents have a dynamical mass given by first the fundamental and secondly the $A_2$ representations. The true mass is likely to lie between these values.

For the top mass there are two key contributions as we can see in \cref{topdiagram}. The top Yukawa coupling,
\begin{equation} \label{topyuk2} y_t = { g^2 Z ~ \tilde{g}^2 \tilde{Z} ~  Z_3 \over M_T^2 ~\Lambda_{UV}^4}  \, ,\end{equation}
is inversely proportional to the top partner mass squared. It is proportional to the $Z_3$ and $Z/\tilde{Z}$ factors.  The $Z$ factors, like the baryon-$\sigma$ vertex in QCD,  are not direct predictions of the holographic framework since they must be generated by couplings beyond the basic quadratic  terms of the holographic action \cref{eq: general_action} and so in principle one can add new couplings. We can though write down holographic terms that are likely to be the dominant contributions and look at their order of magnitude behaviour. As in the previous model we may express the $Z$ factors by

\begin{equation} \label{eq: top_mass_1a}  
Z_3 \simeq \int d \rho ~\rho^3~  { \partial_\rho \pi(\rho) ~ \psi_B(\rho)^2 \over ( \rho^2 +L^2)^2}\, ,
\end{equation}

\begin{equation} \label{eq: top_mass_2a}
Z = \tilde{Z} \simeq \int d \rho~ \rho^3 \partial_\rho \psi_B(\rho)\, ,
\end{equation}
Here $\pi(\rho)$ and $\psi_B(\rho)$ are the holographic wavefunctions for the pNGB and baryon repectively.
They are normalized to give canonical kinetic terms for these states as in \cref{Bnorm}.

We compute the top Yukawa coupling (setting $g=\tilde{g}=1$), from the full set of factors in \cref{topyuk2}.  It is proportional to the $Z_3$ and $Z/\tilde{Z}$ factors which we will set equal. In this

\begin{figure}[H]
\centering
\includegraphics[scale=0.6]{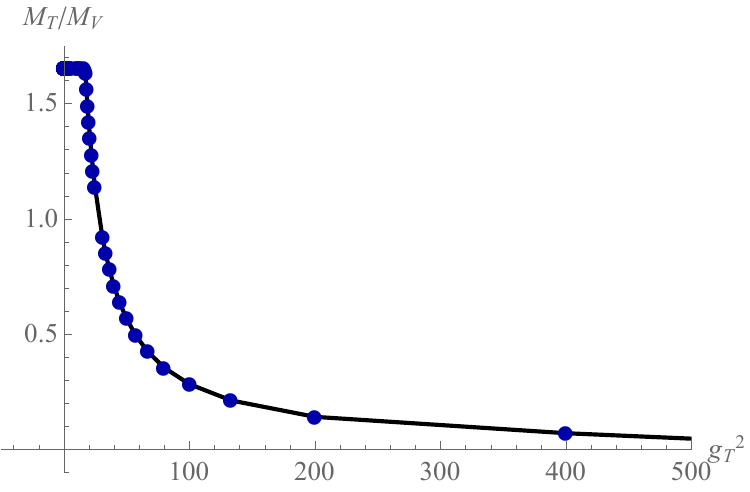} 
\centering
\includegraphics[scale=0.6]{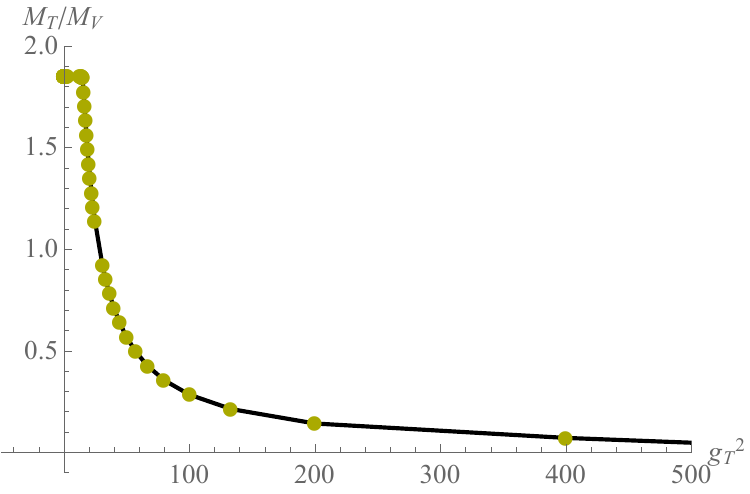} 
\caption{\label{fig: spin_1_2_dt_su4} $SU(4)$  $3F, 3 \bar{F}, 5 A_2$ - We show the effect of adding the double-trace operator \cref{Tphdo2} to the spin-$1/2$ baryon's mass. On the left we use the $L_0(\rho)$ from the  F representation  and on the right $L_0(\rho)$ from the the $A_2$ representation. Note the initial linear behaviour when $g_T^2$ is perturbative but then as it passes a critical value the effect on the mass is much larger.}
\end{figure}

\noindent $SU(4)$ model,  with a cut off on the HDOs, as an example,  of roughly 6 times the vector meson mass, we find the top Yukawa coupling is only of order 0.01 which is far below the value of one needed.

The top Yukawa would be enhanced if the top partners were anomalously light relative to the strong coupling scale (roughly the scale 1 in our \Cref{tab: results_su4_lattice_two_runnings}). As we have described in QCD, it is possible to drive the baryons light by including a HDO - see \cref{fig: double_trace_qcd} for example. In the SU(4) theory we can also look to include a HDO of the form
\begin{equation}  \label{Tphdo2}
{\cal L}_{HDO} = {g_T^2 \over \Lambda_{UV}^5}  |F A_2 F|^2\, ,
\end{equation}
As the operator $F A_2 F$ becomes the top partner field this is directly a shift in the top partner mass. In \cref{fig: spin_1_2_dt_su4} we show the dependence of the top partner mass on $g_T^2$ - we show the effect using both the $F$ and $A_2$ embeddings as $L_0(\rho)$ in \cref{eq: eqn of motion_fermions}. The HDO can indeed be used to reduce the top partner mass - for small $g_T^2$ the effect is linear and small but after a critical value the effect is much larger, as shown.

\begin{figure}[H]
\centering
\includegraphics[scale=0.6]{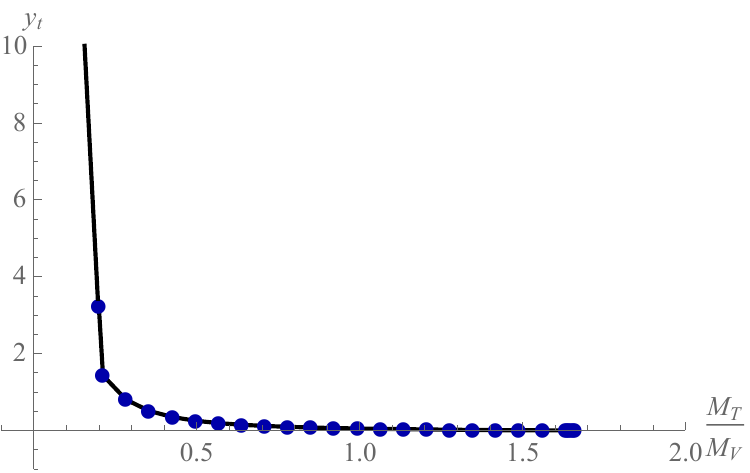} 
\centering
\includegraphics[scale=0.6]{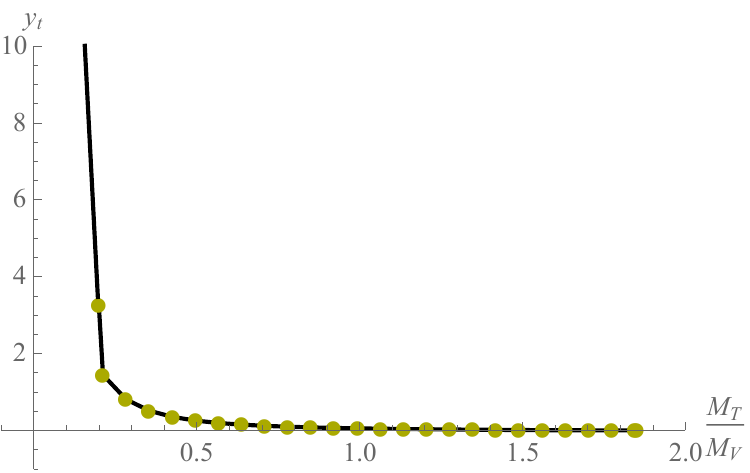} 
\caption{\label{fig: spin_1_2_dt_su4_Z} SU(4)  $3F, 3 \bar{F}, 5 A_2$ -  The top Yukawa coupling, as given by \ref{topyuk2}, is  plotted against the top partner mass units of the vector meson mass. $M_T$ is controlled by adding a HDO as in \cref{fig: spin_1_2_dt_sp4}. We compute on the left with $L_0(\rho)$ for the fundamental quark and on the right we use the $A_2$ $L_0(\rho)$. }
\end{figure}

One must be careful though because as the top partners mass changes so also do the $Z$ factors in \cref{eq: top_mass_1a}   and \cref{eq: top_mass_2a}. In particular as the HDO in \cref{Tphdo} plays a large role it induces a sizable non-normalizable piece in the UV holographic wave function of the top partner. This means that the integrals in the equivalent of the normalization factors in \cref{Bnorm} and directly in the expressions for the $Z$  factors are more dominated by the UV part of the integral. The overlap between different states can change substantially. We therefore plot the full expression for the Yukawa coupling from \cref{topyuk2} against the top partner mass (which changes as we dial $g_T^2$) in \cref{fig: spin_1_2_dt_su4_Z}. We see that the top Yukawa does indeed grow as the top partner's mass falls and can become of order one as the top partners mass falls to about half of the vector meson mass. This suggests, that after fixing the strong coupling scale to a

\noindent sensible large value in the 1-5 TeV range,  in this model we should be able to realize a top partner mass of about 1 TeV  and the required top mass. 

\newpage

\subsection{A catalogue of other composite Higgs models} \label{app: catalogue_results}

Finally, in part to demonstrate the flexibility of the holographic
method and in part as a service to model builders, we  will survey
many of the other gauge theories that have been proposed as composite
Higgs models with top partners. In particular we will calculate their
spectrum and decay constants. We are led by the proposals in
\cite{Ferretti:2013kya} and will identify them by the notation of
\cref{classification}. Here we do not know of any lattice data, so our
results for the meson spectrum and couplings plus the top partner mass
stand in isolation. We hope though they will be of potential use for future work.

All of the models proposed in \cite{Ferretti:2013kya}  that we consider are asymptotically free (they have positive $b_0$ in \cref{running}). However, we find  that some of the models live in the conformal window \cite{Appelquist:1996dq,Dietrich:2006cm} at the level of the approximation of the two loop running results we use. To lie in the conformal window $b_1$ in \cref{running} must be negative.
We then compute the value of  $\alpha_{c}$, which is the value of
$\alpha$ for $\gamma=1/2$, the criterion discussed below \cref{dm},
\begin{equation} \label{critc}
\alpha_c = \frac{ \pi}{3 C_2(r)},
\end{equation}
and the (positive) value of the coupling at the IR fixed point $\alpha^{*}$ where the $\beta$-function vanishes 
\begin{equation}  \label{me}
\alpha^{*} = - \frac{b_0}{b_1}.
\end{equation} 
We classify a gauge theory as lying in the conformal window if $\alpha^{*} < \alpha_{c}$. Such models do not break chiral symmetries and can not make good composite Higgs models. We will note these models below but not compute for them.

We will only compute for models that break chiral symmetries. Of course by adding NJL operators, one can force any gauge theory to break chiral symmetry. Such models, dominated by the NJL term,  have spectra that will depend on the initial condition of the gauge coupling at the UV cut off (a continuous parameter) so are not easily presented. We will therefore only address models where the gauge dynamics drives the symmetry breaking. 

Theories that lie close to the lower (in $N_f$) edge of the conformal window, yet still break chiral symmetry, have a slowly running gauge coupling and are referred to as walking theories - in these theories $\gamma$ runs from zero to one over a substantial regime of RG scale $\mu$, unlike in QCD where this running happens very quickly. The main evidence for walking in the spectrum of the theories we study is that the scalar mass ($S$) falls towards zero because the near conformality tends to flatten the effective potential for the quark condensate - this is the most significant result we find case by case in this section. Where theories break chiral symmetry for a range of $N_c,N_f$, we will display results at the extreme non-walking and walking values of the parameters. Due to their potential interest for model building, we stress  walking theories below, but they are therefore over represented in the sample of theories we present.

In this sub-section we will not decouple the heavier representations from the running coupling - in most cases the two or more representations of matter condense at similar scales (within a factor of 2) and our work on the previous two models suggest the precise form of the decoupling is an interesting but small effect. 
We  will not present as much detail as in  \cref{sec: sp4 model} and \cref{sec: su4_model}, 
   instead we will just display the results for the masses and couplings from the holographic description by theory.  In each case we will normalize to one of the  vector meson state's mass. 
For the numerical analysis  below we  fix $\alpha(0)=0.65$ and require a massless quark in the UV, $m_{q}|_{UV}=0$. 

\subsubsection{Models with exceptional gauge groups}
\label{subsec:pI}	

The first models for which we compute the spectrum and decay constants are those with  exceptional gauge groups that have been proposed in \cite{Ferretti:2013kya}. The gauge group can be either $G_2$ or $F_4$ with matter in the fundamental representation. There are singlet baryons made of three quarks in these cases (see \cite{Ferretti:2014qta} for a detailed discussion). The symmetry breaking pattern with $N_f$ Weyl fermions is $SU(N_f)\to SO(N_f)$. If $N_f \geq 11$ then the SM gauge group can be embedded in the global symmetry and a Higgs doublet and coloured top partners generated. 
In fact it has been argued that these models are not very promising phenomenologically \cite{Ferretti:2013kya} since there is a high number of pNGBs and some of them mediate proton decay.

The $G_2$ group is asymptotically free until $N_f=22$. The theory lies
in the conformal window according to the criteria discussed below
\cref{me} down to $N_f=16$. The $N_f=16$ theory actually has the fixed
point value equal to the chiral symmetry breaking coupling so is
maximally walking and would presumably have a massless scalar meson.
We present results for the extreme cases we can compute, i.e.~for $N_f=11$ and $N_f=15,$ in  \Cref{tab: results_so11}.

$F_4$ theory is asymptotically free until $N_f=16$. The edge of the
conformal window lies between $N_f = 12$ and $N_f =13$ flavours. We
present spectra for the  $N_f=11$ and $12$ cases also in \Cref{tab:
  results_so11} - both of these theories have a slowly running
coupling, resulting in a very light scalar.  The $A$ and $V$
mesons in these present models are more degenerate than in QCD.  

\begin{table}[H]
\centering
{\setlength
\doublerulesep{1pt}   
 \begin{tabular}{cccccc}
  \toprule[1pt]\midrule[1pt]
    Observables 	& $AdS/G_2$			& $AdS/G_2$				& $AdS/F_4$ 	&  $AdS/F_4$	   \\ 
					   	& $11F$				& $15F$					& $11F$ 			& 	$12F$   \\ \midrule
    $f_{\pi}$ 		& 0.0749 				& 0.0797					& 0.0486			& 0.0489	\\
    $M_{V}$ 		& 1* 						& 1*							& 1*					& 1*	\\
    $f_{V}$ 			& 0.456 				& 0.488						& 0.49				&0.501\\
    $M_{A}$ 		& 1.15  					& 1.11						&1.03				&1.03\\
    $f_{A}$ 			& 0.438  				& 0.470						&0.483				&0.494\\
    $M_S$ 			& 0.288 				& 0.114						&0.000431			&0.00039		\\
    $M_{B}$ 		& 1.78 					& 1.76						&1.55				&1.55	\\
   \midrule[1pt]\bottomrule[1pt]
  \end{tabular} 
  }  
  \caption{\label{tab: results_so11} Holographic predictions for the
    spectra and decay constants  of AdS/$G_2$,  $11F$, AdS/$F_4$,  $11F$, and AdS/$F_4$,  $16F$.}
\end{table}

\newpage

	\subsubsection{Models with matter in two representations}
Composite Higgs models with fermionic matter in two representations can only have either a  $Sp(2N)$ or $SO(N_c)$ gauge group and generate the Higgs and top partners. 

There are three possible scenarios with a symplectic gauge group: \vspace{0.5cm}

\begin{tabular}{cl}
$ Sp(2N) ~ 5 S_2, 6F ~~ N \geq 6 $  &  These theories are all in the conformal window.  \\
$Sp(2N) ~ 5 A_2, 6F ~~ N \geq 2$ &  Theories with $N<8$ are below the conformal window, \\
& $\left. \right.$ \hspace{2cm} and break chiral symmetry.  \\
$Sp(2N) ~ 4F, 6A_2 ~~N \leq18 $ &   Theories with $N<5$ are below the conformal window,\\
& $\left. \right.$ \hspace{2cm} and break chiral symmetry. \\
\end{tabular} \vspace{0.5cm}

\noindent We present the spectra for examples of the second and third models that break chiral symmetry. In particular we present for the minimum and maximum number of colours,  in \Cref{tab: results_sp_models} (note the final model with $Sp(4)$ is the one we studied in more detail in \cref{sec: sp4 model}).  The $Sp(14)$ case, which is the slowest walking of these theories, has a very low scalar mass. \vspace{1cm}

\begin{table}[H]
\centering
{\setlength
\doublerulesep{1pt}   
 \begin{tabular}{ccccccc}
  \toprule[1pt]\midrule[1pt]
    Observables 	&  AdS/$Sp(4)$ 			& AdS/$Sp(14)$					& AdS/$Sp(4)$		  						& AdS/$Sp(8)$								   \\ 
          				& $5 A_2,6F$  	 		& $5 A_2, 6F$					& $ 4 F, 6A_2$								& $4 F,6 A_2$							  \\ \midrule
    $f_{\pi F}$  	&0.066						&0.0521							&0.057											&0.115										  		\\
    $f_{\pi A_2}$		&0.113						&0.114								&0.12											&0.149											\\
    $M_{V F}$ 		&0.618						&0.364								&0.61											&0.913										   		\\
    $f_{V F}$ 		&0.304 						&0.229								&0.27											&0.518									   		\\
    $M_{V A_2}$		&1*   							&1*									&1*												&1*												     			\\
    $f_{V A_2}$			&0.494   					&0.851								&0.52											&0.683									 			\\
    $M_{A F}$ 		&0.862 						&0.414								&0.938											&1.13											\\
    $f_{A F}$ 		&0.316   					&0.219								&0.303										&0.507											\\
    $M_{A A_2}$		&1.4	 						&1.02								&1.35											&1.12										\\
    $f_{A A_2}$			&0.507  					&0.843								&0.52											&0.665											\\
    $M_{S F}$		&0.348						&0.000476							&0.33											&0.508									\\
    $M_{S A_2}$		&0.376						&0.000296							&0.38											&0.511										\\
    $M_{B F}$ 		&1.15 						&0.639								&1.13											&1.68												\\
    $M_{B A_2}$		&1.85  						&1.48								&1.85											&1.84											\\
    \midrule[1pt]\bottomrule[1pt]
  \end{tabular} 
  }  
  \caption{\label{tab: results_sp_models} Holographic results for
    masses and decay constants in the $Sp(2 N_c)$ theories with
    two different  matter representations that can trigger chiral symmetry breaking.}
\end{table}

\newpage

In the case of $SO(N_c)$ gauge theories, there is a discrete set of
cases with two quark representations that generate both the SM Higgs
and the top partners in  \cite{Ferretti:2013kya}. The models have
matter in the fundamental and spinor representations, and $N_c$ and
$N_f$ are fixed. Those theories within this set that break
chiral symmetry are 

\begin{table}[H]
\begin{center}
\begin{tabular}{|c|c|c|c|} \centering
$SO(7)~ 5 F,6s$ & $SO(7)~ 5 s, 6F$ & $SO(9)~ 5 F,6s$  & $SO(9)~ 5   s,6F$  \end{tabular} 
\end{center}
\end{table}  \vspace{-0.6cm}

\begin{table}[H]
\begin{center}
\begin{tabular}{|c|c|c|c|} \centering
$SO(10)~ 5 F,6s$ & $SO(11)~ 5 F ,6s$ &  $SO(11)~ 4 s,6F$   &   $SO(13)~ 4 s,6F$ \end{tabular} .
\end{center}
\end{table}

\noindent We display holographic results for the masses and decay
constants for these theories in \Cref{tab: results_so_models_1} as
well as in \Cref{tab: results_so_models_2}. Note that the  $SO(13)~ 4 s,6F$
theory has a very light scalar meson, resulting from the slow
running of the coupling.   The
SO(9) theories are of note since the $F$ fields condense at a higher scale than the spinor fields $s$ so the $F$ bound states are heavier than the $s$ counter parts - here the critical coupling \cref{critc} for the $F$ lies lower than that for the $s$ representation. In the other theories shown,  the critical couplings for $F$ is higher \vspace{0.5cm}

\begin{table}[H]
\centering
{\setlength
\doublerulesep{1pt}   
 \begin{tabular}{ccccccc}
  \toprule[1pt]\midrule[1pt]
    Observables 	 			& AdS/$SO(7)$					& AdS/$SO(7)$		  						& AdS/$SO(9)$						& AdS/$SO(9)$		   \\ 
          				  	 		& $5 F,6s$							& $ 5 s, 6F$									& $5 F,6s$								& $5 s ,6F $	  \\ \midrule
    $f_{\pi F}$  				&0.125								&0.132											&0.115									& 0.121	  		\\
    $f_{\pi s}$					&0.126								&0.119											&0.149									& 0.143			\\
    $M_{V F}$ 					&1.08								&1.07											&0.913									& 0.926	   		\\
    $f_{V F}$ 					&0.58								&0.601											&0.518									& 0.55	   		\\
    $M_{V s}$					&1*									&1*												&1*										& 1*			     			\\
    $f_{V i}$						&0.581								&0.555											&0.683									& 0.653	  			\\
    $M_{A F}$ 					&1.39								&1.33											&1.13									& 1.11		\\
    $f_{A F}$ 					&0.578								&0.593											&0.507									& 0.537		\\
    $M_{A s}$					&1.21								&1.25											&1.12									& 1.14			\\
    $f_{A i}$						&0.571								&0.55											&0.665									& 0.636		\\
    $M_{S F}$					&0.744								&0.687											&0.508									& 0.579		\\
    $M_{S s}$					&0.728								&0.725											&0.511									& 0.568		\\
    $M_{B F}$ 					&1.98								&1.98											&1.68									& 1.71			\\
    $M_{B s}$					&1.85								&1.85											&1.84									& 1.84			\\
    \midrule[1pt]\bottomrule[1pt]
  \end{tabular} 
  }  
  \caption{\label{tab: results_so_models_1} Holographic results for
    the masses and decay constants in the two $SO(7)$ and the two $SO(9)$  theories with two matter representations that can trigger chiral symmetry breaking.}
\end{table}

\begin{table}[H]
\centering
{\setlength
\doublerulesep{1pt}   
 \begin{tabular}{ccccccc}
  \toprule[1pt]\midrule[1pt]
    Observables 	&  AdS/$SO(10)$ 		& AdS/$SO(11)$					& AdS/$SO(11)$		  						& AdS/$SO(13)$							   \\ 
          				& $5 F, 6s$  	 			& $5 F,6s$							& $ 4 s, 6F$									& $4 s,6F$									  \\ \midrule
    $f_{\pi F}$  	&0.11						&0.0811								&0.103											&0.0615									  		\\
    $f_{\pi s}$		&0.147						&0.104								&0.156											&0.0878									\\
    $M_{V F}$ 		&0.876						&0.918								&0.753											&0.57										\\
    $f_{V F}$ 		&0.51 						&0.456								&0.468											&0.322									\\
    $M_{V s}$		&1*   							&1*									&1*												&1*												     			\\
    $f_{V i}$			&0.682   					&0.681								&0.727											&0.694									  			\\
    $M_{A F}$ 		&1.06 						&1.05								&0.878											&0.636										\\
    $f_{A F}$ 		&0.5   						&0.432								&0.455											&0.308											\\
    $M_{A s}$		&1.12	 					&1.04								&1.09											&1.03										\\
    $f_{A s}$			&0.664  					&0.666								&0.708											&0.684										\\
    $M_{S F}$		&0.614						&0.142								&0.404											&0.0453								\\
    $M_{S s}$		&0.578						&0.154								&0.44											&0.0615								\\
    $M_{B F}$ 		&1.61 						&1.33								&1.38											&0.884								\\
    $M_{B s}$		&1.83  						&1.46								&1.82											&1.34										\\
    \midrule[1pt]\bottomrule[1pt]
  \end{tabular} 
  }  
  \caption{\label{tab: results_so_models_2} Holographic results for
    the masses and decay constants for the two $SO(10)$, the $SO(11)$
    and the $SO(13)$ theories with two matter representations that can
    trigger chiral symmetry breaking.}
\end{table}


\noindent than that for $s$ and the F sector is then lighter. 
In addition, we find the following four models to lie in the conformal
window and thus do not display chiral symmetry breaking,
\begin{table}[H]
\begin{center}
\begin{tabular}{|c|c|c|c|} \centering
 $SO(13)~ 5 F,6s$ &  $SO(14) ~ 5F, 6s$ &
$SO(15)~ 5 G,6F$  & $SO(55)~ 5 S_2, 6F$  \end{tabular} \, .
\end{center}
\end{table}

	\subsubsection{Models with matter in three representations}
	
The $SU(4)$ model of  \cref{sec: su4_model} falls into this class. In
addition, there are four $SO(N_c)$ gauge theories  in
\cite{Ferretti:2013kya} with specific matter in the $F$, spinor $s$
and the opposite chirality $\bar{s}$ representations (note the
dimensions of the fundamental and the spin are equal for eight
colour). Three of these break chiral symmetry,

\begin{center}
\begin{tabular}{c|c|c}   $\mathrm{SO}(8)~5 F,3 s, 3 \tilde{s}$ &
                                                                   $\mathrm{SO}(10)~5 F, 3 s, 3 \tilde{s} $ &
$\mathrm{SO}(12)~5F ,3 s,3\tilde{s}  $
 \end{tabular} \,  \end{center}

\noindent and one lies in the conformal window, 
\begin{center}
 $\mathrm{SO}(14)~ 5 F,3 s,3 \tilde{s}$.
\end{center}
\vspace{-2mm}
 We analyze the first three in \Cref{tab: results_so8_so10_so12_so14}.
	
\begin{table}[H]
\centering
{\setlength
\doublerulesep{1pt}   
 \begin{tabular}{ccccc}
  \toprule[1pt]\midrule[1pt]
    Observables 									& AdS/$SO(8)$						&AdS/$SO(10)$								& AdS/$SO(12)$ 																	   \\ 
          												& $5 F,3 s, 3 \tilde{s}$				& $ 5 F, 3 s, 3 \tilde{s}$					& $5 F, 3 s, 3 \tilde{s} $		    \\ \midrule
    $f_{\pi F}$  									&0.117									&0.11											& 0.0796 																				 \\
    $f_{\pi s}$  										&0.123									&0.147											& 0.140 																				\\
    $M_{V F}$ 										&1										&0.876											& 0.608 																				\\
    $f_{V F}$ 										&0.553									&0.51											& 0.356 																				\\
    $M_{V s}$ 										&1*										&1*												& 1* 																						 \\
    $f_{V s}$ 										&0.579									&0.682											& 0.732 																				 \\
    $M_{A F}$ 										&1.24									&1.06											&  0.718  																				 \\
    $f_{A F}$ 										&0.547									&0.5												& 0.341 																				 \\
    $M_{A s}$ 										&1.21									&1.12											& 1.06 																					\\
    $f_{A s}$ 										&0.569									&0.664											& 0.732 																				 \\
    $M_{S F}$ 										&0.817									&0.614											& 0.177 																				 \\
    $M_{S s}$ 										&0.817									&0.578											& 0.324 																				 \\
    $M_{B F}$ 										&1.85									&1.61											&1.08  																					 \\
    $M_{B s}$ 										&1.85									&1.83											& 1.68  																					 \\
    \midrule[1pt]\bottomrule[1pt]
  \end{tabular} 
  }  
  \caption{\label{tab: results_so8_so10_so12_so14} Results for the gauge theories with matter in three representations.}
\end{table}

	\subsubsection{Models with QCD-like breaking patterns}
This variety of composite Higgs models has classes with three and four representations. While we could have presented the three representation models in the previous section we chose to separate them in order to follow the classification of \cite{Ferretti:2014qta}. They each have a symmetry breaking sector for one representation where $SU(N_f)_L \times SU(N_f)_R \rightarrow SU(N_f)_V$. 

In addition to the model of \cref{sec: su4_model}, there are two models with three representations,

\begin{center}
\begin{tabular}{c|c}
$SO(10)~ 4 s, 4 \tilde{s}, 6 F  $ & 
$ SU(4)~ 4 F, 4 \bar{F}, 6 A_2$ \end{tabular} \, ,  \end{center}

Both of these models allow chiral symmetry breaking  to occur. We
display our results for the masses and decay constants in these cases in \Cref{tab: results_electroweak_3}.

Moreover, there are the following models  with four representations:  the isolated model 
\begin{equation} SU(7) ~ 4 F, 4 \bar{F}, 3 A_3, 3 \bar{A_3} ,\nonumber \end{equation}
 two  classes  which break chiral symmetry through the full range of $N_c$,
 
 \begin{center}
\begin{tabular}{c|c} 
 $SU(N_c)~ 4 F, 4 \bar{F}, 3A_2, 3 \bar{A_2} ~~N_c \geq 5~~~~$ & 
 $~~~~SU(N_c)~ 3 F, 3 \bar{F}, 4A_2, 4 \bar{A_2}) ~~N_c \geq 5 $ 
  \end{tabular}  \end{center}

\begin{table}[H]
\centering
{\setlength
\doublerulesep{1pt}   
 \begin{tabular}{ccccc}
  \toprule[1pt]\midrule[1pt]
    Observables 							&  AdS/$SO(10)$				& AdS/$SU(4)$ 							    \\ 
          										& $4 s, 4 \tilde{s}, 6 F$  		& $4 F, 4 \bar{F}, 6 A_2$	 \\ \midrule
    $f_{\pi F}$  							& 0.107  							&	0.0922		   \\
    $f_{\pi i}$  								& 0.156  							&	0.122		  \\
    $M_{V F}$ 								& 0.777 			 				&	0.805		   \\
    $f_{V F}$ 								& 0.470   							&	0.424		 \\
    $M_{V s}$ 								& 1*    								&	1* 		    \\
    $f_{V i}$ 								& 0.723    							&	0.540		    \\
    $M_{A F}$ 								& 0.922  							&	1.05		    \\
    $f_{A F}$ 								& 0.455   		 					&	0.427		 \\
    $M_{A i}$ 								& 1.09  								&	1.29		  \\
    $f_{A i}$ 								& 0.704   							&	0.536		  \\
    $M_{S F}$ 								& 0.311 								&	0.494	  \\
    $M_{S i}$ 								& 0.376 							&	0.488		  \\
    $M_{B F}$ 								& 1.42  								&	1.49		 \\
    $M_{B i}$ 								& 1.81   							&	1.85		  \\
    \midrule[1pt]\bottomrule[1pt]
  \end{tabular} 
  }  
  \caption{\label{tab: results_electroweak_3} Holographic results for
    masses and decay constants in the the $SO(10)~ 4 s, 4 \tilde{s}, 6 F$ and $SU(4)~ 4 F, 4 \bar{F}, 6 A_2$ models. $i=s$ for the former and $i=A_2$ for the latter.}
\end{table}

\noindent and two classes that are only outside the conformal window at large $N_c$,
 \begin{equation} \begin{array}{c}
 SU(N_c)~ 4 F, 4 \bar{F}, 3  S_2 ,3 \bar{S_2}  ~~N_c \geq 5  \hspace{0.5cm} {\rm below~ the~ conformal~ window~ for~} N_c>10   \\  
 \left. \right. \hspace{5cm} {\rm and~ break~ chiral ~ symmetry.~}\\ \\
  SU(N_c)~ 3 F, 3 \bar{F}, 4  S_2 ,4 \bar{S_2}  ~~N_c \geq 8  \hspace{0.5cm} {\rm below~ the~ conformal~ window~ for~} N_c>70 \\  
 \left. \right. \hspace{5cm} {\rm and~ break~ chiral ~ symmetry.}
 \end{array}
 \nonumber \end{equation}
 The first theory with $N_c>10$ and the second with $N_c>70$ are
 clearly very hard to reconcile with any phenomenology. For example,
 the S parameter would be expected to be huge.  However, for lower
 $N_c$ values these models are very finely tuned to the conformal
 window. This leads to a very small scalar meson mass.
 For these classes we just present models for the smallest value of
 $N_c$, for which they break chiral symmetries in \Cref{tab: results_electroweak_4}.

\begin{table}[H]
\centering
{\setlength
\doublerulesep{1pt}   
 \begin{tabular}{cccccccc}
  \toprule[1pt]\midrule[1pt]
      					 	& AdS/$SU(5)$  										& AdS/$SU(5)$ 									& AdS/$SU(7)$									& AdS/$SU(10)$									& AdS/$SU(71)$ \\ 
          				 	& $4 F, 4 \bar{F}, 3 A_2, 3 \bar{A_2} $ 	 	& $4 A_2, 4 \bar{A_2}, 3 F, 3 \bar{F}$	& $4 F, 4 \bar{F}, 3 A_3, 3 \bar{A_3} $	& $4 F , 4 \bar{F}, 3 S_2, 3 \bar{S_2}$	& $3 F , 3 \bar{F}, 4 S_2, 4 \bar{S_2}$	\\ \midrule
    $f_{\pi F}$  	  	& 0.0834 												& 0.0598											&0.0803											&0.0469 											&0.0210 \\
    $f_{\pi i}$  		  	& 0.14	  		 										& 0.153												&0.164												&0.0746											&0.0192\\
    $M_{V F}$ 		  	& 0.67	  		 										& 0.486												&0.628												&0.386												&0.627	\\
    $f_{V F}$ 		   	& 0.372  												& 0.251												&0.378												&0.228												&0.395 \\
    $M_{V i}$ 	 		& 1*  		 												& 1*													&1*													&1*													&1* \\
    $f_{V i}$ 		 	& 0.608 		 										& 0.65												&0.82 												&0.726												&1.49 \\
    $M_{A F}$ 			& 0.845   												& 0.661												&0.741												&0.434												&0.63 \\
    $f_{A F}$ 		   	& 0.368  												& 0.25												&0.37												&0.217												&0.394 \\
    $M_{A i}$ 	  		& 1.19 		 											& 1.15												&1.06 												&1.02												&1 \\
    $f_{A i}$ 		 	& 0.59	 												& 0.628												&0.805 												&0.683												&1.48\\
    $M_{S F}$ 			& 0.338 		 										& 0.13												&0.534												&0.000155											&0.000849\\
    $M_{S i}$ 			& 0.399 		 										& 0.273												&0.439												&0.000479											&0.00140\\
    $M_{B F}$ 	  		& 1.24 		 											& 0.897												&1.16												&0.634												&0.643 \\
    $M_{B i}$ 		  	& 1.84 		 											& 1.83												&1.82												&1.3													&0.952\\
    \midrule[1pt]\bottomrule[1pt]
  \end{tabular} 
  }  
  \caption{\label{tab: results_electroweak_4} Results for gauge theories with matter in four representations. $i=A_2$ for the first two models. $i=A_3$ for the next one. For the final two we have $i=S_2$.}
\end{table}

\section{Phenomenological implications and constraints} \label{sec: pheno_implications}

The above analysis of the spectra of possible composite Higgs models has been a purely field theoretic exercise, without taking into account experimental constraints. Now we briefly consider their experimental impact. We immediately 
note than many of the theories have very large field content 
and this is liable to be in conflict with the precision $S$ parameter \cite{Peskin:1991sw} constraints.

We will briefly summarize some generic phenomenological implications for searches at the LHC  based on
the mass hierarchies in the models presented in the previous sections. The following discussion 
neglects contributions to the masses arising from the gauging of
the SM forces,  analogous to the electric mass splitting between the charged and neutral
pions in the SM. These small differences are only likely to play a role in accidentally very fine tuned cases.
We will concentrate on the models  in \cref{sec: sp4 model,sec: su4_model} where there are 
counterparts of explicit models presented in \cite{Belyaev:2016ftv}. In these models the
global group related to the $A_2$ representations contains  the electroweak SM group
whereas the one related to $F$ (and in the case of $SU(4)$ also $\bar F$) contains $SU(3)_c$.
Thus, given the measured Higgs mass is about
125 GeV and the bounds on heavy spin one resonance are well above one TeV \cite{Zyla:2020zbs}, the
 ratio of the pion mass to the vector meson masses for the $A_2$ condensate shown in 
\cref{fig: pion_masses_sp4,fig: specta_deformed_vaccum_su4_full_A2_analytics} is confined
to small values to be consistent with the data.  On the other, the other pions related to the $F$
condensate can be substantially heavier.

Inspecting \Cref{Sp4table} we find, in the case of the $Sp(4)$ model(s), for $m_q=0$ and negligible contributions from HDOs, that: (i) $A_2$ bound states 
are somewhat heavier than the $F$ counterparts; (ii) The scalars are significantly lighter
than the vector and axial vector states. The fermionic bound states are still heavier than the corresponding
vector bound states. The EW loop corrections mentioned above will tentatively reduce the mass splitting
between the corresponding $A_2$ and $F$ bound states. We also recall from \cref{fig: pion_masses_sp4}
that a finite but small hyperquark mass hardly changes the relative mass ratios yielding
the same overall picture. 

One expects that the scalar mesons will dominantly decay into the corresponding pNGB. There
could also be decays into a pair of top quarks arising from the mixing of the top partners with the
top quark. The next heavier states are the vector mesons as can be seen from \Cref{Sp4table}. 
In the searches for  these states it is usually assumed that
the decay dominantly into SM-particles. However, due to the quite large mass difference between
scalars and these vector states  we also expect sizeable branching ratios for the decay
$V_{A_2} \to S_{A_2} \, S_{A_2}$ leading to a final state with four pNGB which will
decay further.  Thus we expect actually an enhancement of the multiplicity of the SM particles
compared to the standard LHC searches. The top partners, which are the $F A_2 F$ states,
should have about the same mass or might
be even be lighter than the vector state due to the requirement of a large top Yukawa coupling
as discussed in \cref{sec: yukawa_sp4}. Thus we expect that in addition to the standard decays such
as 
\begin{align}
T \to t \, \Pi
\end{align}
(where $\Pi$ is one of the pNGB which belong either to the electroweak or to the strong
sector) there may also be sizeable branching ratios into final states
like
 \begin{align}
T \to t \, S\,.
\end{align}
The subsequent decay of the $S$ into two pNGB results, also in this case, to a more complicated final
state compared to the one used in the standard searches by the LHC collaboration.
This affects for example the phenomenology of the models M5 and M8 presented in  \cite{Belyaev:2016ftv}.

Turning now to the $SU(4)$ models we focus on the main differences compared to the $Sp(4)$ ones. Comparing
\Cref{Sp4table,tab: results_su4_lattice_two_runnings} we notice that in the case of $SU(4)$ the masses of the mesons and baryons depend less
on the underlying hyperquarks' mass compared to the $Sp(4)$ models. Secondly, the scalars are significantly
heavier than in the case of $Sp(4)$ implying that they will be less frequently produced at the LHC
in both, direct production and from cascade decays from heavier states, which affects for
example the LHC phenomenology of model M6 in ref.~\cite{Belyaev:2016ftv}. More generically one
finds from these tables and the ones in \cref{app: catalogue_results}, that the ratio $M_S/M_V$ 
is an important quantity to identify possible underlying gauge groups. However, we note for completeness
that this ratio depends to some extent also on the matter content of the underlying theory.

Finally we note that many of the theories presented in \cref{app:
    catalogue_results} have very slow running resulting in very light scalar
resonances. 
We expect that the loop induced effective potential due to explicit
symmetry breaking effects like the gauging of the SM-group will give sizeable contributions
to their masses. This is beyond the realm of this paper and requires a model-by-model investigation
of the spectrum of the light states. In these cases one needs also to check to which extent direct
searches already constrain them. This will also depend on the precise quantum numbers of these light scalars.

\section{Conclusions and discussion} \label{sec: discussion}

In this paper we have adapted the holographic model of \cite{Alho:2013dka} to describe composite Higgs models, including fermionic bound states as in \cite{Abt:2019tas},  as well as multiple representations of matter. Our holographic approach is inspired by string theory realizations of gauge/gravity duality: The holographic gravity action is based on the top-down DBI action for a probe D7-brane. As a novel feature compared to previous holographic composite Higgs models, the spontaneous symmetry breaking is induced by the dynamics of the gravity theory, just as in the stringy top-down models. 
Within these models, in a phenomenological approach we directly impose the running of the quark anomalous dimension. 
We have used the two-loop running of the gauge coupling, extending it naively to the non-perturbative regime,   to predict the running of $\gamma$. The model then predicts the light meson and nucleon spectrum for given numbers of colours and flavours for chosen groups and representations.

We also included higher dimension operators into the holographic model to describe Nambu-Jona-Lasinio-like interaction terms. We have demonstrated this in two-flavour QCD, where we `perfected' the model in the spirit of lattice QCD by including HDOs at the UV scale where the theory transitions to weak coupling. We have shown that the spectrum can be brought closer to the observed spectrum in this way (see \Cref{tab: results_qcd_perfect_v2}).

After grounding the holographic model with the QCD predictions, we then moved to studying the underlying gauge theory dynamics in composite Higgs models. We studied three theories in particular detail that have associated lattice results - $SU(2) ~ 4 F$; $Sp(4) ~ 4F,6A_2$; and $SU(4) ~ 3F, 3\bar{F},5A_2$. 
The results and comparisons to lattice data are in \Cref{tab: results_su2_lattice,Sp4table,,tab: results_su4_lattice_two_runnings}. The holographic techniques describe the lattice data sensibly. This encouraged us to extended the results beyond known lattice results. In particular, we straightforwardly computed a wider range of observables and crucially had the ability to quench, unquench and change the number of flavours of quarks without the troubles of lattice doubling or sign problems. Indeed in section \ref{app: catalogue_results}, we have surveyed the full set of possible gauge theories for composite Higgs models proposed in  \cite{Ferretti:2013kya}. We expect that this will provide a useful resource for model builders.

In a holographic realization of models with `partial compositeness', 
we have also computed the top Yukawa coupling, using HDOs to impose the required mixing between the top quark and top partner baryons in the strong coupling sector. Extending this approach by  including an additional HDO of the form $|F A_2 F|^2$  both reduces the top partners' masses and raises the structure functions sufficiently to allow for a top Yukawa coupling of order $y_t\simeq1$, consistent with the Standard Model.  This value is obtained  for a top partner mass of half the value of the vector meson mass in the strongly coupled sector. For a choice of the strong coupling scale between 1-5 TeV, this is likely to be compatible with current experimental bounds on these states.

The holographic modelling does depend on assumptions about the IR dynamics. We expect to be able to improve the  tuning of the model to the dynamics  as more lattice results become available, similarly to the results for QCD of \cite{Gursoy:2008bu}.  An example of an IR assumption that we have made is the input for the running of the quark anomalous dimension $\gamma$. We have considered models of  \cite{Ferretti:2013kya} where the
 two-loop ansatz for the running of $\gamma$ places the theory in the conformal window. If the IR fixed point value in these theories turns out to be higher, the ansatz for the running could be easily modified.  The holographic techniques also seem likely to remain useful for chiral models that cannot be studied on the lattice easily  - we anticipate a constructive dialogue with future lattice work.

There is also substantial room for future exchanges with model builders. For example, the addition of NJL interactions could turn models in the conformal window into symmetry breaking theories. This is easily studied with the holographic approach presented. Moreover, models with further matter content that are closer to being conformal in the UV  may be of interest \cite{Kim:2020yvr}. In this case, the HDOs dimensions would reduce, such that enhancements of the top Yukawa coupling might be possible. Generically, all of these ideas  are quick to apply using holographic techniques in any such model. We hope holography will become a common-place tool for model builders.

\section*{Acknowledgements}

We would like to thank Raimond Abt, 
Alexander Broll, Tony Gherghetta,  Jong-Wan Lee,  Yang Liu,  Ioannis Matthaiakakis and Theodoros Nakas for useful discussions. 

NEs work was supported by the STFC consolidated grant ST/P000711/1.

\setcounter{equation}{0}

\setcounter{equation}{0}
\appendix
\newpage
\section{A spinor in AdS and AAdS spacetimes} \label{app: A}
\subsection{Five-dimensional AdS spacetime}
The invariant line element of the spacetime we are working described by the set of coordinates $x^M = (r, x^{\mu}),  \mu = 1,\cdots,4$, such that the  AdS$_5$ is represented by the domain $u > 0$ and the metric is equal to
\begin{align}
ds^2 =g_{AB} dx^A dx^B = \frac{dr^2}{r^2} + r^2 \eta_{\mu \nu} dx^{\mu} dx^{\nu}\, , \label{eq: pureadsmetric}
\end{align}
where we have set the AdS length equal to one. The boundary, $\partial \text{AdS}$, is approached as  $r \rightarrow \infty$. Since we want to consider a spinor in the above background, we need to choose a f{\"unf}bein, $e^A_M$, such that it satisfies 
\begin{align}
g_{MN}=e^{I}_M e^{J}_N \eta_{I J}\, ,
\end{align}
with $\eta_{IJ}$ the $5$-dimensional Minkowski metric. 
The f{\"unf}bein and the non-vanishing components of the spin-connection are given by
\begin{align}
\begin{aligned}
e^{I}_{\mu} &= r \delta^{I}_{\mu}\, , \\ 
e^{I}_{\rho} &= r^{-1} \delta^{I}_{\rho}\, , \\
\omega^{r \nu}_{\mu} &= - \omega^{\nu r}_{\mu} = -r \delta^{\nu}_{\mu}\, . \label{eq: pureadsnormalcoor_spinconn}
\end{aligned}
\end{align}
The Dirac operator in AdS$_{5}$ using these conventions is given by 
\begin{align}
\slashed{D}_{\text{AdS}} = r \gamma^{r} \partial_{r} + \frac{1}{r} \gamma^{\mu} \partial_{\mu} +2 \gamma^r\, . \label{eq: diracoperatorddimension}
\end{align}
In the above, the matrix $\gamma^r$ is the higher-dimensional analog of the chirality operator, $\gamma^5$, in 4 dimensions and the $\gamma$ matrices satisfy the Clifford algebra; $\{\gamma^A,\gamma^B \} = 2\eta^{AB}$. 
Consider the free spinor action in an AdS$_5$ space 
\begin{align}
S_{\text{Dirac}} = \int_{\text{AdS}} d^5 x \sqrt{g} \bar{\Psi} (\slashed{D}_{\text{AdS}} - m)\Psi\, ,
\end{align}
from which one obtains the Dirac equation
\begin{align}
(\slashed{D}_{\text{AdS}} -m) \Psi = 0\, , \label{eq: diracads5}
\end{align}
where we have used the shorthand: $\Psi = \Psi(x^{\mu},r) $. 
From the above we can obtain a second-order differential equation for the scalar function of the holographic radial coordinate. The way we choose to perform the analysis consists of acting with the operator $r \gamma^{r} \partial_{r} + \frac{1}{r} \gamma^{\mu} \partial_{\mu}$ on \cref{eq: diracads5} and then by using the first-order equations of motion that the spinor satisfies to re-express some of the terms. After some algebra we obtain 
\begin{align} \label{eq: finalequationdiracads5}
\left( \partial^2_r + \frac{6}{r} \partial_r + \frac{M^2}{r^4} + \frac{1}{r^2} (-m^2-m \gamma^r+6) \right) \psi = 0\, ,
\end{align}
where we have decomposed the five-dimensional spinor in a plane-wave along the Minkowski directions and an ordinary scalar function which we denote by $\psi$. Note that in principle $\psi$ contains the two eigenstates of the chirality operator. The above equation is in agreement with the equation described in \cite{Iqbal:2009fd}, where the interested reader can find an explicit representation of the $\gamma$-matrices, as well as the analytic solutions of the above equation in terms of Bessel functions. In this work we will not be interested in the analytic solutions of the equations of motion, but rather the scaling behaviour near the conformal boundary ($r \rightarrow \infty$). We give the relevant relations below
\begin{align} \label{eq: asymptotics in AdS}
\begin{aligned}
\psi_{+} &\sim c_{1} r^{-2+m} + c_{2} r^{-3-m}\, , \\
\psi_{-} &\sim c_{3} r^{-3+m} + c_{4} r^{-2-m}\, ,
\end{aligned}
\end{align} 
and in the above relations $c_{1,2,3,4}$ are sources and operators - see the next section for their interpretation. This statement is made exact in the main body of the paper, see \cref{sec: QCD}. 
\subsection{Relations between the sources and the operators}
The Dirac equation solution in \cref{eq: asymptotics in AdS} naively has 4 constants of integration as presented  but these are correctly only 2.  The 2 constants are interpreted as the boundary sources and operators. 
To understand this reduction we begin with the first-order Dirac equation in pure AdS. This is written as 
\begin{align}
\left( r \gamma^r \partial_r + \frac{1}{r} \gamma^\mu \partial_\mu + 2 \gamma^r - m \right) \Psi(x^\mu,r) = 0\, .
\end{align}
We decompose the spinor as $\Psi(x^\mu,r) = e^{i k \cdot x} (\psi_{+} \alpha_{+}+\psi_{-} \alpha_{-})$ where the $\alpha_{\pm}$ are eigenstates of the chiral projector $\gamma^r$ and have values $\pm1$. They are related to one another as 
\begin{align}
\alpha_{-} = \frac{i \slashed{k}}{M} \alpha_{+}\, . 
\end{align}
We can substitute the above decomposition into the Dirac equation and obtain a set of first-order coupled differential equations. Taking into consideration that the chiral eigenspinors are linearly independent the resulting equations read 
\begin{align}
\begin{split}
\psi_{+} &= \frac{r}{M} \left( r \partial_r \psi_{-} +(2+m) \psi_{-}   \right)\, , \\
\psi_{-} &= \frac{r}{M} \left( - r \partial_r \psi_{+} -(2-m) \psi_{+}   \right)\, . 
\end{split}
\end{align}
We can plug the asymptotic expansions, see \cref{eq: asymptotics in AdS} in the above expressions and solve  to relate two of the constants to the other two. The final form of the asymptotics reads 
 \begin{align} \label{eq: final asymptotics in AdS}
\begin{split}
\psi_{+} &\sim c_{2} \sqrt{r} + c_{1} \frac{M}{6} r^{-11/2}\, , \\
\psi_{-} &\sim c_{2} \frac{M}{4} \frac{1}{\sqrt{r}} + c_{1} r^{-9/2}\, ,
\end{split}
\end{align} 
and the identification is $c_1 = \mathcal{O}$ and $c_2 = \mathcal{J}$. 
\subsection{The Dynamic AdS/YM spacetime}
We can now repeat this process in the AdS/YM spacetime. The invariant line element is given by
\begin{align}
ds^2 = r^2 dx^2_{(1,3)} + \frac{d \rho^2}{r^2}\, , 
\end{align} 
from which we can readily obtain the 
 f{\"unf}bein, $e^A_M$.
The f{\"unf}bein and the non-vanishing components of the spin-connection are given by
\begin{align}
\begin{aligned}
e^{I}_{\mu} &= \sqrt{\rho^2 + L(\rho)^2} \delta^{I}_{\mu}\, , \\ 
e^{I}_{\rho} &= \frac{1}{ \sqrt{\rho^2 + L(\rho)^2}} \delta^{I}_{\rho}\, , \\
\omega^{r \nu}_{\mu} &= - \omega^{\nu r}_{\mu} = - (\rho+L(\rho) ~ \partial_{\rho} L(\rho)) \delta^{\nu}_{\mu}\, . \label{eq: dynamicadsqcd_vielbeinespin}
\end{aligned}
\end{align}
The Dirac operator in this spacetime which is asymptotically AdS$_{5}$ (AAdS )using these conventions is given by 
\begin{align}
\slashed{D}_{\text{AAdS}} = \sqrt{ \rho^2 + L(\rho)^2 } \gamma^{\rho} \partial_{\rho} + \frac{1}{ \sqrt{ \rho^2 + L(\rho)^2}} \gamma^{\mu} \partial_{\mu} +2 \frac{\rho + L(\rho) ~ \partial_{\rho} L(\rho)}{ \sqrt{ \rho^2 + L(\rho)^2 }} \gamma^{\rho}\, . \label{eq: diracoperatordynamicadsqcd}
\end{align}
The action for a free spinor in the above geometry is given by 
\begin{align}
S_{1/2} = \int d^5 x ~\rho^3~ \bar{\Psi} \left( \slashed{D}_{\text{AAdS}} - m \right) \Psi\, ,
\end{align}
where again we have used the square root of the metirc determinant from the top-down analyses rather than the one obtianed from our spacetime -as was done in the case of the bosonic sector of the theory. 
It is a straightforward task to vary the above action and obtain the equations of motion. Then, we promote the first-order equations of motion to second-order by acting on them with the differential operator $\frac{1}{r} ~ \gamma^{\mu} ~ \partial_{\mu} + r ~  \gamma^{\rho} ~ \partial_{\rho}$, and in such a way we construct a Klein-Gordon problem in terms of an ordinary scalar function of the holographic radial coordinate $\rho$ that reads
\begin{align} \label{eq: dynamicadsqcdfinaleqnofmotion}
\begin{aligned}
&\left( \partial_{\rho}^2 + \frac{6}{r^2} \left( \rho + L~ \partial_{\rho}L \right) \partial_{\rho} + \frac{M^2}{r^4}  - \frac{m^2}{r^2} - \frac{m}{r^3} \left( \rho + L ~ \partial_{\rho}L   \right) ~ \gamma^{\rho} \right. \\  
&\left.+ \frac{2}{r^4} \left( (\rho^2 + L^2) L \partial^2_{\rho} L + (\rho^2 + 3L^2) (\partial_{\rho}L)^2 + 4 \rho L \partial_{\rho}L + 3 \rho^2 + L^2  \right)   
  \right) \psi = 0\, .
\end{aligned}
\end{align}
As a check of the above relation let us consider the limit $L \rightarrow 0$, in which the Dynamic AdS/QCD metric becomes AdS$_{5}$ and in that limit the equations of motion \cref{eq: dynamicadsqcdfinaleqnofmotion} reduces to 
\begin{align}
\left( \partial_{\rho}^2 + \frac{6}{\rho} \partial_{\rho} + \frac{M^2}{\rho^4} + \frac{1}{\rho^2} \left(6 - m^2 -m \gamma^{\rho} \right)    \right) \psi = 0\, ,
\end{align} 
which is precisely \cref{eq: finalequationdiracads5} under the identification $r \rightarrow \rho$.  \newpage

\section{Group theory factors} \label{app: B}
In what follows, by $G$ we denote the adjoint representation. Other representations are displayed by their Young Tableaux. 
This is a matrix of the group theory factors of the $SU(N)$ gauge group.  
\begin{center}
\begin{tabular}{ |c|c|c|c| } 
 \hline
  R  & T(R) & C$_2$(R) & d(R) \\ 
  \hline
  G & $N$ & $N$ & $N^2-1$ \\
  \hline
$ \tiny \yng(1)$ & $1/2$ & $(N^2-1)/2N$ & $N$    \\ 
\hline
 $\tiny \yng(1) \yng(1)$ & $(N+2)/2$ & $(N-1)(N+2)/N$ & $N(N+1)/2$    \\ 
 \hline
 $ \tiny \yng(1,1) $ & $(N-2)/2$ & $(N+1)(N-2)/N$ & $N(N-1)/2$    \\ 
\hline
 $ \tiny \yng(1,1,1) $ 				   & $(N-3)(N-2)/4$& $3(N-3)(N+1)/2N$ & $N(N-1)(N-2)/6$ \\
 \hline
\end{tabular}
\end{center}
This is a matrix of the group theory factors of the $Sp(2N)$ gauge group.  
\begin{center}
\begin{tabular}{ |c|c|c|c| } 
 \hline
  R  & T(R) & C$_2$(R) & d(R) \\ 
  \hline
$ \tiny \yng(1)$ & $1/2$ & $(2N+1)/4$ & $2N$    \\ 
\hline
 G$= \tiny \yng(1) \yng(1)$ & $N+1$ & $N+1$ & $N(2N+1)$    \\ 
 \hline
 $ \tiny \yng(1,1) $ & $N-1$ & $N$ & $N(2N-1)-1$    \\ 
\hline
\end{tabular}
\end{center}
We present here the matrix of the group theory factors of the $SO(N)$ gauge group.  
\begin{center}
\begin{tabular}{ |c|c|c|c| } 
 \hline
  R  & T(R) & C$_2$(R) & d(R) \\ 
  \hline
$ \tiny \yng(1)$ & $1$ & $(N-1)/2$ & $N$    \\ 
\hline
$ \tiny \yng(1) \yng(1)$ & $N+2$ & $N$ & $(N-1)(N+2)/2$    \\ 
 \hline
G $= \tiny \yng(1,1) $ & $N-2$ & $N-2$ & $N(N-1)/2$    \\ 
\hline
spin(N: even) & $2^{\frac{N-8}{2}}$ & $N(N-1)/16$ & $2^{\frac{N-2}{2}}$\\
\hline
spin(N: odd) & $2^{\frac{N-7}{2}}$ & $N(N-1)/16$ & $2^{\frac{N-1}{2}}$\\
\hline
\end{tabular}
\end{center}
Finally, we include the group theory factors for the exceptional gauge theories $G_2$ and $F_4$ which are relevant to composite Higgs models with matter in a single representation of the gauge group. 
\begin{table}[H]  \centering
{\setlength
\doublerulesep{1pt}   
 \begin{tabular}{ccccccc}  
  \toprule[2pt]\midrule[1pt]
    Groups & $d(G)$ & $T(G)$ & $C_{2}(G)$ & $d(F)$ & $T(F)$ & $C_{2}(F)$ \\ \cmidrule{1-1}
          &  \color{white}{ here's the hidden text}   \\ \midrule
    $F_4$ & 52 & 9 & 9 & 26 & 3 & 6  \\
    $G_2$ & 14 & 4 & 4 & 7 & 1 & 2  \\
    \midrule[1pt]\bottomrule[2pt]
  \end{tabular} 
}
\label{tab: Exceptional_table}
\end{table}
In the above tables, $T$ is half the Dynkin index, $C_2$ is the quadratic Casimir, and $d$ is the dimension of the representation. 
The usual relation holds
\begin{align}
C_2(R) d(R) = T(R) d(G)\, ,
\end{align}
where by $R$ we mean a sepcific representation. We also have that 
\begin{equation}
T (R) = - T(\bar{R})\, .
\end{equation}
\newpage
\clearpage
\bibliography{lit}{}
\bibliographystyle{utphys}
\end{document}